\documentclass[graybox,envcountchap]{svmono}

\usepackage{url}
\usepackage{mathptmx}
\usepackage{helvet}
\usepackage{courier}
\usepackage{type1cm}

\usepackage{makeidx}
\usepackage{graphicx}

\usepackage{multicol}
\usepackage[bottom]{footmisc}

\usepackage{booktabs}
\usepackage{amssymb}

\usepackage[numbers,sort&compress]{natbib}

\usepackage[pagebackref=true,bookmarksnumbered=true,bookmarksopenlevel=4,bookmarksopen=true]{hyperref}
\usepackage{xcolor}
\definecolor{dark-red}{rgb}{1,0.15,0.15}
\definecolor{dark-blue}{rgb}{0.15,0.15,1}

\hypersetup{
  colorlinks   = true,
  urlcolor     = dark-red,
  linkcolor    = dark-red,
  citecolor   = dark-blue,
}

\begin{document}

\author{Fabio Crestani\\ Stefano Mizzaro \\ Ivan Scagnetto}
\title{Mobile Information Retrieval}

\date{}

\maketitle

\frontmatter

\tableofcontents

\mainmatter

\chapter{Introduction}
\label{cha:introduction}

This chapter introduces, in Section~\ref{sec:mobile-inform-retr}, the area of research and development of Mobile Information Retrieval. In Section~\ref{cha:mm} it gives the motivation for this review and explains briefly the methodology used to select the work presented. Finally, in Section~\ref{cha:outline} outlines the organisation of the book.

\section{Information Retrieval with Mobile Devices}
\label{sec:mobile-inform-retr}

Mobile Information Retrieval (Mobile IR) is a relatively recent branch of Information Retrieval (IR) that is concerned with enabling users to carry out, using a mobile device, all the classical IR operations that they were used to carry out on a desktop.
This includes finding content available on  local repositories or on the web in response to a user query, interacting with the system in an explicit or implicit way,  reformulate the query and/or visualise the content of the retrieved documents, as well as
providing relevance judgments to improve the retrieval process.
Examples of mobile devices include mobile phones, smartphones, Personal Digital Assistants (PDAs), smartwatches, and other devices that are connected to the Internet and that do not have a standard-size screen or a keyboard (e.g., optical head-mounted displays, in-car Internet, etc.).
The proliferation of such devices has created a large demand for mobile information content as well as for effective mobile IR techniques that are needed for representing, modelling, indexing, retrieving, and presenting such content that can enable a user to interact with the device to better specify his information need and review the retrieved content.

Mobile IR has increased in popularity immensely in the last ten years.
In fact, since 2015 there are more internet users accessing information through mobile devices than PCs.\footnote{\url{https://adwords.googleblog.com/2015/05/building-for-next-moment.html}} On average, people around the world spent in 2016 about 86 minutes a day using the internet on their phones, compared to 36 minutes on a desktop.\footnote{\url{http://digiday.com/}}
This has brought large attention to how users look for information on mobile devices and perform the classic IR tasks.
In addition, many researchers have started looking at how some characteristics of mobile devices, such as their continued access to Internet, their nature of personal devices, and their sensors capabilities could be used to enhance  IR functionalities.
This resulted in a proliferation of research and development results that this book aims to review to provide the interested reader and researcher with a useful roadmap to be able to extricate himself from the complexity of the field.

\section{Motivations and Methodology}
\label{cha:mm}

Mobile IR is a very fast and emerging area but has a very scattered literature. Many Mobile IR papers appeared outside the classic IR publication venues as until very recently they were even considered outside mainstream IR. Many contributions presented in this book appeared in venues that are associated with other areas of research like Ubiquitous Computing, World Wide Web, Computer-Human Interaction, Intelligent Systems or general Mobile Technology. Only occasionally papers on Mobile IR  appeared in classic IR conferences, most often as short papers or posters. Also, very often  recent results of Mobile IR research could be found outside the scientific literature, in Web page and blogs. This is because work on Mobile IR  was not considered as proper research work by the large part of the IR community. Only in recent years  work on Mobile IR started gaining the attention of researchers, through work presented at workshops (e.g. the MUIA~\cite{Crestani2004}, the IMMoA and the MobileHCI series or SWIRL~\cite{Allan2012}) or in journal special issues (e.g.~\cite{Crestani2006a,Tsai2010}). This motivated us to preparing this book that presents the state-of-the-art of this research area.

The work presented here reflects our own view of the area and our methodological perspective. We take full responsibility of this skewed view. In particular, it gives great emphasis to both user interaction and to techniques for the perception and use of context that, in our view, together shaped much of current research in Mobile IR. This reflects also the choice of papers that have been included in this book. In fact, we are sure that much work that could be considered Mobile IR by other researchers have been left out. This was not an oversight by our part, but was the result of a methodological decision and a reflection of our own view of this area of research. Section~\ref{sec:conclusions} indicates what we acknowledge to have left out and that, with more time and space, could have been included. We leave this to the researchers that will follow us and that will take up the challenge.

\section{Outline}
\label{cha:outline}

This book is structured as follows.
Chapter~\ref{cha:mobile-ir-landscape} provides a very brief overview of IR and of Mobile IR, briefly outlining what in Mobile IR is different from IR.
Of course this chapter can only outline these differences as they will be further explained and exemplified in following chapters.
Chapter~\ref{cha:foundations} provides the foundations of Mobile IR, looking at the characteristics of mobile devices and what they bring to IR, but also looking at how the concept of relevance changed from standard IR to Mobile IR.
Chapter~\ref{cha:docum-coll} presents an overview of the
 document collections that are searchable by a Mobile IR system, and that are somehow different from classical IR ones;
available for experimentation, including collections of data that have become complementary to Mobile IR,
including, for example, collection of movement data, points-of-interest (POI), mobile Apps, etc.
Similarly, Chapter~\ref{cha:user-needs} reviews mobile information needs studies and users log analysis, with the aim to provide a detailed analysis of how the information needs of Mobile IR users are different from  classical IR needs.
The chapter provides many literature pointers, hopefully useful for  researchers that are approaching this area of research.
Chapter~\ref{cha:user-interface} reviews studies aimed at adapting and improving the users interface to the needs of Mobile IR.
Chapter~\ref{cha:context-awareness}, instead, reviews work on context awareness, which studies the many aspects of the user context that Mobile IR employs. As it will be clear,  context-awareness is a fundamental notion for Mobile IR that could only be briefly reviewed in that chapter.
Chapter~\ref{cha:evaluation} reviews some of evaluation work done in Mobile IR, highlighting the distinctions with classical IR from the perspectives of two main IR evaluation methodologies: users studies and test collections.
Finally, Chapter~\ref{cha:conclusions-outlook} reports the conclusions of this review, highlighting briefly some trends in Mobile IR that we believe will drive research in the next few years.

\chapter{From IR to Mobile IR}
\label{cha:mobile-ir-landscape}

Mobile Information Retrieval is a new area of research within the general area of Information Retrieval. This chapter highlights how Mobile Information Retrieval was born out of the mobile phone revolution and explains the similarities. Also, after a very brief overview of Information Retrieval (Section~\ref{sec:IR}), it addresses the main differences between standard Information Retrieval and Mobile Information Retrieval (Section~\ref{sec:MobileIRvsIR}).
Finally, in Section~\ref{sec:mir-landscape}, it provides an overview of the Mobile Information Retrieval literature landscape, highlighting the directions that will be further explored in following chapters.

\section{Information Retrieval}
\label{sec:IR}

This section briefly introduces Information Retrieval, the area of research that is at the base of this book. It briefly summarises its main differences with Database technology and looks at issues related to evaluation and interactivity that will reappear again and again in this book.

\subsection{Brief Introduction to Information Retrieval}

\emph{Information Retrieval} (IR) is the branch of computing science that aims at storing and enabling fast retrieval to a large amount of textual or multimedia information, such as for example text, images, speech, etc.
considered relevant by a user~\cite{Rijsbergen1979}.
The objects handled by an IR application are usually called \emph{documents}, and the software tool which automatically manages these documents is called \emph{Information Retrieval System} (IRS).
The task of an IR system is to help a user to find, in a collection of documents, those that which contain the information the user is looking for, or in other words, providing help in satisfying the \emph{user's information need}.

IR is an established technology that has been delivering solutions to users for more than 40 years and yet it is still an active area of research. This suggests that although much work has been done, much remains to be accomplished.
In fact, in over 40 years, researchers in IR have developed and evaluated a bewildering array of techniques for indexing and retrieving text. These techniques have slowly matured and improved through a large number of small refinement rather than there having been any significant breakthrough.

\begin{table}
\centering
\caption{Differences between IR and DB technology.}
\begin{tabular}{l l l}
\toprule
\hspace{3cm}                 & Information Retrieval \hspace{1cm} & Database \\
\midrule
Matching & Partial match & Exact match \\ Inference & Induction & Deduction \\
Model & Probabilistic & Deterministic \\ Classification & Polythetic &
Monothetic \\ Query Language & Natural & Artificial \\ Query
Specification & Incomplete & Complete \\ Items Wanted & Relevant &
Matching \\ Error Response & Insensitive & Sensitive \\
\bottomrule
\end{tabular}
\label{tab:differences}
\end{table}

Frequently people not familiar with computer science confuse IR with database (DB) technology.
Table~\ref{tab:differences} summarises some of the major differences between the two fields.
These differences have been identified and described in a book that is a classic of IR~\citet[p.
2]{Rijsbergen1979}.
Without going through the full table, the most fundamental difference between IR and DB is that IR needs to retrieve documents that are \emph{relevant} to a user information need expressed by a query, rather than just matching the query, as for DB  systems.
The difference is fundamental, as relevance is a very complex concept that goes beyond the simple matching between some query terms and terms found in the document (see Section~\ref{sec:relevance} for a more complete discussion of the implications).
One of the consequences of this is that the matching between a user's information need and the content of a document is considered uncertain and the IR system can only try to estimate it, leaving to the user the final decision about its actual relevance.  This means that the IR system retrieves documents (mainly) in a probabilistic way (or using some other way to handle the uncertainty of the matching), while a DB system retrieves documents (mainly) in a deterministic way.\footnote{The use of the word ``mainly'' here indicates that the boundary between the two areas is becoming fuzzy, with some IR systems working in a more deterministic way and DB systems working in a more probabilistic or ``uncertainty conscious'' fashion.}
Thus, an IRS retrieves documents that are likely to be considered relevant by the user, that is, likely to satisfy the user's information need.
In other words, DB facts retrieved in response to a query are normally considered to be a true and complete answer to the query, while in IR the perceived relevance of a document varies dramatically across users, and even for the same user at different times.
Of course this characteristic of IR has some consequences.
First, user's queries submitted to an IR system are usually more vague.
They are usually in the form: ``I want documents about xyz.''.
Instead, users of a DB want facts, such as: ``I want the price of the product abc''.
Second, the evaluation of an IRS is related to its ``utility'', that is, how helpful the system is to a user.
This is clearly not a well specified measure.
DB  systems, on the other hand, are evaluated in accordance with well specified and standardised performance measures.

\begin{figure}[tbp]
  \centering
  \includegraphics[width=10cm]{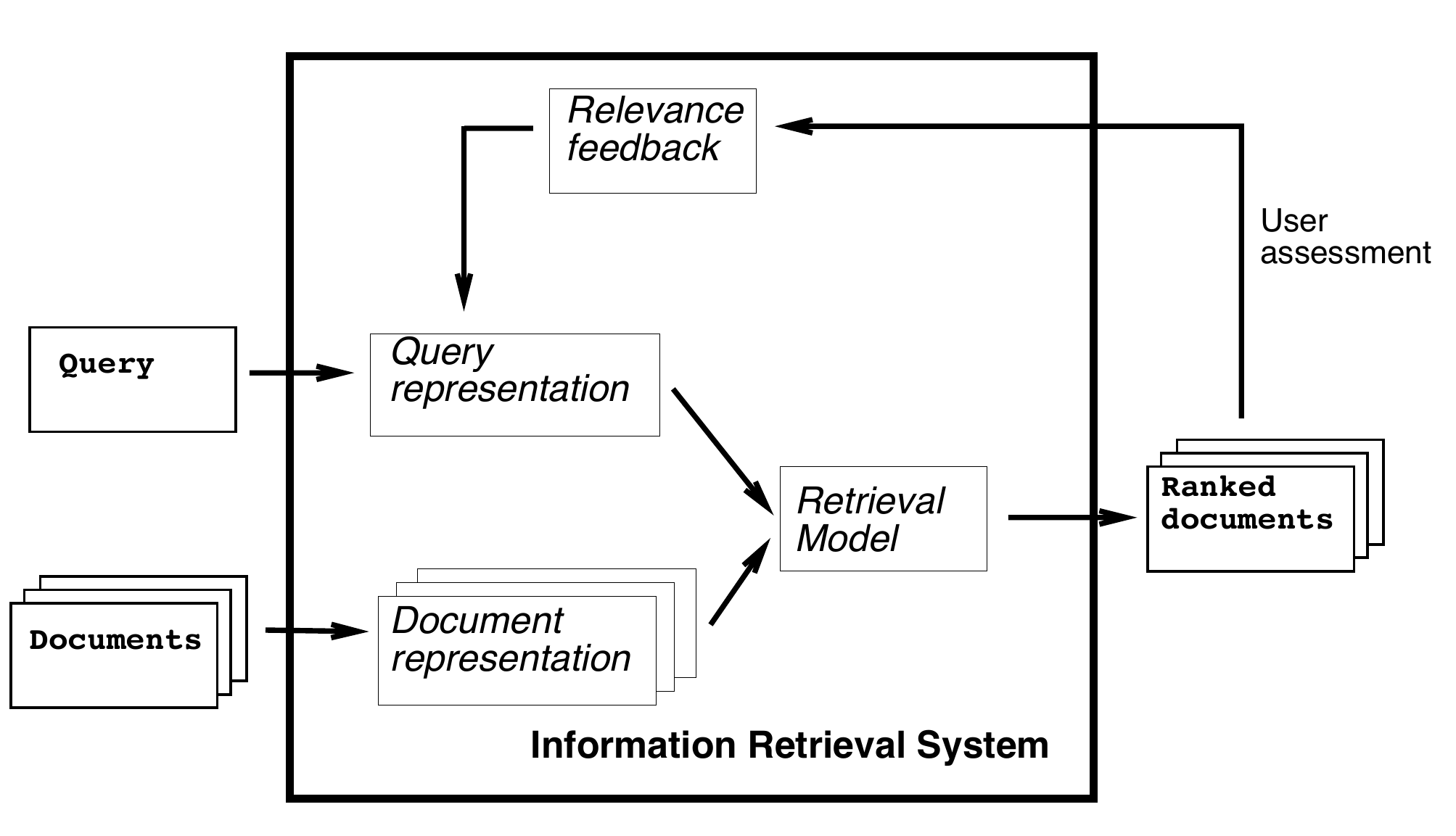}
  \caption{A schematic view of a classical Information Retrieval
        system.}
  \label{fig:ir-view}
\end{figure}

In the classic approach to IR, a schematic view of which is presented in Figure~\ref{fig:ir-view}, one of the main problems is the representation of the documents' informative content.
This is normally tackled by assigning descriptors, most often called \emph{index terms}, to the document.
This process is called \emph{indexing}, and it is done almost invariably automatically.
The representation of the document informative content is one of the most important problems of IR and much efforts is being spent to develop better representations and better forms of indexing.
Although in the early stages of IR indexing was mostly a simple lexical analysis with term weighting, most IR systems now use  more complex indexing techniques.
The most recent IR systems, in particular, use some very complex indexing techniques employing a large number of features related to the content of the document and to its \emph{context}.
Here for context we mean, generically, additional information about the document, like for example the authors' names, the date in which the document was created, its location within the collection, the list of documents that point at it, etc.
Each  system has its own formulation of context that is often kept confidential.

Once a suitable representation structure has been provided, an IRS faces the problem of evaluating the similarity between query and document representations.
This is achieved by a \emph{retrieval model} that uses features of the content of documents and query and, sometimes, information from the user's context to evaluate an overall degree of relevance of each document to the query~\cite{Harman1992}.
Retrieval models are a very important topic of research in IR and we cannot even start to treat them although superficially here.
There is a number of very good books~\cite{Manning2008} and articles~\cite{Crestani1998} dealing with the topic and we can only cite the few of them that we consider more representative.

Finally, one cannot leave the topic of IR without a reference to the \emph{World Wide Web} (commonly referred to WWW or the Web) and Web search engines.
The Web is a global information system that provides hypertextual access to resources on the Internet. This is achieved via a common syntax of addressing network resources, a common protocol for the transfer of data from a Web server to a Web client, and a mark-up language for writing web pages.
The Web client, like for example Firefox, Explorer or Chrome, is responsible for communicating with servers to retrieve the necessary web documents.
The Web server is responsible for making local documents available to other software systems.

The number of available sources of information on the Web has constantly increased over the last few years: for example, during 1993 the number of Web sites  passed from 50 to 500, but by the end of March 2014 the number of indexed Web pages reached over 4.76 billion.\footnote{See \url{http://www.worldwidewebsize.com/} for an updated estimate.}

Web search engines are tools that enable a user to find Web pages that satisfy a user query.
Web search engines use technology developed in  IR  that has been extended to take into account the peculiarities of the Web and Web pages~\cite{CMS09}.  This is quite a consolidated technology now, like in the case of most large commercial search engines (e.g., Google, Bing, and Yandex).

From the point of view of the work reported in this book we will assume no difference between a standard IR system and a Web search engine.
Both can be at the ``answering'' end of a Mobile Information Retrieval system.
In fact, in both cases they will just provide a set of results in response to user query, being them found in a repository or on the Web.

\subsection{Information Retrieval Evaluation}

One important area of research  worth mentioning in this brief review of IR is concerned with the \emph{evaluation of IR systems}.  Much effort and research has gone into studying the problem of evaluation in IR.
Following the approach proposed by \citet{Rijsbergen1979}, in trying to evaluate an IRS one has to try to answer at least two questions:

\begin{enumerate}

  \item What to evaluate?

  \item How to evaluate?

\end{enumerate}

In response to the first question one might note that a researcher could be interested in evaluating the speed of the retrieval process, or the level of interaction an IR system allows, for example.
There are various aspects of IR that a researcher could be interested in evaluating, as we will see many times in this book.
However, the main feature we will consider here are related to the \emph{effectiveness} of the retrieval process.

The second question instead needs a more technical answer. We will see in this book that there are many ways to evaluate an IR system, in particular if the system is designed for a specific type of interaction and for specific types of documents.
However, the two best known measures of effectiveness used in general in IR are \emph{recall} and \emph{precision}.
They are by far the measures of effectiveness most commonly used in IR literature, with many other measures derived from them.

\begin{table}
\centering
\caption{Derivation of precision and recall evaluation values.}
\begin{tabular}{|l|c|c|c|} \hline
documents:       & \emph{relevant} & \emph{not relevant} & \\ \hline
\emph{retrieved}  & $A \cap B$    & $\neg A \cap B$       & $B$ \\
\hline \emph{not retrieved}& $A \cap \neg B$ & $\neg A \cap \neg B$ &
$\neg B$\\ \hline                & $A$ & $\neg A$                & \\
\hline
\end{tabular}
\label{tab:precision-recall}
\end{table}

In order to have clear the meaning of  recall and precision, their definition, as described in~\citet{Rijsbergen1979}, is here reported. It is helpful to refer to Table~\ref{tab:precision-recall}, from which recall and precision can easily be derived.
They are defined as:
\[
Precision = \frac{ \mid A \cap B \mid }{ \mid B \mid }
\]
\[
Recall = \frac{ \mid A \cap B \mid }{ \mid A \mid }
\]
\noindent where $ \mid A \cap B \mid $ is the number of documents that are both relevant and retrieved, $ \mid B \mid $ is the number of retrieved documents, and $ \mid A \mid $ is the number of relevant documents.

The evaluation of precision and recall  is only possible if one has complete knowledge of the relevant documents present in the collection.  This is not possible in most experimental cases.
Therefore, in order to enable an evaluation of Ian experimental R systems a considerable amount of resources have been spent in building \emph{test collections}~\cite{Sparck-Jones1976}.
These are collections of textual documents that come with a set of queries and lists of documents in the collection that are known to be relevant to the queries.
The availability of these relevance judgements enables the evaluation of precision and recall values for diverse indexing and retrieval strategies in a controlled environment.
This makes it possible to compare the results of different IR techniques and draw conclusions that can be extended to experimental or operative cases.

\begin{figure}[tbp]
  \centering
  \includegraphics[width=8cm]{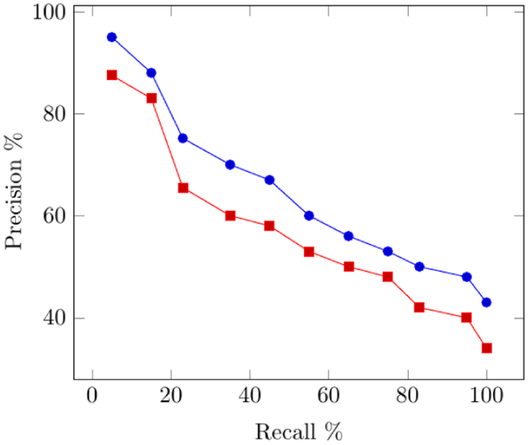}
  \caption{An example of a recall-precision graph comparing two different methods.}
  \label{fig:rec-prec-graph}
\end{figure}

In the evaluation of an IR system, precision and recall values need to be evaluated for every query submitted to the system.
However, the evaluation of these values depends on cut-off points in the ranked list of documents retrieved in response to the query.
Therefore, a better way of displaying these measures is through a {\em recall-precision graph}.
An example of such a graph is depicted in Figure \ref{fig:rec-prec-graph}, where precision values are reported corresponding to standard values of recall (e.g., 10\%, 20\%, etc).

To measure the overall performance of the system on a set of queries, it is necessary to produce as many graphs as the number of available queries and then combine them in some way.
This is often done using the ``macro-evaluation'' approach, which consists in averaging the individual precision values corresponding to the standard recall values over all queries.
For a more in depth explanation of IR evaluation techniques see~\citet{Rijsbergen1979}.

\subsection{Interactive Information Retrieval}

The previous presentation of IR hides one important aspect that has been prevalent in most IR systems designed in the last few years: the importance of \emph{interactivity}. Interactive IR (IIR, for short) systems are, according to the accepted but simplistic definition reported in~\citet{Kelly2009},  ``systems performing IR with users''. In fact, IR has changed over the last few years, becoming more and more interactive, that is enabling the user to formulate his query from the information need through a process of inspecting search results and reformulating the query in an incremental manner~\cite{Robins2000}.  This should enable the system to better understand what the user wants, incrementally improving the results presented until the user is finally happy with them.   The consequences of moving from IR to IIR affects all areas of research in IR, like for example interfaces (e.g., \citet{Lingnau2010}), modelling (e.g., \citet{Ingwersen2005}), and evaluation (e.g., \citet{Kelly2009}).  The incorporation of users into an IR system has been identified as an important concern for IR researchers, in particular for the study of users' information search behaviours
and interactions~\cite{Borlund2003}.

\section{Mobile Information Retrieval}
\label{sec:MobileIRvsIR}

\emph{Mobile Information Retrieval} (henceforth Mobile IR, for short) can be simply defined as IIR on mobile devices~\cite{Church2007}. However, this definition is clearly very simplistic as it hides the enormous complexity of the task.
In fact, while the general IR task of retrieving documents in response to a user query can be considered roughly the same, Mobile IR can be considered from a number of different perspectives that show how the differences between IR and Mobile IR are learger than they might appear.

An obvious difference is given by the  specifics of the device employed to submit the query and view retrieval results. This difference goes well beyond the fact that the keyboard is small (but the microphone is often rather good), the screen is also small (but it is often a touchscreen), internet access is costly (but the cost is very rapidly decreasing), and the documents the user is viewing are rather different from standard web documents (but this difference is becoming almost negligible). The main difference is related, in our view, to the kind of information needs the user tries to satisfy with a mobile search. We will return to this difference again and again in the context of this book. For now it suffices to say that it has been shown that Mobile IR users often have very different information needs from users of standard IR systems. In fact, the type of documents a Mobile IR user is interested in viewing on a mobile device in response to a query is often quite different from the type of documents a user of an IR system is interested in. Equally importantly, a Mobile IR user often does not have the time or the inclination to submit a complex query to retrieve such documents. Thus the Mobile IR system has to use additional information to that provided by the query to better understand the user information need. This might include location information, temporal information or any additional kind of information the Mobile IR system has available through the increasing number of sensors the mobile device is equipped with. Of course, the use of such information that we  call ``contextual'' for the time being (this concept will be better explained later on) requires indexing and retrieval models that are more complex than those used in standard IR. These models have to gather and use at retrieval time this additional information whose combination with the original user query requires special user and context models. These models are more complex than standard models because they have to be able to deal with the uncertainty this information adds to the original user query.  The uncertainty  is originated only partially from the different sensors capturing the information, that are less accurate for a number of reasons (e.g., size, cost). It is also due to the complexity of combining properly their readings to better clarify the user information need.

This brings us to the difference in the concept of relevance that a Mobile IR user has with respect to an IR user~\cite{Coppola2004}.
This difference is very important and substantial.
In fact, it motivates the differences in information needs, queries, documents to be presented to the users, evaluation methodology, and other.
It is also rapidly widening, with Mobile IR growing much faster than IR.
The foundations of these differences are analysed in Chapter~\ref{cha:foundations}, and further scrutinised in Chapters~\ref{cha:docum-coll} and~\ref{cha:user-needs}, while 
Chapters~\ref{cha:user-interface} and~\ref{cha:context-awareness}  explain how they can be exploited to provide better services to Mobile IR users. Finally,  Chapter~\ref{cha:evaluation} addresses the consequences of these differences on Mobile IR evaluation.

\section{The Literature Landscape of Mobile IR}
\label{sec:mir-landscape}

To conclude this chapter, we  provide a general view of the Mobile IR field. The Mobile IR area is growing very rapidly and by the time this book will be published the filed will have expanded even more. So, this book is not meant to be an up-to-date review of the area of research, but an advanced introduction and an overview of the Mobile IR literature that is meant to help the reader to find his path in the area. This chapter will be an abstract overview, partially to present a road map to what is to follow in the book and partially also to justify our editorial choices of what is included in it.

We decided to present this overview in the form of tagclouds.
A \emph{tag cloud} (also known as word cloud) is a visual representation for text data, typically used to depict keyword metadata (tags) or to visualise free form text.
Tags are usually single words, and the importance of each tag is shown with varying font size so that the larger the font size, the more important the tag is.
The clouds give greater prominence to words that appear more frequently in the source text.
This format is useful for quickly perceiving in a visual way the most prominent terms in a collection.
While this was difficult until just a few years ago, if not almost impossible for very large collections, nowadays there are tools that make it easier to do, once a set of keywords with different terms frequencies are provided.
Among the different tools available we decided to use \emph{Wordle},\footnote{\url{http://www.wordle.net/}} a tool for generating ``word clouds'' from text.

\begin{figure}[tbp]
  \centering
  \includegraphics[width=\textwidth]{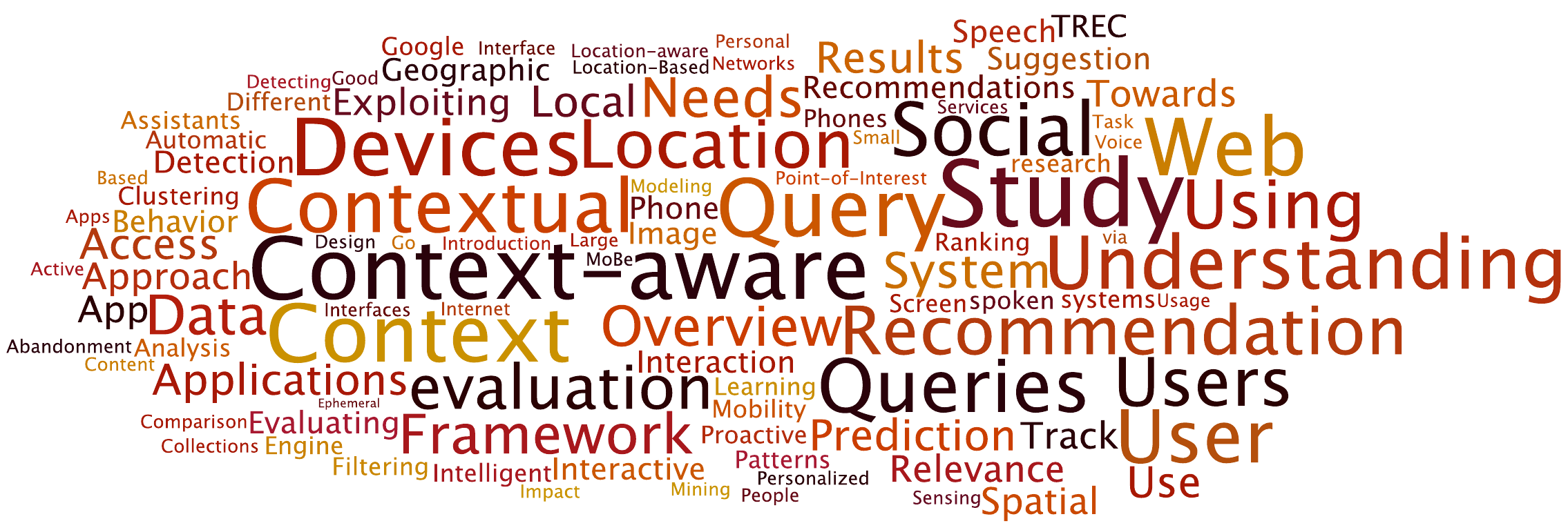}
  \caption{Tagcloud of paper titles (without ``mobile'', ``information'', ``retrieval'', and ``search'').}
  \label{fig:tagcloud-titles}
\end{figure}

Figure~\ref{fig:tagcloud-titles}  reports the most common terms used in the Mobile IR literature. We extracted these terms from the titles of the papers and books about Mobile IR included in the bibliography at the end of this book. Only the top 100 most frequent terms were included, removing some very common  words, like ``mobile'', ``information'', ``retrieval'', etc. These words were removed not because they were considered uninformative, but because they were far too frequent in the titles and not really representative of topics dealt with in this book. As you can see some  terms stand out. For example, the terms ``devices'', ``understanding'', ``user(s)'', ``needs'', ``study'', ``evaluation'', ``use/using'', all refer to what we consider the foundations of Mobile IR. To this topic we devoted
Chapters~\ref{cha:foundations} and~\ref{cha:user-needs}. In addition, given its importance to the field, we have devoted a full chapter to evaluation and user studies:
Chapter~\ref{cha:evaluation}.
Although the terms ``documents'' and ``collection'' seems to have only a marginal importance, we decided to devote to the new kinds of collections and documents of Mobile IR the whole of Chapter~\ref{cha:docum-coll}.

The terms ``devices'', ``query/queries'', ``interface(s)'', ``interactive/interaction'', ``system/systems'' are all related to the interface between the user and the system and the study of the interaction between the two. 
Again, this can be considered part of the foundations, but it is certainly most specifically directed to Mobile IR and not to the general field of interactive IR. 
We have devoted Chapter~\ref{cha:user-interface} to this topic. 
The terms ``context-aware'' and ``context'' or ``contextual'', express clearly the prominence of this concept in Mobile IR. 
Closely related are the terms ``local'', ``location'' and ``spatial''. 
This brings us deeper into the Mobile IR field, taking advantage of the user context and using it to best retrieve documents relevant to him within this context. 
We have devoted Chapter~\ref{cha:context-awareness} to this important topic. 
Finally most other terms are related to specific applications of Mobile IR, like ``clustering'', ``filtering'', and ``recommendation'' (and the term ``application'' itself). 
These will basically occur through the whole book.

\begin{figure}[tbp]
  \centering
  \includegraphics[width=\textwidth]{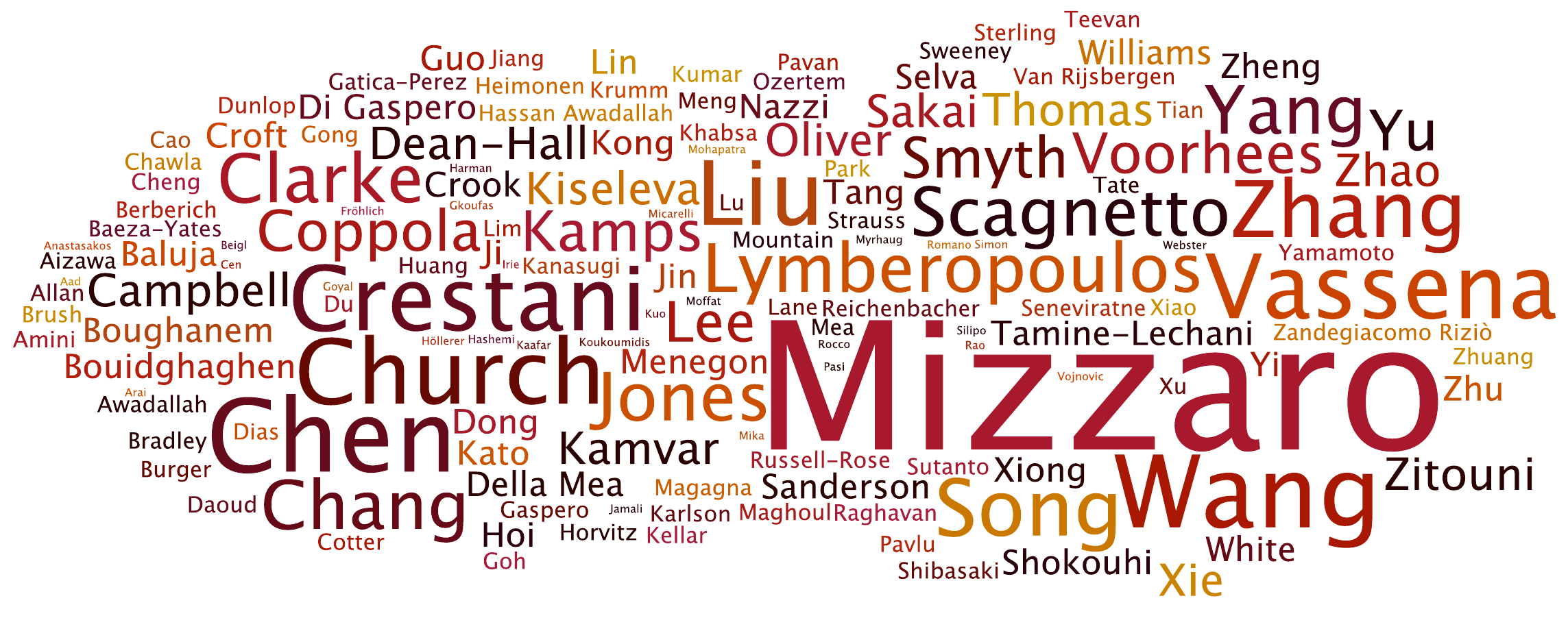}
  \caption{Tagcloud of author names.}
  \label{fig:tagcloud-authors}
\end{figure}

The second tag cloud, reported in Figure~\ref{fig:tagcloud-authors}, instead presents an overview of what we considered the 100 most prolific authors in this field.\footnote{We prefer to use the term ``author'', rather the researchers, because we know of many other researchers in this area that for different reasons are not concerned very much with publishing. These are researchers working at companies for whom the most important output is not a paper, but a product or perhaps a patent. This book covers only partially their work.} The surnames were taken from the list of references at the end of this book. Leaving aside the three authors of this book and their co-authors (that are among the most prominent authors in the tag cloud for obvious reasons), other important surnames are Church,  Wang, Chen, and Olivier. The reasons these authors are the most cited in this book will become apparent as the reader progress through the book. In brief, these authors have been working on Mobile IR for a long time and have produced a large part of the relevant literature.

Of course both tag clouds provide a  personal view of the most important terms  and the most prominent authors in the field: that of the three authors of this book. We hope nobody will be offended by this choice and, of course, we take full responsibility for this and for any other possible error and/or omission.

\section{Conclusions}
\label{sec:conclusions-2}

This chapter provided a very brief introduction to IR, IIR and Mobile IR. It also highlighted the differences and the commonalities between IR and DB and between IR and Mobile IR. It finally provided a graphical overview of the topics dealt in the book and of the names of the authors cited throughout the book and in the list of references. We expect this to be useful to the reader to address the range of topics and depth of our review that we did not intend to be complete or exhaustive, but only an detailed overview of this exciting area of research.

\chapter{Foundations}
\label{cha:foundations}

Mobile IR can be quite easily mistaken as ``squeezing IR on a small screen''. Actually the situation is more complex, and there are some conceptual and foundational aspects that deserve to be properly addressed.  In this chapter we start with more general and introductory material about the history of mobile devices, their technological development, and some social aspects in Sections~\ref{sec:device} and~\ref{sec:role-technology}. Then we turn to more technical details.  Section~\ref{sec:relevance} describes the studies on how the concept of relevance changes in the mobile world.  Section~\ref{sec:MobileIR-models} presents the research on IR models adapted to the new scenario.

\section{The Mobile Phone Device}
\label{sec:device}

The history of mobile devices, although rather short, is probably too rich to be analyzed in detail here, and it is well described elsewhere, including detailed Wikipedia pages. We briefly summarize the most important concepts only, and focus on what will be useful in the following.

The first mobile phones were demonstrated in the 70s and released in the 80s, but a large scale adoption started in the 90s.  The first models suffered from several limitations: very small and low resolution screens, cumbersome input/output capabilities, low bandwidth, minimal computational power, and short battery life. This has constantly changed, with increasing higher resolution screens, more convenient input/output by means of better hardware and software keyboards and touch screens, wider bandwidth and capability to connect to Wi-Fi networks, more computational power, and improved battery life.

Of course the history is not without unsuccessful attempts; for example, the first attempts of web enabled phones, based on WAP (Wireless Application Protocol), have had a rather short life. And in the first years it was clear that the technological limitations were just temporary. For these reasons, several early experiments in Mobile IR, and in mobile computing in general, have been performed using PDA (Personal Digital Assistants), that did not suffer from the limitations of earlier mobile phones --- we will see examples in the following.

Indeed, the technological development has been impressive: today the screens have a resolution higher than the resolution power of the human eye; the sequence of network protocols with increasing bandwidth (GPRS, EDGE, 3G, 4G, LTE to name a few), allows extremely high data transfer rate when the network coverage is adequate; battery life is rarely an issue; and, concerning computational power, as it is often stated, today everyone can have in his own pocket more computational power than what could fit in a room just some decades ago.

To focus on the recent ten years or so, one important turn point has been the launch of the \emph{iPhone}, commercialized in 2007. Leaving aside the advertising slogans presenting it as a combination of music device, revolutionary phone, and internet communicator featuring a touch screen, the iPhone has changed the way people use their phones, and made possible more advanced, sophisticated, and powerful interaction modalities, as well as novel services.

Another important aspect is that mobile devices are truly \emph{multimedia} devices: the availability of high resolution cameras, sound input/output devices, and video has increased the amount of multimedia material to be searched (sometimes tagged on the basis of the context).

It is also important to understand that the mobile devices are equipped with several other \emph{sensors}, like GPS and location detection, direction, movement, temperature, touch, light, noise, etc. The importance of this aspect comes from the fact that the mobile device is truly ``in the real world'', and this has important consequences. For example, being in the physical world enables location-based (or, more generally, context-based) search, that can be raised to a level unreachable with classical desktop devices. Also, the interplay among all these sensors is quite complex \cite{Cheng2010}, and perhaps not yet understood; for example, of course location can enhance image recognition when trying to understand where a photo has been taken and what is its subject; but conversely the location can be derived by a photo taken with the device, and/or maybe its tags. A further and related aspect to be mentioned is that the mobile device is not an isolated device, and always less so; rather, there is a clear trend towards an increasing \emph{integration} with both desktop computers and other devices and sensors like, e.g., the ``glasses'' or the armbands capable of monitoring motion, gesture, and all kind of lifestyle, fitness, and health data. This wide array of sensors poses challenges and offers opportunities; probably the potential is still unexplored and will be an important research topic in the next years.

Also, the meaning of the term ``mobile device'' itself has changed over the years: from mobile phone, through PDA, and today including \emph{tablets} with much improved input output capabilities by means of larger touch screens.

Finally, besides being in a ``physical environment'', the device and its owner are also in a ``social environment'', in several respects. Mobile devices are used extensively to access social networks (e.g., Facebook, Twitter, etc.), sometimes with a strong location attention (e.g., Foursquare). The first studies are being published that show how to infer the current activity of the user from sensors data. It is theoretically possible to understand whom the user is with.

In the following we will see many concrete examples of how the mobile device features described here cause deep differences between IR and Mobile IR.
For example, in Section~\ref{sec:mult-coll} we discuss how the availability of cameras and sound input/output devices makes multimedia collections a natural environment for Mobile IR, where photos and videos are created and searched; in Chapter~\ref{cha:user-interface} we analyze more in detail the effects of size, screen, and keyboard on a new kind of user interface; in Section~\ref{sec:pois-collections} we discuss how being in the real world leads to searching a new kind of collection, consisting in Points of Interest (POIs); and in Chapter~\ref{cha:context-awareness} we emphasize how the many sensors can be exploited for location- and context-aware search.

\section{The Role of Technology and Society}
\label{sec:role-technology}

When looking at its brief history, it is clear that Mobile IR is a technology grounded field: Mobile IR exists because the mobile devices exist, and the field has followed the technological development. Indeed, in the case of mobile devices, the technological development is so fast paced that sometimes commercial products seem even ahead of the results of scientific research. This is perhaps already true to some extent for Web search, but it is even more important for Mobile IR: one paradigmatic example is the introduction of the iPhone that has changed the way people use their mobile phones. The device, however, although crucial is not the only element, as the services that have been and are being proposed change the way people use the device as well; think, for example, of SIRI, Google Now, Foursquare,
etc. Technology must be interpreted in a broad sense. It is also likely that future development will depend on new technologies as well.

Moreover, it would be shortsighted to consider Mobile IR simply as ``adapting IR to the latest fancy mobile phone'', let alone ``squeezing the IR user interface to a small screen and keyboard''. As already mentioned, the development is not only on the device, but on networks, bandwidth, software, and services.  Therefore, to fully understand the field it is important to take into account the latest relevant technological developments on devices, products, and services.

Finally, the technological development is related to changes in the society and in human habits. Although the first handheld mobile phones appeared in the 80s, there has been a constant growing adoption rate since the 90s that has led to today situation, where there are probably as many active mobile phones on the Earth as human beings. This has direct consequences, like the fact that, since 2015 there are more queries to search engines by mobile devices than by desktop computers.

\section{The Concept of Relevance}
\label{sec:relevance}

Relevance is a crucial notion for IR, and it has been studied extensively in the IR literature. Some attempts have been made to study at a conceptual level this foundational notion from a Mobile IR standpoint, highlighting the differences between the classical and the mobile environment.

Some features of relevance are examined by \citet{Coppola2004}, that extends to the mobile world Mizzaro's framework of relevance \cite{Mizzaro1998}. Mizzaro proposes four dimensions to characterize the various types of relevance:
\begin{itemize}
\item information resources,
\item user problem,
\item time, and
\item components.
\end{itemize}
The first dimension models what the user is searching for, and makes explicit the fact that relevance can deal with the information resources at three levels of abstraction: a document containing information; a surrogate of the document (e.g., title, keywords, and abstract); and the information received by the user, as she perceives it. The user problem (second dimension) has four levels of abstraction: her real information need; the information need as she perceived it; the information need as she can express it in natural language; and the query as it is expressed in the system language. The third dimension `time' refers to the steps in which the information flows with the interaction from the moment in which the user's real information need arises to the moment in which it is satisfied. The fourth dimension `components' lists the aspects that compose the first two dimensions: the topic the user is interested in; the task or activity she aims to perform; and the context in which everything happens.

\citet{Coppola2004} modify and extend Mizzaro's framework. One of the changes is that, in the first dimension, one further level of abstraction is added, thus getting four resources: the actual entity (or thing). In the mobile world, relevance addresses real/physical world entities and not only documents. A related change is on the second dimension, where a corresponding `thing need' is added. These two changes depend on the increased importance of the real world (which is not surprising at all, being the device ``in the physical world'', see Section~\ref{sec:device}), and the same motivation underlies two further remarks made by the authors: the increased importance of time (related to the third dimension) and of context (fourth dimension).  Another observation related to context is that, in the Mobile IR scenario, context is both more complex and more easily derivable than in the classical IR case. This position, and the examples presented in the second part of the paper, inspired the future work by the same research group \cite{Coppola2005,Coppola2005a,Coppola2005b,Coppola2010,Mizzaro2011} (discussed at length, with similar approaches, in the following chapters).

The same concept of relevance is discussed (under the name of Geographical relevance) in another conceptual paper by  \citet{Raper2007}, specifically from a background, and a standpoint, in geography. Some considerations are similar: information needs, named geographic information needs, are discussed; geographic information objects are studied; geographic relevance is defined on this basis; time is often mentioned; a detailed model of geographic relevance is proposed; etc. However there are also some differences, in part due to the different background (geography): a taxonomy of geographic information needs is proposed along two dimensions (intentions vs. tasks and geo-representation vs. geo-context); it is remarked that, as it is customary in geography, a model of geographic space can be defined using either geo-centric coordinates or places and landmarks; space is often mentioned together with time, suggesting that space/time might be the proper concept to take into account; and more in general, whereas \citet{Coppola2004} --- implicitly --- assumed that the various entities in the model are not influencing each other, in Raper's paper the need to understand such influences is explicitly mentioned. One example in this respect is the space/time concept.  Another example is discussed specifically by \citet{Reichenbacher2011}, who remark the importance of the spatial organisation of geographic information objects: they have a ``spatial layout [that] can have historical, geographical, economical, or social reasons'' \cite[p.~69]{Reichenbacher2011} (like, for example, objects in a cluster or hierarchically organized), and that should be taken into account when determining their relevance.

The last paper on this topic is by \citet{DeSabbata2015}, that build on the papers by \citet{Mizzaro1998} and \cite{Coppola2004} and further extend their models along the following four directions.
\begin{itemize}
\item In the first dimension, it is specified that sometimes there is no document at all and the system internal representation of the sought item is a descriptor of a thing in the real world, rather than a descriptor of a document that refers to the thing.
\item The time third dimension is modified: the basic observation is that as the user is moving, the time needed to reach a specific point in the surroundings will change as time goes on, maybe even rapidly. This calls for an integrated space-time dimension that extends the time one.
\item It is noted that there are different representations of the world, including that perceived by the user, that represented in the system, that documented in stored information, and the real one. This is accounted for by proposing to add a further world dimension to the framework.
\item A different components fourth dimension is proposed, that includes topic, activity (which replaces task and  is further analyzed in more detail), user's preferences, social aspects, mobility, and context.
\end{itemize}

One important take home message that we can draw is that being the device ``in the physical world'' (Section~\ref{sec:device}) makes also real world entities searchable.
Another important lesson is that relevance in Mobile IR is truly multidimensional, and beyond-topical features
like time and location
must be taken into account,
probably more than in the classical desktop IR case.

A more practical study of relevance in Mobile IR is by \citet{Verma2016}, who compare relevance judgments on desktop computers and mobile devices.
Results of a crowdsourcing experiment show that gathered relevance labels differ in the two cases, whereas inter-rater agreement is similar.
It also is found that mobile judges are faster.

\section{Mobile IR Models}
\label{sec:MobileIR-models}

The above observations on the nature of mobile relevance are the starting point of more pragmatic work like \citet{Bouidghaghen2011}, where a model is proposed to deal with combinations of relevance criteria. Although the approach could be more general, it is particularly suited to Mobile IR. The relevance criteria included in Bouidghaghen et al.'s model are topic, user's interest, and user's location. Each of them receives its own value: topic is measured using BM25, interest by a tf.idf schema, and location by a geographical weighting function based on  the frequency (a sort of tf) of both location and sub-locations in the document. The three values are then combined by means of ``prioritised aggregation operators''. These operators are defined on the basis of user's preference, expressed as a rank, on the criteria, but they do not simply compute some weighted average; rather, their definition is based on: (i) the preference rank of the criteria, (ii) the weight of the previous more important criterion, and (iii) the satisfaction degree of the document with respect to the previous more important criterion. An experimental evaluation on an in-house built collection shows the effectiveness of the approach.

The issue of how to exploit in local search new features beyond the textual ones used in Web search is discussed also by \citet{Berberich2011}.
They derive distance and popularity features from external sources and show that such features are effective in improving effectiveness.

Another technical contribution is by \citet{OHare2013}, who study the problem of location identification from arbitrary text.  To this aim, they exploit a core IR technique and develop statistical language models for locations, that are defined as square cells of 1 by 1 km, 10 by 10 km, and 100 by 100 km. That is, each location (square cell) is considered a document, and its terms are the Flickr textual tags of the photos geotagged in that area. The term distribution defines the language model of that location.  The models are then tested to predict the location of (other) geotagged Flickr photos by using as queries only their textual tags, and on another dataset (CoPhIR) as well. Several classical language model approaches are adapted and used, like maximum likelihood estimate, prior location probability (to take into account location popularity), and smoothing (to refine the term distributions of a location on the basis of its neighbourhoods). The experimental evaluations show that reasonable accuracy can be obtained at various granularity levels. For example, 1 by 1 km locations can be predicted with a 17\% accuracy, and 3 by 3 km ones with 40\%. The experimental results also confirm that smoothing can be effective.

Other researchers focus on assigning positions to queries, and exploiting the position to improve retrieval.
Although the relevance to Mobile IR of this research issue is perhaps more marginal, and it can be useful for search in general (indeed the term ``mobile'' does not even occur in some of the papers), it deserves anyway a short mention at least.
\citet{Backstrom2008} develop a probabilistic model to assign to each query (they work on the complete Yahoo! query logs) a center plus a spatial dispersion.
\citet{Yi2009} aim at discovering implicit user's local intent in queries. They develop a classifier able to effectively categorize queries at the city level.
A similar approach is followed by \citet{Lu2010} who focus on implicit local intent queries, i.e., queries that are clearly searching for things in a particular location but without any explicit location mention.
They develop a probabilistic based model to automatically classify the queries that have an implicit intent.
An attempt to improve retrieval effectiveness of those queries is made by first inferring user's location from the IP address and then either (i) expanding the query by adding the location, or (ii) re-ranking the retrieved documents on the basis of the location.
In an experimental evaluation, both approaches improve retrieval effectiveness, with re-ranking being significantly more effective.

\section{Conclusions}
\label{sec:conclusions-3}

After a brief mention of general issues related to devices, technology, and society, this chapter has discussed how the notion of relevance needs to be adapted to the Mobile IR world, highlighting some important changes.
We will see consequences of these conceptual approaches in the following (for example, in Section~\ref{sec:pois-collections}, and in Chapters~\ref{cha:context-awareness} and~\ref{cha:evaluation}).
We can say that the discussed papers on relevance contain cues, hints, and examples that, although at a conceptual level, are perhaps still to be fully understood and exploited.
In the last section we have discussed how Mobile IR affects the foundational topic of IR models, and presented a sample of the Mobile IR models that have been proposed.

\chapter{Documents}
\label{cha:docum-coll}

User generated content is one of the main differences from IR to Web IR, and the difference is even more dramatic when comparing IR and Mobile IR.
This chapter discusses the changes, from classical IR to Mobile IR, concerning the document collections.
We first briefly highlight the changes in ``traditional'' collections, namely text (Section~\ref{sec:text-only}) and multimedia (Section~\ref{sec:mult-coll}).
We then focus on collections that are specific to Mobile IR:
we present the literature on Apps retrieval (Section~\ref{sec:apps-collections})
and we discuss the important links between Mobile IR and movement data, points-of-interest, and Internet of Things (Section~\ref{sec:pois-collections}).

\section{Text}
\label{sec:text-only}

Since classical IR is indeed mostly about textual documents, one might claim that there is nothing new in Mobile IR when looking at text retrieval.
However, it is also clear that the widespread adoption of mobile devices has lead to the production of short and quick texts (tweets, SMSs and texting, facebook posts, etc.): brevity is admittedly a common feature of mobile textual content, due to physical and ergonomical limitations of devices.
This might not be considered a huge difference (as, for example, Apps retrieval might be, as we will see in the following of this chapter), and perhaps it can also be claimed that it is not so relevant to Mobile IR (since, for example, Twitter can be used also on a desktop and it has very little in it that forces one user to use it in a mobile environment).
It is anyway worth to briefly mention some of the studies in this area.

A practical consequence of the brevity issue is that mobile content has a very limited vocabulary, which can severely hinder the effectiveness of mobile search engines.
A common solution to this problem it text enrichment.
For example, \citet{Church2007a} propose to enrich mobile pages with new terms extracted by means of a standard web search engine (like, e.g., Google).
Then, the enriched text is used to index the original mobile page with the reasonable expectation that future queries submitted by users will benefit from the enriched terms.
More precisely, a mobile page is first converted into a set of queries, which are then submitted to the standard web search engine.
The new enriching terms are extracted from the top ranking search results of such queries.
\citet{Pavan2016} exploit text enrichment on the basis of news collections to improve tweet categorization.
Twitter is a widely used source of data for experiments on short texts, but it is not the only one.
\citet{Almeida2011} address SMS spam filtering and also provide an SMS dataset.
\citet{Tian2010} focus on organizing into conversations a set of SMS messages, by means of clustering plus LDA.

\section{Multimedia}
\label{sec:mult-coll}

As for (short) text retrieval, also multimedia retrieval is a quite well studied problem, and although mobile devices have introduced some novelties, it is difficult to claim that Mobile IR has brought a revolution in multimedia retrieval.
However, of course the new environment is clearly multimedia, as current mobile phones allow easy production and perusal of photos, sound/music, and videos.
Again, we provide a non-exhaustive list of some of the most significant approaches.

The presence of a decent quality photocamera in many smartphones/tablets/etc.\ has promoted the collection and the sharing activity of pictures among users of mobile devices.
Such attitude has been exploited by several social networks, e.g., Instagram, Pinterest, Google Photos, and today we can access, search and browse large datasets of user-contributed pictures from mobile devices (as well as from ordinary computers, of course).
Whence, there is the need of new and specific retrieval systems for ``mobile images'', coping both with networking and hardware limitations of mobile devices.
For instance, \citet{Yang2014} propose a novel approach
that
leverages on the fact that often users take multiple shots of a given scene; hence, the mobile device is first searched for photos visually similar to the query image.
Then, the latter, together with the relevant photos found in the device, is used to mine ``visual salient words'', which are ordered according to their contribution in order to reduce the noise and the computational complexity of spatial verification.
Moreover, such ordering allows the client to transmit only part of the salient words to the server, in case of limited bandwidth and connection instability, a common scenario for mobile networking.
Finally, spatial verification is performed server-side with the purpose of re-ranking the retrieved results.

Another peculiar feature of mobile photograph, called ``ephemeral photowork'', is described by~\citet{Brown2016}.
The authors' thesis is that the use of photos is not limited to interactions through online social networks, it is also based on ``face-to-face conversations around the devices themselves''.
Indeed, the gist of such ephemerality is the practice of quickly producing, sharing and consuming photos often in the same place where they are taken.
Hence, pictures in mobile devices do not seem to be carefully chosen and saved for future reuse, but to answer an immediate need of discussion or contact with other users.

A more insightful  approach is that by \citet{Yan2010}, who show how crowdsourcing can be exploited not only for IR evaluation, but also for providing a real time mobile image search service exploiting the workforce of a crowd.

Finally, multimedia is sometimes combined with contextual retrieval.
For example, \citet{DeSilva2009} propose a system to retrieve photos related to a previous travel on the basis of spatio/temporal queries and \citet{Braunhofer2013} describe and evaluate a location-aware music recommendation mobile application.
These will be described in more detail in Chapter~\ref{cha:context-awareness}.

As anticipated, this is a rather short and not exhaustive account of multimedia Mobile IR; we now move to more core-Mobile IR by describing more in detail collections made of Apps, movement data, and points of interest.

\section{Apps}
\label{sec:apps-collections}

Apps definitely form an important kind of collection in the ``mobile ecosystem''. In particular, there are some well characterized subfields of research related to interesting and peculiar aspects of Apps, which we will discuss in this section:
\begin{itemize}
\item search engines and recommendation systems;
\item security and privacy related issues;
\item content filtering frameworks;
\item automatic tagging systems.
\end{itemize}

The public release of App development SDKs, both official and derived/third-party, for the most popular smartphone platforms together with the availability of Apps marketplaces (e.g., Apple App Store, Google Play etc.)
paved the way for a daily overwhelming flood of hundreds of new Apps, produced not only by companies, but also by enthusiast users worldwide.
Such collections are just too big to be managed and searched manually, by listing all the Apps: users cannot afford this information overload.
Moreover, official search tools only list recommended, popular and novel Apps: finding an unpopular or old App is becoming almost impossible.
Hence, a part of the research community has devoted his efforts to studying and developing search engines or recommendation systems for Apps.
For instance, \citet{Mizzaro2014} introduce AppCAB: a recommendation system for mobile Apps retrieval, exploiting the information extracted from the user's context description (in the format generated by the Context-Aware Browser, see Chapter~\ref{cha:context-awareness} and~\citet{Coppola2010}), to find the right applications for the user's needs in that specific context.
In particular, using Apps metadata taken from the online Apple App Store, an experimental evaluation of several variants of the system is carried out.
The results clearly indicate the specific nature of this retrieval task, where taking into account the category of the App and of the current context improves effectiveness over using \textit{tf.idf} only.
Other tunings take into consideration the numbers of Apps in each category (to avoid privileging the most popular categories), the App average rating, the presence of a query term into the App title and Apps belonging to a ranked list of categories describing the current context.
Experimental results show that the adopted corrections to a standard retrieval practice are indeed effective in the domain of mobile Apps.

The problem of Apps searching and retrieval is rigorously studied for the first time by~\citet{Park2015}: the authors build a dataset crawling 43,041 App descriptions and 1,385,607 user reviews from the Google Play store, and they evaluate state-of-the-art retrieval models (namely, Okapi BM25, Query Likelihood Language Model, Combined Query Likelihood, LDA Based Document Model, and BM25F) together with their proposed solution which exploits topic modeling mixing users reviews with App descriptions.
Such choice is made in order to ``fill the gap'' between the vocabularies of users (making queries) and of developers (providing Apps descriptions).
According to the authors, the generated Apps representations should allow the search engine to achieve high effectiveness, and this is confirmed by their experiments, where their solution outperforms the above mentioned state-of-the-art retrieval models.
Experiments are carried out using realistic queries based on Android forums and relevance assessments are made by means of a crowdsourcing service.

The issue of properly categorizing mobile Apps is strictly related to the task of recommending them to users, since a good categorization may lead to better recommendations.
However, mobile Apps provide limited contextual information in their names (and, if an App is not published in a public store, then its name is the only source of information we have).
Hence, \citet{Zhu2012} propose a categorization system exploiting an enrichment strategy leveraging both on ``Web knowledge'' (i.e., information provided by a web search engine like, e.g., Google) and ``real world contexts'' (extracted from device logs about App usage).
The idea of exploiting
enrichment comes from state-of-the-art work in the categorization of Tweets (mentioned in Section~\ref{sec:text-only}), due to similar issues of brevity and sparsity.
After the enrichment step, the authors use the Maximum Entropy model to combine the effective features in order to train a mobile App classifier.
The latter achieves very effective results during their experiments carried out with data about 443 mobile users, 680 unique mobile Apps and more than 8.8 million App usage records.

Searching and recommending Apps is not merely a matter of matching user's interests and App's functionality. Indeed, privacy and security may become an important concern: the user may be interested in preserving her privacy preferences and the security of both her device and data. The solution proposed by~\citet{Zhu2014} exploits the permissions requested by Apps to the underlying operating system in order to assess their security; more precisely, it applies a random walk regularization over an App-permission bipartite graph. Hence, the security risk of Apps can be automatically learnt without any predefined risk function, yielding an App hash tree as result. The latter and the modern portfolio theory~\cite{Wang2009} for recommendations are then used by the proposed system to find a match between the given mobile users’ security preferences and App categories, providing a final ranking of the candidate Apps with respect to both Apps’ popularity and users’ security preferences. The experimental data were collected from Google Play
(Android Market) in 2012. The dataset used in the experiments consists of 170,753 Apps (with 173 unique data access permissions) taken from 30 App categories.

Predicting permissions of a new mobile App according to its functionality and textual description is the main aim of the framework presented by~\citet{Kong2015}. Again, the goal is to help the user preventing security risks, by means of a classifier addressing the potential over-claimed permission request issues in Android Apps. The proposed supervised approach has a training part which includes three key steps: (i) App textual description and functionality extraction by crawling the Google Play website,\footnote{The semantic meaning of different words is inferred using a deep learning technique named word2vec, \url{https://code.google.com/p/word2vec/}.} (ii) App permission request extraction using Androguard,\footnote{\url{https://code.google.com/p/androguard/}} and (iii) training a multi-label classifier via structure feature learning. The resulting automated permission
prediction classifier is then evaluated on 173 permission requests from 11,067 mobile Apps across 30 categories. The experimental phase indicates a consistent improvement of
performance (3\%--5\%\ in terms of F1 score), w.r.t. the other state-of-the-art methods (i.e., Simple Linear Regression, K Nearest Neighbors, Support Vector Machine, Multi-label Latent Semantic Indexing~\cite{Yu2005}, Multi-label Dimension reduction via Dependent Maximization~\cite{Zhang2010b}, and Multi-label Least Square~\cite{Ji2010}).

Privacy issues are instead the main concern of~\cite{Liu2015}, where the authors propose a solution to systematically incorporating privacy preferences into a personalized recommendation system. They introduce three variants, according to three privacy levels: the first one considers only ten resources (e.g., contact, message, location etc.) as private; the second level adds to the same ten resources their related operations (e.g., reading and writing a message) for a total of 23 permissions; the third and final level is the most fine-grained, since it incorporates all the resources belonging to the previous levels and many more for a total of 72 permissions. Then, they experimentally compare their proposed solution with four state-of-the-art previous models in the literature which do not take privacy concerns into account (i.e., Singular Value Decomposition, Probabilistic Matrix Factorization, Non-negative Matrix Factorization, Poisson Factor Model, and~\cite{Gopalan2013}). The experiments are carried out on a dataset with 16,344 users, 6,157 Apps, and 263,054 rating observations. All the variants of the proposed solution outperform the four classic models, showing that users generally prefer to consider level I and level II resources as private assets.

Automatically filtering mobile Apps which can be considered inappropriate for children (e.g., featuring mature themes like sexual content and violence) is the focus of the framework proposed by~\citet{Hu2015}. Indeed, maturity rating policies implemented in App stores can be costly and  inaccurate, requiring manual annotations and periodic checking. The authors exploit machine learning techniques to automatically predict maturity levels for mobile Apps. More in detail, they use deep learning techniques to infer semantic similarities between words extracted from App descriptions. Such similarities are used to specify novel features which can be fed to Support Vector Machine classifiers in order to capture label correlations with Pearson correlation in a multi-label classification setting. Then, the framework is evaluated against several baseline methods, using datasets collected from both Apple App Store and Google Play store.

Another common task, when dealing with mobile Apps repositories, is spam detection. A mobile App can be considered a spam application when some conditions are met, e.g., it does not have a specific functionality, App description and/or keywords are unrelated, the developer is publishing similar Apps several times and across different categories. Popular App stores have their own policies against spam, but the detection and removal activities are usually carried out through human intervention. Instead, \citet{Seneviratne2015} introduce an Adaptive Boost classifier for automatically identifying spam Apps by exploiting Apps metadata. During their experiments, in a first phase, the Google Play store is crawled starting from a previous dataset of 94,782 Apps (obtained from the lists of Apps from approximately 10,000 smartphone users, namely, volunteers, Amazon mechanical turk users, and users publishing their lists of Apps in social App discovery sites), yielding a final dataset of over 180,000 Apps. Subsequent crawlings are performed in order to detect and label Apps removed from the store. A random subset of those Apps is subsequently manually analyzed by three expert reviewers, in order to identify the reasons behind the removal process. It turns out that the main reason for App removal is spam, accounting for approximately 37\%\ of the removals with a high agreement between reviewers. The heuristics used in this phase for manual spam detection are then converted into features for automatic spam detection in the second phase of experiments. Such features are used to build a binary Adaptive Boost classifier, where spam Apps are used as positives and Apps from a top-$k$ set as negatives.\footnote{$k$ is the cardinality of the set of spam Apps. Top Apps (for number of downloads, user reviews etc.) are quite likely to be non-spam} Taking 80\%\ of the dataset as training set and the remaining 20\%\ for testing, the classifier, for small values of $k$,\footnote{Obviously, the larger $k$ is, the higher is the probability that spam Apps are taken as non-spam.} can achieve an accuracy over 95\%\ with a precision between 85\%--95\%\ and a recall between 38\%--98\%.

Overall, many issues related to recommendation systems, Apps search engines etc. could be simplified in presence of a tool being able to detect ``semantically similar Apps''. Indeed, in~\cite{Chen2015b} the authors address such problem, introducing the \emph{SimApp} framework. SimApp has two stages: the first one if a set of kernel functions measuring App similarity for each modality of data (i.e., App name, category, description/update texts, permissions, images, content rating, size, and user reviews), while the second one is an \emph{online kernel learning} algorithm which can infer the optimal combination of similarity functions of multiple modalities, i.e., the optimal combination of weights. An empirical evaluation is carried out with 21,624 Apps from 42  different categories taken from Google Play store, showing that SimApp improves the precision
scores by more than 20\%\ w.r.t. baseline algorithms.

Besides the continuous growing of the number of Apps published in popular stores, a natural consequence is the related increment of the number of installed Apps in the user's smartphone. Hence, there is the possibility that the user is not completely aware of all the installed Apps and finding the most suited App for a particular task in a given situation (by browsing and filtering through folders) can be tedious and/or challenging. In~\cite{Kim2014} a ``conditional log-linear model'' is used, in order to predict the ``best'' application that the user is most likely going to use at a given situation. The model is a \emph{discriminative} one: this fact makes it more reliable when the features to analyze are correlated. The experimental evaluation shows good results and it is carried out using data taken from the Nokia Mobile Data Challenge~\cite{Laurila2012} against other common algorithms in the literature, namely, Most Recently Used, Most Frequently Used, Weight Decay, Na\"{\i}ve Bayes, and K Nearest Neighbors.

App tagging systems can be used to correctly indicate App functionalities, categories, etc. Programmers/maintainers of Apps search engines and recommendation systems, users browsing App stores and mobile advertisers can greatly benefit from this kind of activity and this is the focus of~\citet{Chen2016}, who introduce a mobile App tagging framework for annotating mobile Apps automatically. The approach is based on machine learning techniques. More precisely, given a novel ``query'' App to be tagged, there is a first step where online kernel learning algorithms are used to search a large repository for the top-$N$ most similar Apps. Then, the most relevant tags for annotating the query App are ``mined'' from its own text data and the text data of the top-$N$ most similar Apps found in the previous phase. The experimental evaluation has been carried out with a training set of 15,000 random Apps sampled from the Google Play Store. The \emph{query} dataset, providing Apps to be tagged, has been sampled, for a total of 1,000 Apps, from alternative Android markets (i.e., SlideME, Samsung Galaxy Apps and PandaApp) where Apps are provided with tags (these are taken as the ground truth). Finally, the retrieval data set consists of 60,000 Apps sampled again from the Google Play store. The results of the experiments are quite good, but, according to the authors, there are some limitations due to the limited sizes of both the retrieval database (60,000 Apps) and the ground truth dataset (1,000 Apps).

\section{Movement Data, POIs, and IoT}
\label{sec:pois-collections}

As we discussed in Section~\ref{sec:relevance} the context of the user is an important aspect affecting relevance in Mobile IR.
In particular, the latter leverages on real/physical world entities besides documents.
Nowadays, even low-cost smartphones are GPS-enabled devices; hence, they are a potentially unlimited source of movement data.
Of course, such raw data must be processed before producing useful information, and the typical output of this processing phase is a set of \emph{Points of Interest} (POIs), i.e., specific locations that can be useful or interesting.
Alternative sources of POIs are location-based social networks (LBSNs) like Foursquare, which help users to search and discover places, according to their interests and check-ins to the platform.
\citet{Zheng2008} started in 2007 the project GeoLife: a location-based social-networking service,  which allows users to share travel experiences by means of their GPS trajectories. Moreover, GeoLife can discover the most interesting locations, travel sequences (see, e.g., \cite{Zheng2009}) and expert users for each geographical region. Finally, it can act as a recommender system mining individual location histories and finding similarities between users. From a scientific point of view it is a very valuable source of information for evaluating new systems, since it allows free access to a huge database of (anonymized) GPS data.

Similarly to other kinds of user's related document collections, POI datasets have given rise to the necessity of retrieval/recommendation/prediction systems.
For instance, \citet{Scellato2011} exploit a novel approach based on time of the arrival and time that users spend in relevant places, to develop a framework for location prediction, called NextPlace.
A dual approach is instead pursued by~\citet{Liu2013}, that exploit the transition patterns of users' preference over location categories to build a novel POI recommendation model.
Their evaluation experiments show an improvement of the accuracy of location recommendation.
Another successful attempt in the direction of providing more accurate POI recommendations is introduced by~\citet{Cheng2013}.
They exploit personalized Markov chains and region localization to take into account the temporal dimension and to improve the performance of their system.
More recently, \citet{Gao2015} introduce a novel approach to improve the effectiveness of existing POI recommendation systems of location-based social networks.
Instead of focusing only on ``spatial, temporal, and social patterns of user check-in behavior'', they consider the relationship between user behaviour (in particular user check-in actions) and the content information provided by social networks.
Their experimental evaluation shows indeed the significance of such relationship and its contribution in improving the effectiveness of POI recommendation.
Social correlations together with geographical and categorical correlations are also the focus of \citet{Zhang2015} where the GeoSoCa framework is introduced.
The latter learns the above mentioned correlations from the history of user's check-ins and outputs POIs recommendations ranked by a relevance score.
The experiments are carried out over two large-scale real-world check-in data sets extracted from two popular LBSNs, namely, Foursquare and Yelp, achieving results superior to other state-of-the-art POI recommendation systems.

One of the most common user's needs, exploring a new city or location, is to find information about places and/or activities/events in the surroundings. Usually such information can be obtained by specialized web portals and services; however, the results returned by these engines are often generic and inaccurate. A more adequate solution is represented by, so called, local experts, i.e., individuals who possess the right knowledge about a location. Local experts are determined according to their check-in profiles: for instance, if a user frequently makes a check-in in a given location, we can assume that he may be familiar with that place. Thus, there is an ongoing research activity devoted to the development of models and methods for automatically retrieving local experts and, as a consequence, local knowledge. For instance, \citet{Li2016} focus their work on developing some probabilistic models of local expertise (about ``POI topics''\footnote{Knowledge about a single location, e.g., business hours of an office.} and ``category topics''\footnote{Knowledge about all locations in a specific category, e.g., positions of all the gas stations in the town center.}) based on geo-tagged social network streams. In particular, they exploit geo-tagged streams extracted from Twitter and check-ins in Foursquare, in order to build and evaluate some models supporting three heuristics (and their possible combinations) about user's check-in profiles, namely, visiting frequency, diversity and recency. The evaluation results prove that their approach models local expertise quite well (better than both random and refined baselines methods in the literature).

As another final theme, we mention that with the advent of the Internet of Things (IoT), it is now possible to address and tag real objects, ``extending'' some techniques, models and services to the physical word. For instance, \citet{Knierim2012} introduce Find My Stuff (FiMS): a search engine for physical objects (e.g., mobile phones, keys, sunglasses etc.) provided with ZigBee or passive RFID tags, allowing users to search hierarchically (starting from last known positions and going through connected locations) for misplaced things in indoor environments. Search results provide cues for possible locations of lost object, as anticipated above, leveraging on relative positions of objects themselves and smart furniture (exploiting ZigBee modules). The system is also suited to changing environments (like homes or offices) without requiring manual intervention or frequent maintenance.

\section{Conclusions}
\label{sec:conclusions-4}

We have seen the differences between IR and Mobile IR when looking at documents and collections. Whereas text and mutimedia collections can be seen as having many similarities (although differences do exist), Apps and data directly derived from the physical world like movement data, POIs, and IoT data are quite different, and likely they need novel approaches.

\chapter{Users and Information Needs}
\label{cha:user-needs}

This chapter presents a large set of experimental studies aimed at better understanding aspects of mobile search that differ from classical desktop search.
The different role of users and their information needs is certainly a major aspect.
Studies based on log analysis are presented in Section~\ref{sec:logs}, while more classical users and diary studies are presented in Section~\ref{sec:diary-studies}.
The chapter finishes with some comparisons and remarks in Section~\ref{sec:conclusions-5}.

\section{Log Analysis}
\label{sec:logs}

\begin{table}[tbp]
  \centering
  \begin{tabular}{llllrll}
\toprule
Reference&Pub.&Data&S.&Size&Where&Notes\\
\midrule
    \citet{Kamvar2006}& 2006 &2005 & SE& 1M(?)& U.S. & 1st study\\
    \citet{Church2007}& 2007&2005&  T&  0.4M& EU &Several SEs\\
\addlinespace
    \citet{Kamvar2007}& 2007&2007 & SE & 1M & U.S. & Follow up of \cite{Kamvar2006}\\
    \citet{Baeza-Yates2007}& 2007&2006 & SE &1M(u) & Japan &Larger scale\\
    \citet{Church2008}& 2008&2006&  T& 6M & EU & Click-throu \\

\citet{Gan2008} & 2008 & 2006 & SE & 36M& U.S.& AOL geographic queries\\

    \citet{Yi2008}& 2008& 2007 & SE & 40M &  Int. & Larger scale\\
    \citet{Vojnovic2008}& 2008& 2007& SE & 0.5M & U.S. & Dynamics\\
    \citet{Kamvar2009}& 2009& 2008& SE & $<$1M & Int.& 3 devices, iPhone, users\\
\addlinespace
    \citet{Lane2010a}& 2009& 2010& SE & 80K & U.S. & Context\\
    \citet{Church2011a}& 2011& 2009& P & 1.4M & Spain& Follow up of \cite{Church2007,Church2008}\\
    \citet{Yi2011}& 2011& 2010& SE & 20M & U.S. & Trends, Voice queries\\
    \citet{Koukoumidis2011}& 2011& 2009 & SE & 200M  &Int.& Cacheability\\
    \citet{Zhuang2011}& 2011& 09-10& SE & 75M  &U.S. & Location. Time\\
    \citet{Lymberopoulos2011}& 2011&~~-- &SE& 2M&U.S.& Location\\
    \citet{Nicholas2013}& 2013&09-11 &P& 150M&EU& Europeana\\

    \citet{West2013} & 2013 & ~~-- & O & 1M &Int. & Primary location\\
    \citet{Song2013} & 2013 & 2012& SE &32M& U.S.& iPhone \& iPad\\

\addlinespace

    \citet{Dong2014} & 2014& 2008 &T & 1G& Anon. &Social connections\\
    \citet{Oentaryo2014} & 2014& 2012& O& 24M& Singapore &City-wide mobility\\

    \citet{NorouzzadehRavari2015} & 2015 & 2014 & O & 0.6M& U.S.+U.K.& POI searching\\

    \citet{Fan2016} &2016 & ~~-- & T&10M(u) & Bangladesh &Mobile ads\\
    \citet{Song2016} & 2016& 10-13& T& 1.6M(u)&Japan&City-wide mobility\\
    \citet{Poonawala2016} & 2016& ~~-- & T+O& 3G& Singapore&Mobility, last mile\\
    \citet{Graus2016} & 2016 & 2015& O& 0.5M& U.S.& Tasks, Cortana\\
\bottomrule
  \end{tabular}
  \caption{Summary of the log studies described in the text. Pub. and Data show the year of the publication and of data collection; S. is the source of data (SE = Search Engine; T = Telephone operator or company; P= Portal; O = Other); Size is the approximate number of queries (or records) in the dataset; and Where is the location of the users. (u) stands for users, -- for unreported data.}
  \label{tab:logs}
\end{table}

Several researchers have tried to understand more about Mobile IR and Mobile IR users by analysing large scale log data, derived either from the mobile versions of commercial web search engines, or from the logs of mobile phone operators.
These studies are listed in Table~\ref{tab:logs}, and presented in the following three subsections, in chronological order.

\subsection{First Studies}
\label{sec:first-studies}

The first study of a commercial web search engine log is by \citet{Kamvar2006}.  The data comes from the logs of Google mobile search, which at that time featured two slightly different versions, one for  mobile phone and one for PDA.   The data was collected in  2005 and is made up by ``1 million hits'' (it is unclear how many of these hits  indeed correspond to search queries, and how many instead correspond to other actions like click on a result,  a ``next page'' link, and so on). To avoid queries by bots, only the requests coming from a single U.S.\ phone carrier are selected, thus making the data  U.S.\ based. Some results on query length are presented that are found to be remarkably similar to those found in the literature related to standard desktop searches, with mean and median query length at 2.3 and 2.7 for phone and PDA, respectively. Concerning query categories, some differences are highlighted with respect to classical desktop search: for example, adult queries are found to be more common on phones (and, to a much lesser extent, on PDAs); queries on local services look more common as well. An analysis of query distributions shows that phone queries are more homogeneous than classical desktop ones (the first $1000$ unique phone queries account for 22\% of all queries, whereas this figure is 6\% for desktop). Some results on sessions are also reported: for example, the average number of queries per session ($1.6$) is smaller than previously reported desktop values ($2.02$--$2.84$); and also the number of page views in the results page is lower.  The paper also discusses the time taken to input a query, which of course depends on the 12 keys keypad that is much less common today,  on results page perusal, and on user persistence in submitting subsequent queries in the same category.

Other studies are based not on the logs of a web search engine, but on mobile provider / operator specific search services. Using a mobile operator data has some advantages. For example, data contains not only queries, but also other user activities like browsing. Also, queries are directed to many search engines, not only one. Finally, the sample of the user population does not present the bias of using a specific search engine, although of course it is biased in a different way.

The first study of this kind is by \citet{Church2007}. The data comes from an European mobile operator, collected during 24 hours at the end of 2005 and containing over 30 million requests, of which 400 thousand are search queries, generated by more than 600 thousand unique users.  We summarise only some of the many results of this interesting study. Queries, although limited when compared to browsing, do anyway play an important role: $94\%$ of the sessions are browsing-only (i.e., do not contain search queries), but the $6\%$ sessions that include search queries tend to be longer in terms of both time and requests and to cause a higher data transfer.  When focusing on users, browsing-only users (i.e., those that do not issue any search query) are $92\%$, but the remaining $8\%$ search users account for around $20\%$ of the data. The search engine most used is Google ($76\%$). An analysis of the devices used shows that  more sophisticate devices are used in search sessions than in browsing-only sessions.

Other analyses about query length, usage of advanced search features like boolean operators, query similarity, query reformulation, and query categories highlight a somehow larger difference than those found in the previous search engine log study by \citet{Kamvar2006} and in other similar studies. For example, queries are found to be shorter: $2.06$ terms per query on average, although when considering only Google queries the value increases to $2.22$, more similar to the other studies. The top 500 most frequent queries are  classified  into 16 predefined categories. Differently from  other studies, the classification is done manually. Again, the most popular category is for adult related content, and the most striking difference with respect to the other studies is in the  Search category, related to finding other search engines, that accounts for  $8\%$ of the queries.

A parallel is drawn with the studies analysing the early days of web search: many similarities are found, and often justified by the ``infancy'' of mobile search; some differences are ascribed to the input/output limitations of mobile devices available at that time (2005).

\subsection{Follow-ups and More Complete Studies}
\label{sec:follow-ups-more}

\citet{Kamvar2007} presented an interesting follow-up study to their previous work \cite{Kamvar2006}. The data is again U.S.\ based, using a collection from Google logs in early 2007. The methodology and results are similar to the previous study \cite{Kamvar2006}, although some differences are noted and interpreted as the emergence of new  trends. For example, improvements of network speed and of keyboards are likely to be the main reasons for the shorter time required to enter a query, more users are clicking on the results page probably due to better information presentation and again, faster network connections,  queries are less homogeneous, and sessions are longer (i.e., having more queries).

\citet{Baeza-Yates2007} perform a search engine log study focusing on the Japanese market. Japan is an interesting case study for the peculiar features of its language and alphabets, and because of the high penetration of mobile Web. The study is based on a Yahoo! log of one million mobile and one hundred thousand desktop unique queries collected in 2006. The main results are generally similar to those of previous studies \cite{Kamvar2006,Kamvar2007} although specific for Japanese. Query length, when measured in words, is similar to  previous studies; indeed, contrary to what one might expect, mobile queries tend to be  slightly longer than desktop ones. For example, two word queries are more frequent than one word ones on mobile (differently from \citet{Kamvar2006}), and vice versa on desktop. The reason is conjectured to be that users prefer to spend more time in typing a query than in the reformulation phase. When measured in characters, query length is shorter on mobile than on desktop (median lengths are 7 vs. 9). 

The follow up study by \citet{Church2008} is the first to examine click-through behaviour, i.e., how users click on the results page. Similarly to the previous study by the same authors \cite{Church2007}, the data has been collected over a week in 2006, coming from an European mobile operator, and it is made up by more than 700,000 search sessions corresponding to more than 6 million search queries.  Query length, query reformulation, query categorisation, and term overlap across queries are analysed with the same methodology of the previous study, and the results are similar, although query reformulations are found to happen more rarely. Query variety is found to be lower than that reported by \citet{Kamvar2006,Kamvar2007}. Turning to the click-through analysis, which was a novelty for Mobile IR, results show that click-through rates are quite low: there is no click on the results page for as much as $75\%$ of users, $59\%$ of sessions, $88\%$ of Google queries, and $76\%$ of unique Google queries. Although some of the cases can be explained by users finding their answer directly on the results page, these high values clearly show that Mobile IR was not effective at that time.

\citet{Yi2008} present a complete and systematic analysis of a Yahoo! query log collected in August and September 2007 and made up of an even larger amount of data, divided into two parts: 20 million English queries submitted from U.S., plus another 20 million English queries from Australia, Canada, India, New Zealand, and U.K. The paper presents a large amount of detailed results, that we can only briefly summarise here. Results on query length and uniqueness are overall consistent with previous studies. The U.S.\ data set features  longer queries (11\% more words per query). Consistently with  the study of \citet{Baeza-Yates2007}, the most common query length is $2$ words, whereas \citet{Kamvar2006} found it to be $1$. Head queries (those with high frequency and/or repetitions) are found to be shorter than tail queries.  It is interesting to note that most results on query length agree in all the studies \cite{Kamvar2006,Kamvar2007,Baeza-Yates2007,Yi2008}, and they are also similar to the data for desktop search: where $2.5$ words are used on  average.

\citeauthor{Yi2008} also study query variety, and they find that U.S.\ queries have less variety than  international ones. The repetition rate of queries follows a power law: it is low for most of the queries, but there are very few queries that are repeated many (tens of thousands) times.  Query categorisation is analysed, along both the usual topical dimension (with similar results to previous studies; for example, ``personal entertainment'' is the most frequent category) and the \emph{intent} dimension. With respect to the latter, it is noted that queries with ``local intent'' are around $10\%$, and navigational queries  are about $5\%$, definetely fewer than the $17\%$ found by \citet{Kamvar2006}. Finally, since the log is generated by three different user interfaces (XHTML, Java, SMS), a breakdown of the categories  is presented, and some differences are highlighted. It is also remarked that the differences might well depend not only on user interface features, but on different user and device populations associated with each interface (e.g., the Java interface could be used only in Java-enabled phones).

\citeauthor{Vojnovic2008}'s  \cite{Vojnovic2008} analyses the data from the log of a non specified U.S.\ based search service, made up of about 500 thousand mobile queries collected during the week 1-8 April 2007. Besides some limited results on query statistics (called ``semantics'' in the paper) and categories (called ``topics'') that generally do not differ from previous studies, the paper mainly focuses on \emph{temporal dynamics}, the most interesting feature. The results show that there is a clear diurnal periodicity in query submission, with bursts of activity for each user.  Users do not behave differently during weekends and working days and mobile search is not, in 2007, a daily activity. The distribution of time between two subsequent queries tends to follow a power law up to one day, and then to decrease exponentially, even for individual users, with sessions being quite short both in time (90\% are under 5 min.) and in number of queries (median is 1 and $90\%$ have no more than 3 queries);

The paper by \citet{Kamvar2009} presents yet another study based on Google logs. Data is gathered during Summer 2008, taking into account more than 800 thousand English only queries, submitted by more than 40 thousand users. One novel aspect in this study is the comparison of three different devices: conventional Mobile phone, iPhone, Desktop computer. This is the first search engine log study that explicitly addresses and recognises the influence of the iPhone device.
The usual three analyses of query length, query classification, and query variety show similar results to previous studies. However, it is noted that the iPhone behaviour is more similar to Desktop's than Mobile's, with respect to all three kinds of analysis. When analysing the sessions, the highest number of queries per session is found for  Desktop, followed by iPhone, and then Mobile. The session data confirm the iPhone-Desktop convergence: indeed, when categorising sessions the iPhone turns out to be even more similar to Desktop than when categorising single queries.

The authors also aim at a direct comparison of \emph{users} on the three different devices, and therefore attempt to focus on users and not only queries. Some sophisticated analyses of the variety of user needs on the basis of the categories of their tasks show that Mobile users are much less variable than Desktop and iPhone, which turn out, again, to be very similar.
Local queries are also discussed, and contrary to the expectation it is found that they have a rather marginal importance in the logs. It is conjectured that this depends on users using other apps (for example, Maps) when seeking for local information.

\subsection{Third Wave}
\label{sec:third-wave}

Studies in the more recent third wave focus on themes that are rather different from the previous ones.

\citet{Lane2010a} analyze about 80 thousand local search queries from Mobile Bing Local submitted over the first six months of 2009. It is found that the clicked results depend quite significantly on several contextual aspects: time of the day, day of the week, user's own profile, weather, and location (distance of the user from the clicked location). No significant correlation is found between proximity and relevance, though.

The short paper by \citet{Church2011a} presents similar analyses to the work by \citet{Church2007,Church2008}. It is based on a query log data derived from a mobile portal in Spain in late 2009 and containing around 1.4 million Spanish queries. The results include: query length is 1.6 terms on average, shorter than the other studies; queries categorised as ``adult'' are much less frequent, whereas navigational queries towards social websites account for a much higher percentage (around $40\%$); over $70\%$ of the queries are navigational (i.e., aimed at reaching a specific website); and query variety is rather low.

A study presented in the short paper by \citet{Yi2011} highlights some changing trends: query variety increases; head (frequent) queries become shorter and tail (rare) queries become longer; and the frequency of entertainment (especially adult) queries decreases substantially at the expense of local and commercial queries. Notice that these trends are very similar to those found in the early days of Web search, and that it is then reasonable to hypothesise that mobile queries, after an ``infancy'' are going to follow the same path. Voice queries on mobile devices are also analysed and it is found that they are longer and have a slightly different categorisation from typed mobile queries.

\citet{Koukoumidis2011} while aiming at studying the cacheability of mobile search for their PocketSearch service,
present a log analysis of 200 million queries of the mobile version of Bing,\footnote{\url{http://m.bing.com}} gathered over several months in 2009. The results show that both mobile queries and clicks on search results feature a high rate of repetition, at both community and individual levels. At the community level, just around 6000 most popular queries and 4000 most popular search results are responsible for around $60\%$ of all the 200 million queries and of the clicked search results, respectively. At the individual level, when a user submits a query, that will be a repeated one around $70\%$ of the time. It is also notable that users search for the same page by means of different queries. These results are of course interesting to support the query and search results caching on the mobile device, but they are also interesting to understand, more in general, Mobile IR users and their needs.

Some further log analysis studies are focused on location and context. \citet{Zhuang2011}, while aiming at designing a term query system (described in Section~\ref{sec:context-based-mobile}) briefly analyse a large query log, collected from October 2009 to March 2010 and made up of more than 75 million queries. Besides the usual query statistics, that do not differ significantly from other studies, also geographical and temporal aspects are singled out, and it is found that queries are more pervasive in large cities, especially in commercial areas, and during the day.

\citet{Lymberopoulos2011} aim to understand how the click behaviour of users is affected by the locations of the user and of the clicked result. The log is derived from approximately 2 million queries submitted in the U.S.\ over three months (the exact period is not detailed in the paper) through a mobile phone app to a major search engine (also unspecified). Results show that users  tend to click on results that are located within a short distance (e.g., at the U.S.\ level, $30\%$ of the clicks are within a 2km radius). However, quite often users seem to be willing to travel longer distances, usually 5-15km but in some cases even 80km. Also, there is quite a variation over different U.S.\ states, probably due to geographic features, traffic, and population density.
On the basis of these results, a Mobile IR model taking location into account is developed and evaluated.

\citet{Nicholas2013} focus on a single website, the Europeana portal and search engine\footnote{\url{http://www.europeana.eu}} dedicated to European cultural heritage (e.g., museums, libraries, archives). The log is derived from 2 years of Europeana usage in 2009-2011. Several aspects are analysed. Users are categorised into ``one-shot'', ``normal'', and ``heavy'', and their behaviour into ``bouncing'' (i.e., immediately leaving the site), ``checking'' (trying to find a specific fact), and ``exploring'' (some sort of casual browsing).  The differences due to the device type are highlighted: not only the classic desktop/laptop vs. mobile, but also looking at iPad that seems to be  more similar to desktop/laptop than mobile phones. The analysis of time of the day and day of the week of queries show that mobile users peak at night and during weekends.   As a final note, this study, when compared to the previous ones, is perhaps more oriented towards information-seeking than search, and it explicitly mentions that log studies provide complementary insights when compared to user studies (which we survey in the next section).

\citet{West2013} analyse the log of a navigation application.
They find that being at a primary / dominant location (home or workplace), as well as being at the distance from it, has an effect on search behaviour. Queries are issued with a higher frequency when the user is at the primary location, or closer to it.
Also, queries about the primary locations, or closer ones, are less frequent than queries about unfamiliar and distant locations, and
queries can be effectively used to predict future locations of the user.

\citet{Song2013} analyse Bing logs to compare search behaviour on iPhone, iPad, and Desktop.
  Reported results include: query length for mobile devices is slightly higher than what was previously reported (and, the authors conjecture, still evolving); query categories vary across devices; location of the issued query varies more for phone than tablet.
  Also, on iPhone and iPad, click-through rate is increased when improving the standard desktop ranker by adding some domain-specific features.

\subsection{Related Studies}
\label{sec:ir-related-studies}

Finally, we also mention some further studies that focus on aspects that are perhaps more eccentric but that can be considered still a part of Mobile IR in the broader sense; we think that they deserve to be included in this discussion.

Social strategies exploited by mobile users to maintain their social connections are the main topic of the work by \citet{Dong2014}. The data source consists of mobile phone data, namely, data taken from a real-world mobile network of more than seven million  users and over one billion   calls and SMS.

Mobile advertisements' logs are another type of user generated content which is extremely valuable for publishers.
Indeed, they typically establish contracts with ad-networks, in order to display ads on their web pages, and to earn commissions depending on the traffic driven to the advertisers.
More in detail, the latter place bids for their ads to be added to web pages, and they pay only when the ads are both displayed (i.e., there is an exposure) and clicked (i.e., chosen) by users.
Such cost-per-click (CPC) scheme is the most common paradigm in mobile advertising.
Hence, publishers, in order to maximise their income, are interested in displaying the ad with the highest expected revenue, i.e., the ad’s bid price times the probability it is clicked or clickthrough rate (CTR).
Ad-networks obviously collect logs of user clicks on ads to compute the publishers' commissions.
Those logs can then be used to estimate CTR, as in the work by \citet{Oentaryo2014}. In this paper they introduce a generic latent factor framework for mobile advertising, that incorporates importance weights and hierarchical learning, that is called Hierarchical Importance-aware Factorisation Machine (HIFM).
In particular, the experiments they carry out on real world datasets prove that their approach is effective in cost-varying scenarios and cold-start cases (i.e., new pages and ads), outperforming state-of-the-art methods (namely, TensorALS and TimeSVD++).

\citet{NorouzzadehRavari2015} distinguish location search (aimed at finding POIs, and usually reaching them) from local search (aimed at finding information about locations in the neighbourhood), focusing on the former.
In their log analysis of a common GPS application, they find that users' behaviour when carrying out location search is slightly different form Web search: it features shorter queries (measured  as number of terms) and sessions (measured  as number of terms and time); effectiveness seems high, on the basis of clickthrough data and actions indicating that a route is being selected.
Some differences between iPad and iPhone as well as some analysis of temporal dynamics are also reported, with no surprising results.
This paper can be seen as a concrete example, besides those already discussed in Section~\ref{sec:pois-collections}, of what anticipated in Section~\ref{sec:relevance} on the nature of relevance being more oriented towards the real, physical world.

\citet{Fan2016} apply a spatio-temporal collaborative filtering system  to call records to infer human mobility information on a city-wide scale.
Another city-wide scale study is reported by \citet{Song2016} whose focus is producing simulations and predictions about human mobility and transportation mode.
A solution for modelling public transport data in order to propose a system  able to address first- and last-mile problems as well as route choice estimation within a densely-connected train network is presented by \citet{Poonawala2016}.

\citet{Graus2016} use a sample of six months of logs of user-specified time-based reminders from Cortana  to identify common categories of tasks and to organise them in a taxonomy.
Moreover, they link some temporal patterns to the type of task, time of creation, and terms in the reminder text: the interesting outcome is that such findings become very useful in prediction tasks.

\section{User Studies, Surveys, Interviews, and Diary Studies}
\label{sec:diary-studies}

More qualitative and user-oriented research aims at understanding Mobile IR users motivations, experience, satisfaction, etc.\ by means of surveys, interviews, and diary studies. Here we still focus on studies aimed at understanding needs and users, leaving the discussion on evaluation studies to Chapter~\ref{cha:evaluation}.

Some studies are quite general: they do not focus on Mobile IR but  on mobile usage. A  paper of this kind is by \citet{Sohn2008}, that reports on a diary study involving 20 subjects asked to keep a  recording of their information needs over two weeks. The study interprets ``information need'' in a general way, and is not restricted to information needs that are satisfied  using a search engine. It is more a mobile information seeking than a Mobile IR study.
The results include a categorisation of information needs, an analysis of when the need is addressed, what contextual factors originated the need ( participants reported that as much as 72\% of their needs were originated from contextual factors), and the source of the information satisfying the need. Some ingenious ways used by people to find information are discussed and some design implications are derived, including the necessity to take into account  context (that we discuss in Chapter~\ref{cha:context-awareness}).

Still from a general information seeking perspective, \citet{Cui2008} perform a user study and derive a taxonomy of mobile Web activities. The taxonomy confirms in part previous research, and includes four kinds of activities: information seeking, communication, transaction, and personal space extension.

A diary study more focused on Mobile IR is by \citet{Church2009}, that aim to analyze the goal and intent of searchers. The classical categorisation into informational, navigational, and transactional needs that had been found on the Web does not find a confirmation for Mobile IR. Instead, information need intents are classified into informational, geographical (in turn, divided into local explicit, local implicit, and directions), and personal information management. Other interesting data include: most of the diary entries were generated when the users were mobile (i.e., away from home and work); location and temporal contexts were found to have an influence on information needs of mobile users, as well as on current activity of the user; when analysing the topic of the diary entry, the most frequent topics are local services, travel \& commuting, general information, and entertainment.

Most of the kinds of needs identified in the previous three studies \cite{Sohn2008,Cui2008,Church2009} are classified by \citeauthor{Tate2012} \cite{Tate2012} into a two dimensional matrix of mobile information needs, as  shown in Table~\ref{tab:needs-tate}.

\begin{table}[tbp]
  \centering
  \begin{small}
    \begin{tabular}{p{2.8cm}p{1.8cm}p{1.8cm}p{1.8cm}p{1.6cm}}
      \toprule
      &\textbf{Casual}&\textbf{Lookup}&\textbf{Learn}&\textbf{Investigate}\\
      \midrule
      \textbf{Informational} &Window Shopping &Trivia& Information Gathering& Research\\
      \addlinespace
      \textbf{Geographic} &Friend \newline Check-ins& Directions & Local Points \newline of Interest &Travel \newline Planning\\
      \addlinespace
      \textbf{Personal  Information Management}& Checking \newline Notifications& Checking \newline Calendar& Situation Analysis& Lifestyle Planning\\
      \addlinespace
      \textbf{Transactional}& Acting on \newline  Notifications& Price \newline Comparison& Online \newline  Shopping& Product \newline Monitoring\\
      \bottomrule
    \end{tabular}
  \end{small}
  \caption{The classification of mobile information needs (adapted from \citet{Tate2012}).}
  \label{tab:needs-tate}
\end{table}

Several studies focus on contextual aspects that affect Mobile IR, like location,  time, activity, and the social context.
\citet{Amin2009} for example, presented a web-based diary study on location-based search behaviour using a mobile search engine.  The diary tool collects users' detailed mobile search activity, such as users' location and users' diary entries. This method enabled to capture users' explicit behaviour (the query made),  implicit intention (the motivation behind the search) and the spatial, temporal, and social context in which the search was carried out. The results of the study showed that people tend to stick closely to regularly used routes and regularly visited places, like for example home and work. This has been shown to enable a better characterisation and classification of users' information needs for improved retrieval.

\citet{Hinze2010} describe a diary study for understanding needs. They analyse how location (e.g., home, a friend's home, work place, being mobile on the road as in a car, and ``going out'', as in shopping) influence the query, and in turn how the retrieved results influence the next activity of the user. Both variables are measured by asking the participants to directly record in a paper diary such influences on a five level scale. Although half of the diary entries are recorded at home, and around half of them are not related to location and activity, the reported results  show a rather strong effect of location on both queries (i.e., terms and query categories) and results (i.e., type and amount of information sought).

Another diary study, by \citet{Chen2010a}, focuses on needs during leisure traveling. Even in this different scenario, location, time, and activity are found to influence  quite strongly the user needs. When comparing directly the intents with the study of \citet{Church2009},  the distribution is slightly different, with the geographical category much more frequent in the leisure traveling situation. These results have obvious implications on the design of Mobile IR applications for tourists, like mobile guides.

\citet{Sterling2011,Sterling2012} analyses some reports  about when and where mobile search is performed (``on the go'', walking, at home, watching TV,  etc.). Although perhaps these two studies are outside the classical scientific literature, they summarise some findings that are rather common to several other papers. \citeauthor{Sterling2012} stresses that, perhaps unexpectedly, mobile search is often performed at home, in the evenings, during the weekend, or while watching TV. Also, mobile searches on a phone are often performed in front of a computer, and people tend to search more by phone than by computer. Finally, around 50\% of mobile queries are local, i.e., have a local intent.

\citet{Church2011} again study  mobile Web in general but, when describing the results of their diary study with 18 participants, they also focus on Mobile IR. Their results confirm those already reported above, in particular query length is about 3 terms per query and mobile search tends to have more importance or urgency than classical search.

The short paper by \citet{Teevan2011} focuses on mobile local search and reports the findings of a survey of 929 people.
Several contextual aspects are singled out as important, including: location (being at home vs.\ on the move), movement (stationary vs.\ on the move), and time (often needs are rather urgent).

One aspect that is common to several of the above described studies, and that is interesting to highlight, is the {\em social nature of Mobile IR}.
For example, an ongoing conversation is sometimes the reason to start a search \cite{Sohn2008}, and mobile Web is named a ``conversation enhancer'' \cite{Cui2008}.
Also, it is reported that a high fraction of searches is performed while in the presence of other persons, ranging from 63\% \cite{Teevan2011}, through 65\% \cite{Church2011} up to more than 75\% \cite{Amin2009}.

Many other papers are specifically aimed at studying social aspects of search.
For example, \citeauthor{Dearman2008}'s four weeks diary study \cite{Dearman2008} is not specifically about Mobile IR, neither it is oriented towards mobile in general, but it addresses some related social issues: the 20 participants shared information, especially with friends and family members; social networks are explicitly mentioned as means to answer information needs; other people too are indicated as information sources, especially those having weak ties with the searcher.

\citeauthor{Tan2012}'s diary study of 24 tourists \cite{Tan2012} is more oriented towards Mobile IR, although it has a general information seeking approach and it is specific to the tourism domain.
It is found that social interactions, communications towards both acquaintances and strangers is required and takes place.
The study by \citet{Church2012} is more specific, on social mobile search, and more complete, as
it reports findings of a survey of about 200 participants and a two week diary study with 20 participants.
It  confirms that location is a crucial contextual factors, perhaps even more important than what found in previous studies \cite{Church2011,Teevan2011}: social mobile search is most often used in familiar places, like at home (26\% in the survey, 29\% in the diary study) or at work (17\% and 24\% respectively). The distribution of information need types (classified accordingly to a previously proposed taxonomy by \citet{Dearman2008}) is quite different from other; for example, ``Trivia \& Pop Culture'' is the most frequent (31\%, 37\%), whereas it counted for less than 5\% in \citeauthor{Dearman2008}'s study. Also, it is confirmed that social search is often originated from conversations, and that search results tend to be shared and they often have an effect on the future plans and activities.

It should be noticed, however, that not all studies agree with these findings. In fact  some other results make the situation not so clear cut: for example, \citet{Chua2011}, on the basis
of their 20 participants involved in a one week diary study plus individual interviews, find that users make more use of the mobile phone when alone than when surrounded by other people.
However, overall, it seems clear that, although a mobile phone is clearly a personal device, it is used also with social implications.

\section{Conclusions}
\label{sec:conclusions-5}

This chapter  surveyed the papers studying Mobile IR users and their needs, focussing on the two approaches of logs analysis and user-oriented studies. The two approaches, described in Sections~\ref{sec:logs} and~\ref{sec:diary-studies}, feature some differences, and are complementary: of course, logs have the advantage of the amount of data; and the disadvantage that they can not provide insights into the user's experience,  intent and motivations behind the search. These, instead, can be studied in more details through a user-oriented approach.

Trying to predict what might happen in the near future, we can remark that there are not many studies yet aiming at collecting search data (queries, clicks, environmental and contextual data from sensors, etc.) directly on the phone.
Although there are some exceptions \citet{Woerndl2011}, this line of research will certainly grow in the next years, as it will allow to build a more accurate model of the users  and of their information needs.

\chapter{User Interface}
\label{cha:user-interface}

One of the most important distinctive characteristics of Mobile IR and differences between Mobile IR and standard IR is the user interface.  It is for this reason that much research has been devoted to study and improve it, going well beyond the areas of IR and Mobile IR. Also, much research in Mobile IR has been devoted to take advantage of these differences or to emphasise how to take advantage of them. This is apparent also from the terms reported in the tag cloud in Chapter~\ref{cha:mobile-ir-landscape}, related to the devices and to the interaction between the user and the device. However, in this chapter we will not deal with issues related to the human-computer interaction (HCI) between the user and mobile devices. We will simply highlight the differences between standard IR and Mobile IR from the point of view of  the user interface, putting the emphasis on the consequences that this has on the Mobile IR area.

This chapter is organised as follows. In Section~\ref{sec:intro-user-interface} we will discuss the differences  a mobile interface imposes on the user interaction. We will show how mobile interfaces have changed over the past 20 years and highlight the consequences this has on mobile information access. In the following two sections we go in more details into the two main modalities of interaction: input (Section~\ref{sec:input}) and output (Section~\ref{sec:output}), reporting research on how Mobile IR makes use of their peculiarities.
Section~\ref{sec:query-reform} addresses the specific issue of query reformulation and
finally Section~\ref{sec:conclusions-6} will report the conclusions of the chapter.

\section{A Different User Interface}
\label{sec:intro-user-interface}

As we have already anticipated in Section~\ref{sec:device}, one of the most striking differences between standard IR and Mobile IR is  the user interface~\cite{Schofield2002}. Mobile phones have gone thorough a lot of changes in the last 20 years, going from the ``size of a brick''  to the ``size of a pack of chewing gum'' and consequently changing the modality of interaction with users. In fact, while mobile phones of the first generation were nothing else than standard desk type phones with a battery and an aerial that enabled the user to carry them around, the latest generation of mobile phones are instead miniaturised computers with wireless communication functionalities.

\begin{figure}[tbp]
  \centering
  \includegraphics[width=11cm]{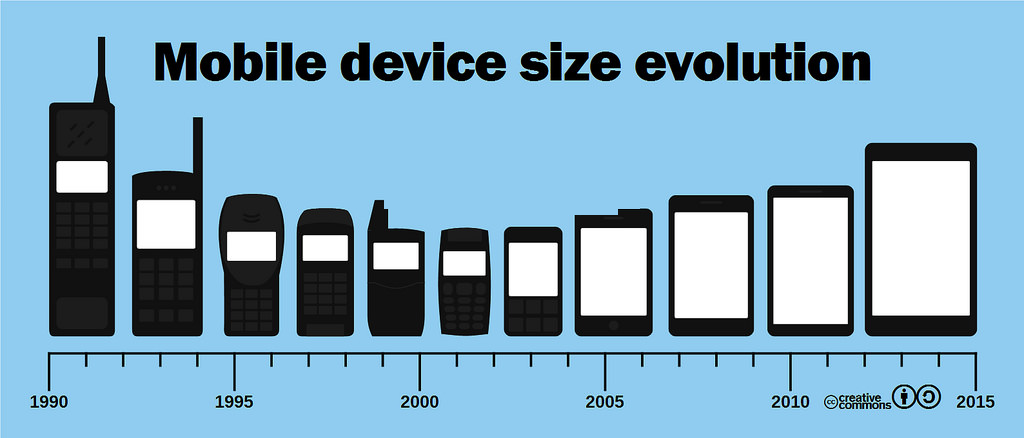}
  \caption{The evolution of the size of mobile phones and of their screens.}
  \label{fig:mob-size}
\end{figure}

The size of mobile phones has gone through a funny shape curve, resembling a ``U''. The first mobile phones were big and heavy, mostly for technical reasons. For several years the trend has been to reduce their size, until they reached really minute proportions, that made almost impossibile for them to have a functional screen or keyboard. However, most recently their size has been progressively enlarging, as they take more and more functionalities. Figure~\ref{fig:mob-size} exemplifies this trend.
Recently mobile phones are increasingly taking the shape of a high resolution touchscreen, with a microphone, a speaker and two videocameras, one on each side of the device. While this form makes them more and more similar to tablet computers, on the other side it extends considerably the interaction mode between  user and  device. In the rest of this chapter we will look more in details into the input and output functionalities of the mobile phone and the effect that these have on Mobile IR research.

\section{Input}
\label{sec:input}

One important difference between IR and Mobile IR is the range of inputs that Mobile IR enables. This is due to the fact that nowadays users input much more on their phones than just the phone number of the person they would like to call. In fact, input is very important for Mobile IR as it is the main way the user indicates his information need to the system. In addition, as most IR is currently interactive, it is also the way the user communicates with the search system, selecting the results to be displayed and refining his query. The modality of input is therefore very important, but has to be as little user-engaging  as possible, as the user is likely to be doing other things while searching on the phone. Thus, current mobile phones enable a wide modality of interaction styles: a user can type (or write with a stylus) his input on the screen, speak to it, or let it sense the situation in which the system has to ``guess'' what the user might be interested in.  In the next sections we will review the effects that these input modalities have on Mobile IR research.

\subsection{Text Based Input}
\label{sec:text-input}

The most classic form of input for a mobile phone is text. With this term we mean the input of text written by the user on the screen.

Mobile phones have gone through a gradual evolution on this front, passing from a series of advances that enabled to move from text inputed using the standard mobile phone keypad to using voice only.
The road was not easy and a number of different technologies were experimented to deal with the ambiguity generated by the user on a 12 keys keypad.
Different keypads (e.g., Dvorak, FogPad, FastTap), handwriting recognition (e.g., unistroke, Graffiti) and different technologies for disambiguating text entry (e.g., T9) were experimented with.
We will not discuss these input technologies here as they have almost disappeared from current mobile phones.
We refer the reader to the article by \citet{Dunlop2009} that provides an impressive review of mobile text entry research and adoption over the past 30 years.
Currently, most high-end mobile phones make use of touch screens (or, more rarely, stylus interaction) and full keyboards, that almost completely eliminate the needs for these technologies.
Still, the relative small keyboard and the fact that sometimes the user employs only one hand to write (as the other holds the phone) limits the speed and amount the user could write.
This compounds the user to write less and less and submit shorter queries (as we discussed in Section~\ref{sec:logs}).
This tendency, already studied for web search engines, is also
apparent for mobile phones.
An extreme consequence is that the query itself tends to somehow disappear
from mobile phone interactions with search engines and new approaches to capture the user information need will have to be devised (see Section~\ref{sec:zero-term-queries}).

An important component of almost all current search engines and certainly of all mobile phone search systems is Query Auto-Completion (QAC).
QAC is a technique for predicting and suggesting a user how to properly complete a query that is only partially written.
Studies in Japan (Yahoo!) and USA (Bing) have shown that QAC techniques are very useful as users tend to submit longer queries on mobile devices than on desktops to avoid having to query again and thus spend even more time on it.
We cannot review here the large number of approaches proposed for QAC on mobile devices.
We simply cite those that we believe are most original.

Standard QAC models rank a list of suggested queries given a prefix according to relevance scores based on various signals, such as, for example, historical query frequency, previous users queries, user profile, time, or user location.
While standard QAC techniques have been shown to help considerably mobile users, \citet{Vargas2016} propose a novel technique of QAC  that is specifically geared to mobile search.
Classic QAC operates by suggesting whole-query completions.
However, the authors showed that this is sub-optimal for mobile search, given the limited screen size and the unavailability of users to editing queries using the small keyboard.
The proposed term-by-term QAC, that suggests to the user one term at a time, instead of whole-query completion, proved to be more efficient in particular with regards to the query examination effort.

Another novel approach to the QAC problem is proposed by \citet{Zhang2016}, who exploit signals related to the recently installed or recently opened applications to suggest terms matching input prefixes.
For example, if a users recently installed or opened the \emph{Real Madrid} app on the smart phone, then when he starts typing ``real'' his query intent may be implied by that app, rather than real estate-related queries, as an historical query log might likely suggest.
Such AppAware model was shown to consistently and significantly boost accuracy of QAC on mobile devices.

\subsection{Spoken Input}
\label{sec:spoken}

The second most popular modality of input for mobile phones is speech.
In fact, with the recent progress in speech recognition \cite{Jurafsky2000,Allan2002}, this modality has been progressively becoming more and more popular with users and it seems to be destined to overcome or substitute almost completely text input.
This is particularly true for Mobile IR, where the most important form of input is already a spoken query.
In fact, nowadays most Mobile IR systems accept spoken input and are in fact optimised for such type of input.

The use of spoken queries is considered one of the main reasons for the increasing length of queries submitted to Mobile IR systems.
\citet{Kamvar2008} studied this trend in detail, comparing it with multitap text entry and written queries for PDA.
They showed how simpler modalities of user input, like spoken input, empowered the user to submit longer queries (in terms of number of words but not necessarily in terms of useful content), using less time to enter the query and giving him more time to inspect the search results.
These same results can also be found in the paper by \citet{Chang2002}, although they are more difficult to quantify, given that the experimentation was carried out in Chinese.
A more detailed analysis of this trend and an extensive comparison between spoken and written queries is presented by \citet{Crestani2006}.
This paper presents a user study of the quantitative and qualitative differences between written and spoken queries.
Users were presented with 120 TREC topics and asked to generate both written or spoken queries to retrieve relevant documents.
The results showed that users generated longer queries in spoken compared to written format, that is using more unique terms.
Not surprisingly, this did not take more time to the user, as it is intuitive to see that the spoken modality is faster than the written one.
There was also not substantial differences related to the complexity of the topics.
However, while the difference in length between written and spoken queries was substantial (and more than that observed by \citet{Kamvar2008}), an analysis of the part-of-speech of the queries showed that spoken queries contained a lot more prepositions, adverbs, pronouns, and articles, in proportion, than written ones.
Considering that these types of words are less informative than nouns and adjectives, this showed that this increase in length was not completely ``usable'' (remember that articles and propositions are considered stopwords in IR).
Nevertheless, considering the increase in the length of the queries and the increase in real numbers (and not in proportions) of nouns, adjectives and verbs, these were still more than sufficient to enable an increase in retrieval performance (if this was not affected by speech recognition errors).
Similar results were reported by \citet{Zhang2010}, that carried out a much more formalised and statistically strong study (although with only six participants) on a specific platform (iPhone).
More recently, the results of \citet{Guy2016} also supported these findings.
The study, that compared spoken and typed-in queries using the log of a commercial search engine's mobile interface, is currently the most substantial, given the provenance and size of their datasets.
The results, using naturally spoken queries, supported the previous findings related to the type of languages used by spoken queries.
In addition, it also found that, although spoken queries tended to focus on topics that required less interaction with the device's screen, they were still expressed in a language that was easier than that used in Q\&A.
In fact, while voice queries were longer and used a richer language than written queries, they often used a language that was far from natural language.
This suggested that the language of spoken queries was a separate type of language, in-between traditional text queries and natural language questions.

This increase in length of queries, however, does not come without its problems, related to speech recognition in a noisy and highly heterogenous environment. The heterogeneity, in fact, can be easily attributed to users voices, background noise, and different topics and could have devastating effect on the speech recognition. \citet{Crestani2003}  looked at the limitations of current spoken input from the perspective of an IR system and  showed that while speech recognition on the query was still a problem, there were a number of techniques that could be used to improve it. In fact, while the trend to have longer queries was confirmed also by \citet{Yi2011}, these served to express more diverse information needs that were certainly more complex than those initially submitted to Mobile IR systems. In addition it showed that the interests and information needs of mobile users have substantially changed compared to just a few years ago, being now more directed toward retail, local, automotive and finance types of queries. Thus, while the speech recognition process is considerably affected by the out-of-vocabulary problem that these types of query generate, affecting the query recognition, there are known techniques that can help, as we will see later  in the section.

In fact, while the typical query submitted to a Mobile IR in spoken form is still short, it is already longer than the equivalent written query. Besides, with the use of contextual information and more advanced form of speech recognition we could see that better results are easily within reach. \citet{Silipo2000}, for example, looked at the use of prosodic stress to identify the most important terms to characterise the content of a spoken sentence. The idea was to compare words selected in a spoken document or query using standard IR term weighting, with words selected using high prosodic stress. The results showed that there was a large overlap between the two sets, but the words that appeared only in one of the two sets were almost equally important to characterise the content of the document. In fact, when the two sets were joined,  the content of documents could be better identified.

Other non-verbal features have been investigated too. \citet{Picard2010} advocated  the use of sentiment and emotion detection using wearable technology to improve the understanding of a user information need. \citet{Jayagopi2010}, instead,  used mobile sociometric sensors to recognise the user conversational context in order to enhance online meeting support. The use of non-verbal cues was used to characterise the entire group and to discriminate the context of the conversation by the aggregation of participants nonverbal behaviour, seen on a temporal perspective. So, it is not a big step to consider the use of these sensors also to better understand the intention of a user in a typical query session. Such a use was  advocated by \citet{Feng2011}.

A study of users' satisfaction with intelligent assistants using voice interaction, namely, Siri, Cortana, Alexa, etc. is by \citet{Kiseleva2016}. The study showed how important contextual information was to properly understand a query and   how essential maintaining context thought a conversation was.  In a user study aimed at measuring user satisfaction over a range of typical scenarios of uses, the notion of satisfaction was found to vary across different scenarios, but the overall user satisfaction could not be reduced to a simple query-level satisfaction alone.

Finally, a different kind of IR, related to query by humming, is reported by \citet{Rho2011}.
The target of the paper is music retrieval, where human voice is used to produce a short clip of singing, whistling or humming to give a rough approximation of the music requested.
This form of query was processed in a very complex way to produce a representation to be matched against audio clips contained in a large archive.
Such a system was especially useful when the user did not have detailed information about the songs, such as its title or singer's name, but just remembered a short segment of the music.
The use of such a system, with little feedback from the user, enables to obtain performances that are significantly better than state of the art music retrieval systems.

\subsection{Mobile Phone Sensors}
\label{sec:mobile-phone-sensors}

Mobile phones have some inherent physical limitations compared with other computing devices (e.g., desktops, laptops, or tablet PCs).
They have a small screen, a tiny keyboard, limited processing power, etc.
On the other hand they have one great advantage over them: they are full of sensors.
In fact, advancements on sensor technologies has made them small and cheap enough to be embedded in great numbers into most modern mobile devices.
Sensors such as GPS, accelerometer, digital compass, temperature and light readers, are now commonly embedded in mobile phones.
If we consider also other standard components of a mobile phone, like for example the microphone or the mobile phone aerial, and think of their possible use as additional sensors, then we can see that mobile phones have a huge array of sensing capabilities.
A review of the state of the art of phone sensing is outside the scope of this paper, but we recommend the interested user to see the paper by \citet{Lane2010}, where a survey of sensors capabilities of current mobile phones is reported.

In recent times, researchers started working on innovative applications of these sensors that are different from those they were initially designed for.
This enabled to enhance the functionality of standard mobile phones.
For example, accelerometers were first introduced to enhance the user interface in order to determine the orientation of the phone.
However, accelerometer data can also be exploited in order to recognise user's different activities such as walking or running.
The combined use of the sensors gave rise to a new area of research called \emph{mobile phone sensing} where phones were used to sense the user surroundings and enhance user activities in a number of different domains.

The sensing technologies of current mobile phones also provide a new opportunity for Mobile IR.
Phone sensors data can be used to sense the user surrounding environment and better characterise the user context.
In fact, they are an ideal device to capture the user context as they are always carried around by users and can continually sense changes in the context.
Despite this recognised potential, many Mobile IR applications do not make use of all these sensors. In Chapter~\ref{cha:context-awareness} we will see how sensors are used to implement context-awareness in mobile applications.

\subsection{Zero-Term Queries}
\label{sec:zero-term-queries}

\emph{Zero-Term Queries} are queries that are directly triggered by some user action, rather than started by the user writing some terms in a query box of some search engine.
Instead, they are triggered by some user action on some application that can be considered implicitly related to some search for information.
An example of this is when the user opens the calendar to check the action to be taken on that day.
In this case it is the context in which the user is placed (e.g., the time, the location) that is used to generate the query and produce some results automatically.

Therefore, Zero-Term Queries systems  must be proactive, i.e., they must not rely on explicit manual input in order to start fetching and presenting potentially relevant information to the user.
Hence, an effective inference of the user's context (in order to correctly model his information needs) and a good ranking of the results are two prominent features of this kind of applications, i.e., failing one of those would hinder severely the overall effectiveness.
Many proactive applications, once launched, operate in background, providing the user with information cards to satisfy his information needs in different domains (e.g., news, entertainment, food, places etc.).
\citet{Shokouhi2015} carry out an analysis about the way users interact with such cards in Microsoft Cortana, highlighting some interesting patterns.
For instance, similarly to reactive search results (i.e., results triggered by manual queries in search engines), the top ranked results (i.e., info cards) yielded by proactive search account for the majority of the clicks.
Another similarity with reactive search results is about the dependence of the usage (i.e., click or viewport duration) of cards from temporal and position conditions: some topics are searched for only in certain moments of the day or in some places.
From the point of view of searched topics, users tend to be routinary, and this fact is confirmed by their consistent behaviour across different platforms (e.g., mobile, desktop).
In the second part of their experiments the authors take advantage of the above mentioned findings in order to improve card rankings.
In particular, they observe a noticeable improvement when using the reactive search history of users (i.e., their search logs) to rearrange the rankings of retrieved cards.

Thus, mining users' logs can provide much useful information for this kind of systems, in order to correctly model user interests.
This is precisely the main purpose of the experiments carried out by \citet{Yang2016}.
They start by collecting information from multiple user logs of different platforms (including reactive search engines and clicks and viewports on an intelligent personal assistant software), to proceed further by defining different classes of user interest features: (i) card type based implicit feedback features (IF), (ii) entity based user interests features (EF), and (iii) user demographics features (UD).
Then, information card URLs and types are used to extract features for modelling cards.
Finally, all those features are exploited to train a learning to rank model based on LambdaMART, which has proven to be very effective and it is used by many other researchers for personalised ranking tasks.
The resulting ranking system is called UMPRanker (User Modelling based Proactive Ranker) and it is triggered by context events which induce the ranking of the information cards by the model, pushing them to the user’s device.
The experimental evaluation proves that their UMPRanker outperforms a strong baseline approach (very close to the one used by \citet{Shokouhi2015}) in three different configurations, i.e., when using IF, IF+EF, and IF+EF+UD features.

\citet{Alidin2012}, to compensate for the lack of access to the content of different applications, propose a more advanced combined use of the different sensors of the mobile phone. For example, the combined use of  GPS and accelerometer data enables  the system to detect the kind of activity the user is performing and use this as a first step toward a ``just-in-time'' Mobile IR  system. Such a system would enable the user to receive required information just in time, that is when they are needed and before the user asks for it. This approach is an extension to the mobile environment of the just-in-time IR approach proposed by \citet{Rhodes2003}.

A rather different approach, but still within the Zero-Term Query class, has been proposed by  \citet{Yu2011,Yu2011a}, that present a system for helping the user to recover from a failed first query submitted to a mobile location search system. A novel ``active query sensing'' system helps the user forming a second query for location search by sensing the surrounding scenes. An offline process for analysing the saliency of the views associated with each geographical location, based on score distribution modelling is proposed. This is used to predict  visual search precision of individual views and locations and to estimate the view of an unseen query, suggesting the best subsequent view change.

Again, somewhat different from Zero-Term Query and more similar to ``query prediction'' is the work presented by \citet{Lee08} that employs profiling methods for monitoring user actions  to build content-based and location-based user profiles to be used for Mobile IR and mobile advertisements. On the same line is the work reported by \citet{Jang2010}, that used  the term ``proactive search'' to describe a similar approach aimed at predicting what the user might ask before he explicitly does it. However, the approach is somewhat different, as it builds an offline hypertext where the user can navigate, aided by the system, to locate the information items of  interest. The navigation is directed by the system through the provision of candidates anchors that are built based on the users' spoken words and on the perceived context.

The Zero-Term Query approach should not be confused with query recommendation, though. While both do not require a query, the Zero-Term Query approach  is mostly related to the user context, while mobile query recommendation is mostly related to the user's search intent captured from the combination of previous queries and context. \citet{Zhao2015}  introduce a system of mobile query recommendation that uses the user-location-query relation with a tensor representation. Through the use of a tensor function learning approach the system is  able to predict users' search intents and thus effectively recommend queries. It is easy to see how such an approach could be used to generate a Zero-Term Query.

Finally, this section should contain reference to a lot more work. However, as the Zero-Term Query approach relies a lot on the user's context to generate queries, we will review other approaches in the relative sections of Chapter~\ref{cha:context-awareness}.

\subsection{Intelligent Personal Assistants (IPAs)}
\label{sec:IPAs}

Nowadays, the sales decrease of traditional PCs in favour of smartphones and mobile devices has considerably shifted the investments of big actors (Apple, Google, Microsoft, etc.)
towards mobile technologies and services.
As a result, one of the emerging trends both as a research topic and as software commercial products is that of the Intelligent Personal Assistants (IPAs): the most popular are Siri from Apple, Cortana from Microsoft, and Google Now from Google.
They are essentially butler programs running in background and having the primary goal to anticipate users' needs, automatically retrieving information (in the form of \emph{information cards} according to the technical jargon used in this research field) from online services, search engines, social networks etc.
Hence, they appear as the perfect case study for the most part of the research results and frameworks about context awareness in Mobile~IR that we will discuss in Chapter~\ref{cha:context-awareness}.
Indeed, behind the user's interface and the non-trivial issues related to natural language processing, there is a ``black box'' which tries to exploit as much information as it can from user's input, activities, preferences, location etc.\
(the general rule of thumb is ``the more data you feed a digital assistant, the better it works''~\cite{IPAsWar2014}), i.e., from something which looks very akin to the notion of context discussed in the literature referenced in Chapter~\ref{cha:context-awareness}.
Besides, sometimes the IPA is proactive: this makes the link with Zero-Term Query systems clear.
In the following we will try to glance at some research issues related to Mobile IR raised by IPAs.

\begin{figure}[tbp]
  \centering
  \includegraphics[width=12cm]{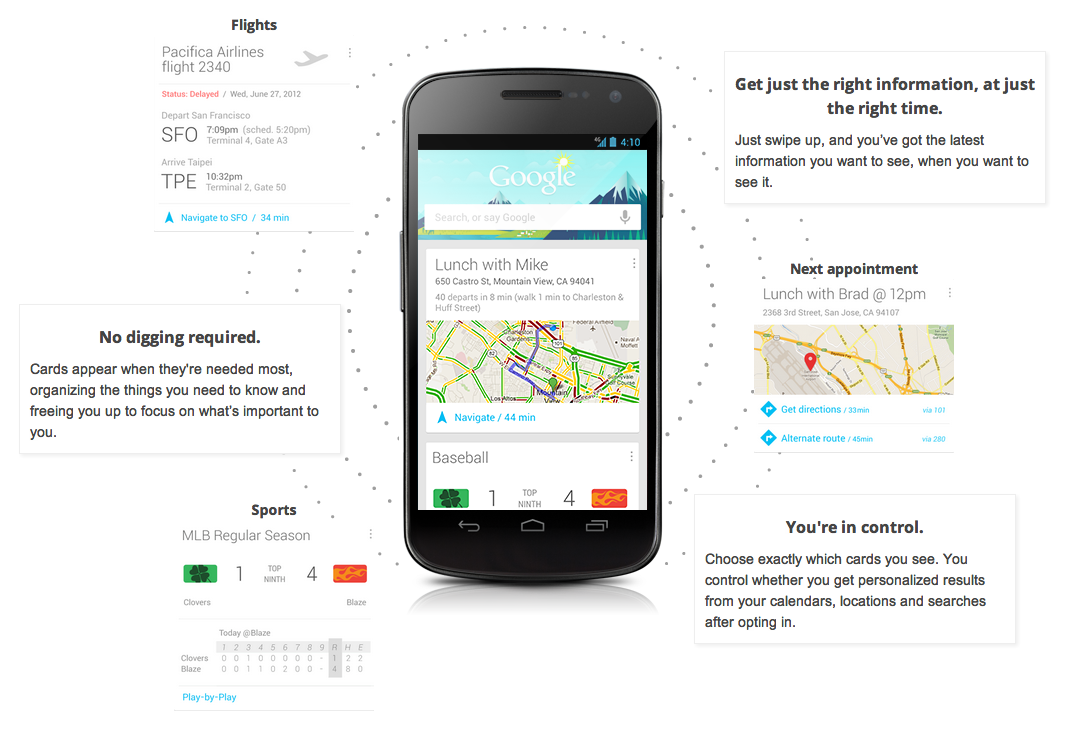}
  \caption{The Google Now interface.}

  \label{fig:google-now}
\end{figure}

Perhaps the most well known example among IPAs is the application Google Now, since it is available for many platforms.
It was launched in July 2012 for the Galaxy Nexus smartphone and made available for iOS on April 2013.
Google Now passively delivers information to the user predicting what the user wants, based on the user search habits and other information it knows about the user.
When the user opens the Google mobile search application Google Now starts by delegating requests to a set of web services, using a natural language user interface to make recommendations, and (this goes beyond the zero query scenario) to answer questions or queries.
Figure~\ref{fig:google-now} presents the interface of Google Now.

Google Now also recognises repeated actions that a user performs on the device (e.g., repeated calendar appointments, search queries or common locations) to display relevant information available online to the user in the form of ``cards''.
For example, cards could display information about flights, currency, hotel, location reminders, places, time at home, or weather.
The system leverages the Google's Knowledge Graph,\footnote{See: \url{https://www.google.com/intl/es419/insidesearch/features/search/knowledge.html}} a system used to assemble more detailed search results by analysing their meaning and connections.
While this is a very good service, acting in a clean and intuitive manner, some criticisms could be made on how this information is obtained and on how much data and information Google actually has about the user routines and daily lives.
We are confident that Google does not use this information for other purposes, but in the wrong hands this information could be of potential damage to the user.
Nevertheless, it is obvious that with all the sensors it is equipped with and with its ability to track and acquire hints (if not more) about the users needs from the analysis of applications' content (e.g., calendar, email, search engine), a mobile phone has the ability to accurately infer what the user might want or is interested in.

Obviously, IPAs are not limited to mobile devices, but they have a broad range of applications also in desktop systems, where proactiveness is still an important feature. For instance, \citet{Luukkonen2016} introduce a method for proactive IR, in order to automatically suggest background information to the user, when he is carrying out a writing task (i.e., it is writing a document). Again, information needs and user history (i.e., the user context) are taken into account, incorporating also text input prediction by means of a long short-term memory (LSTM) network.

\section{Output}
\label{sec:output}

As device technologies improve and advance, so do the services that they provide. Combined with the wealth of electronic information currently available, additional digital services add to the problem often referred to as \emph{information overload}. Frequently associated with the information consumer, information overload describes the effects of having more information available than the user can readily assimilate. It is not only the quantity of information  that contributes to the problem, but also the way in which information is presented. Thus, the design and presentation of content is of particular importance when accessing information in a non traditional setting, like for example in Mobile IR.

In this section we review research on the way information can be presented on mobile devices, keeping in mind their limitations in comparison with desktop and laptop computers.  We  start by looking at how standard text based information can be presented, and move into ways to reduce it,  summarising it, visualising it, or presenting it through maps or augmented reality.

\subsection{Text Based Presentation}
\label{sec:text-output}

A lot has been written on how little information could be presented on small screen devices.
Many of these studies were done in the last century and therefore dealt with the small size and limited resolution of mobile device's screens at the time.
In Section~\ref{sec:intro-user-interface} we already discussed how the screen size followed a ``U-shape'' curve, first becoming smaller and smaller, in the same direction of the size of the device.
Later, from 2004, the device started becoming again larger and larger, due mostly to the increasing size of the screen (see Figure~\ref{fig:mob-size}), which also enabled to display more information thanks to increasing resolution.

These technological changes had an effect on what information could be presented on the device in response to user query or information access interaction and on the utility of results presented. So, while studies carried out with small screens reported a clear drop in performance attributable mostly to the small size of the screen, they also showed that with a slightly larger screen and better resolution the performance drop was less damaging. \citet{Jones2002}, for example, compared searching on devices with different screen size and different interaction modalities. Compared with searching on a large-sized desktop screen, they showed how using a search engine on a Personal Digital Assistant (PDA)  was slightly less effective due to limited amount of information that could be displayed and the somewhat limited interaction style. For those same reasons they showed that  searching on a Wireless Application Protocol (WAP) phone was very ineffective.

Since 2010 the use of WAP started fading and as of 2013 it had largely disappeared. Most modern handset internet browsers now support full HTML, CSS, and most of Javascript, and do not need to use any kind of WAP markup for webpage compatibility. The list of handsets supporting HTML is extensive, and includes almost all handsets currently on sale. Most major companies and websites have since retired from the use of WAP and it has not been a mainstream technology for web on mobile for a number of years. In addition, screen have become much larger and their resolution has increased considerably, so much that text could be presented on mobile devices better than it was presented on desktop screen just a few years ago.

We can
conclude this short section saying that although small differences still remain between what could be presented on a mobile device compared with a desktop screen, these differences have become smaller and smaller.
So, unless somebody compared a mobile screen device with a large 24-inches curved desktop screen that could be found on many desks, there is currently no much different between what could be presented as text on a mobile device compared to a desktop computer.

\subsection{Summarisation}
\label{sec:summarization}

To meet the demands of anytime, anywhere information access, information has to be delivered in a form that can be readily and easily digested whilst on the move. Automatic summarisation can be employed to condense a textual document, presenting only the important parts of the full text, thereby reducing the need to refer to the source document. Therefore, at a document level, summarisation may be considered as a mean of reducing overheads in digesting information.

The intended use and consequent type of summarisation employed is an important characteristic, while another is the length of the summary. Summaries can be classified as ``indicative'' (of the content of the source document) if they provide a quick overview of the context, or ``informative'', if they provide a  detailed description of some aspects of the information contained in the document. A summary can also be described according to its orientation, being ``document-based'', that is containing information from the document author's perspective, or ``query-based'', that is containing information directly tailored to the interests of the user that submitted the query.

The work reported by \citet{Sweeney2006} builds on a technique called ``query-biased summarisation'', proposed some time ago by \citet{Tombros1998}, to present on the user's mobile phone with summaries that are related to the query that retrieved them. In addition,  the length of the summary is particularly important for mobile information access, given restrictions in screen displays and the associated navigation costs of scrolling vertically, or ``paging'', to view the content. The paper tries to find the optimal query-biased summary length for news articles. It reached the conclusion that about 7\% of the document length (of a British newspaper article) is  sufficient to provide the user with the required informative content to be able to make a mostly correct decision about the relevance of the document.

A similar approach is presented by \citet{Chen2007} with the difference that the focus is on spoken documents.
The technique is, at its core, the same.
An extractive spoken document summarisation algorithm automatically selects indicative sentences from a document according to a certain target summarisation ratio, and then sequences them to form a summary.
The main difference is given by the use of an HMM-based probabilistic generative framework to improve extractive spoken document summarisation by using information from relevant documents for each sentence of a spoken document to be summarised.
The method works also where there is no prior knowledge about the relevant set for each sentence.
In that case a local feedback procedure is employed by taking the sentence as a query and sending it to IR system to obtain a ranked list of documents.
It is assumed that the top K documents returned by the IR system are relevant to the sentence, and are thus treated as the relevance set of the sentence.
The probability of a document being generated by a sentence is modelled by an HMM, while the retrieved relevant text documents are used to estimate the HMM parameters and the sentence prior probability.
The results of experiments on Chinese broadcast news show that the new methods achieve noticeable performance gains over the standard HMM approach.

However, relevance is a multi-faceted concept and one important aspect of relevance, in particular for news, is novelty.
In fact, information that is already known by the user (i.e., not novel) is usually not relevant.
\citet{Sweeney2008} consider summarisation with novelty detection, where information is not only summarised, but also an attempt is made to remove redundancy.
Whilst the combination of summarisation paired with novelty detection is not a new concept, the work focuses on the mechanism of delivering the novel information to mobile devices.
The focus was on the concept of ``show me more'', where given an interest on a document's topic a user might wish to see additional information on that same topic and therefore trigger the system to generate and deliver new summaries.
The paper measures the effectiveness of two different approaches: one that adds the novel information to that already present (building a new and longer summary) and one that presents the novel information alone, assuming the user remembers what he read in the previous summary.
Thus, the paper also measured the value of providing information ``in context''.
Unfortunately, this new way of presenting information was not found to be statistically more effective than the standard way.
However, it did certainly save valuable screen space in document presentation.

Current mobile phones have a much higher screen resolution than only a few years ago, enabling the presentation of more text.
However, much attention has been recently devoted to graphical information.
Unfortunately, while the resolution has indeed increased considerably, the size of the screen of most mobile phone is not large enough to enable a properly readable presentation of this information.
Thus, in order to display web pages designed for desktop-sized monitors, some small-screen web browsers provide single-column or thumbnail views.
Both have limitations.
Single-column views affect page layouts and require users to scroll significantly more.
Thumbnail views tend to reduce contained text beyond readability, so differentiating visually similar areas requires users to zoom.
\citet{Lam2005} present a different approach, called Summary Thumbnails.
This is a thumbnail view enhanced with readable text fragments.
It helps users identify viewed material and distinguish between visually similar areas.
The thumbnails can be scaled arbitrarily, allowing them to fit the screen size of any target device.
In addition, the font size can be adjusted independently, which allows adapting Summary Thumbnails to the requirements of a number of difference scenarios.
Thus the system combines the benefits of thumbnail-based web browsing with the benefits of readable text.

A complementary approach is presented by \citet{Irie2011}, where the aim is to summarise an image collection. Recent studies have tried to develop a framework for image collection summarisation that extracts a smaller set of representative images from the original set. Most existing methods take into account the relevance and the coverage of each image to build the summary.  However, there are considerable issues related to the usability of these summaries. The paper introduces two additional factors when generating summaries: compactness and legibility. Their meaning is obvious. The proposed solution relies on a two-stage optimisation method. Given a keyword query and display size, the first stage ranks the images by taking relevance and coverage into account. The second stage takes into account compactness and legibility to determine the number and sizes of images included in the final summary, in order to satisfy the display size constraint.

\subsection{Results Visualisation}
\label{sec:clustering}

We have seen how summarisation helps the presentation of long text on the small screens that most mobile phones have. The presentation on long or complex information (e.g., images, videos) on a small screen is a very difficult issue that is bound to lose some of the details of the information presented. For this reason all search strategies adopted on mobile phones are precision oriented, that is they aim at finding a few highly relevant documents, maybe leaving aside other documents that are only slightly less relevant. This is a strategy used also by almost all search engines, but on mobile phones it has been focused even further to reduce the information to be presented to the user. This resulted in a number of different approaches that rather than presenting the usual ranked list of documents, either presented just one (or very few) good document or presented relevant documents in a summarised form, in order to help the user locate that one good document containing the information sought.

The first class of attempts is, obviously, the riskier, as it is very easy to completely misunderstand the user query and show a non-relevant document. Leaving aside the case when an answer, rather than one or more relevant documents is sought, like for example in \citet{Lagun2014}, the case with only one document presented to the user increases considerably the chances of user dissatisfaction. An attempt that is partially taking this view by showing a readable summary of some relevant web pages is reported in \citet{Lam2005}. This paper introduces summary thumbnails, that is enhanced thumbnails that present users  some readable text fragment of a web page that helps them identify relevant viewed material and distinguish between visually similar areas of longer documents (given the size of the mobile phone screen). This visualisation technique was found to be more effective in supporting web browsing  than single-column displays that for a while dominated the commercial market of small screen devices.

The second class of approaches that aimed at summarising a whole set of documents and present them in a more concise way was more popular.
One effective approach involves the clustering of results, so that results that are similar are grouped within the same cluster and only the cluster representative is presented, avoiding redundancy and keeping the presentation space utilised to the minimum.
This is the approach followed by \citet{Carpineto2009}.
This paper, building on a Web clustering engine based on concept lattices, presents two Mobile IR prototypes with search results clustering.
The first works on a mobile phone, while the second works on a PDA.
The extensive evaluation carried out shows that mobile clustering engines can be faster and more accurate than the corresponding mobile search engines, especially for subtopic retrieval tasks.
They also found that although mobile retrieval becomes, in general, less effective as the search device gets smaller, the adoption of clustering helps expanding the usage patterns beyond mere informational search while mobile.
The work presented in \citet{Mizzaro2012} goes one step further, using tag clouds as a way of visualising the results of the (same) clustering search engine.
A tag cloud is a set of terms organised spatially and graphically (in terms of fonts and colours) to visually highlight the most important terms.
Tag clouds are very common: they are being used quite often on the Web to show the tags used to annotate web resources, to summarise the main topics of a Web site, and so on (for example, we used them in Section~\ref{sec:mir-landscape} to present the topics covered by this book).
There are several kinds of tagclouds, that can differ for the selection of terms, the graphical aspect, and the auxiliary information shown.
In this paper a tag cloud is used to show the labels of the clusters.
The rationale for this approach is that a tag cloud can show the same labels and use less space than the classical tree-like visualisation, although admittedly in a less organised way.
The experimental user study showed that tag cloud visualisation is an effective visualisation alternative to clustering, particularly suitable to small screen devices.

Finally, \citet{Machado2009} report another approach to clustering and summarisation of query results. The main advantage of the approach is that it combines a new clustering algorithm with a new summarisation technique. The  new clustering algorithm is  called Clustering by Labels and it is specifically designed for web page results. The algorithm does not depend on pre-defined categories and, as such, can be applied ``on the fly'', as web results are retrieved from any domain, any language or any search engine.  As for the summarisation technique, the novelty stands on the use of the SENTA multiword extractor (proposed by one of the authors of the paper) that allows the real-time processing and language-independent analysis of web pages that is needed for the semantic treatment of the text  to be extracted, combined in a summary and visualised. Notice however, that the summarisation is still of the standard type, that is informative, extractive and document-based. Other approaches using clustering exploit the characteristics of these methods to work on any kind of media. So \citet{Moreno2011}  cluster images, \citet{Lam2005} cluster thumbnails, and \citet{Ying2012} cluster geospatial data. Of these perhaps the most interesting approach is presented in \citet{Moreno2011} where web image search results are clustered using a technique that expands the originally query-based Ephemeral Clustering and compare it with a query log based approach.  Without going into the complex details of this clustering algorithm, the method used showed more compact results compared to clustering query logs, although query logs showed a better user acceptance rate.

\begin{figure}[tbp]
  \centering
  \includegraphics[width=8cm]{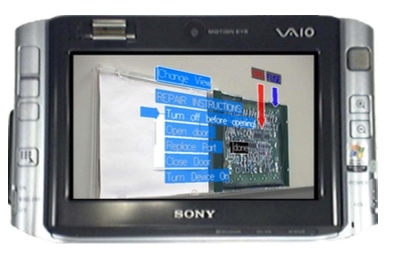}
  \caption{Search results displayed using augmented reality (from \citet{Mooser2007}). }
  \label{fig:aug-reality}
\end{figure}

A separate treatment should be devoted to maps and augmented reality visualisations. In fact, being the mobile device in the real world (see Chapter~\ref{cha:foundations}), result presentation could make use of the real world. Two main approaches have been explored: maps, used for positioning retrieved results on a topographical map so that they could be displayed in their geographical context, and augmented reality, exploiting the camera of the phone for enriching an image of the world with the retrieved data/documents. The best  of the second class of approaches is reported in \citet{Mooser2007} where an augmented reality user interface exposes disparate data sources through a single application server. The proposed system uses a multi-tier architecture to separate back-end relevant data retrieval from front-end graphical presentation and UI event handling. This enables to show Mobile IR results in a natural augmented reality scenario, as it can be seen in Figure~\ref{fig:aug-reality}.

Finally, recent research also focuses on cross-device search and results visualisation. Cross-device search comprises the ability to forecast the device that is going to be used for the next query, thus letting search engines proactively retrieve device-appropriate content, improving query disambiguation and results presentation. An example in this direction is reported by \citet{Montanez2014}, where search and results presentation on four different devices (laptop, smartphone, tablet and console) over a huge query log is analysed. The paper shows that people use different devices to search for different content at different times of day, having an impact on the content sought. The authors also showed how exploitable patterns emerged for device transitions with previous device signalling with a very high probability the next device, even when the devices differ. In fact, device-specific and temporal aspects of cross-device search were able to successfully predict the next device from which a user will query.

\section{Query Reformulation and Query Expansion}
\label{sec:query-reform}

In this section we  look at different ways in which  information presented to the user can be used for relevance feedback or query reformulation in interactive Mobile IR systems and thus employing the output as a new input for a new query process.

The first issue we need to deal with is the amount of time and complexity related to submit a query or to reformulate it.
Typing text on a standard 9-key mobile phone is difficult and time consuming: it has been shown that to type the average 15 letters long query that users of the mobile Google search engine (in 2007) took 30 key presses and approximately 40 seconds~\cite{Kamvar2007}.
Nowadays, most current mobile phones are actually smartphones and can display on their screens a full keyboard, although with small keys.
However, even typing text on a full keyboard is time consuming, although far less when the user can use text prediction techniques such as eZiType, iTap or T9.
It is nonetheless difficult since the user has to choose the right predicted word among those suggested, with a continuous switching of attention between typing and checking the screen.
For this reason, mobile users rely heavily on suggestions, when these are provided, as we have indicated in Section~\ref{sec:text-input}.

Two sets of techniques that are heavily used in interaction with mobile phones as query reformulation and query expansion.
These two techniques are quite different.
\emph{Query reformulation} is concerned with enabling the user to reformulate an already submitted query by providing additional clues and suggestions for a new query, like for example new terms to add to the query.
\emph{Query expansion} is instead concerned with expanding, mostly in a fully automatic way, the originally submitted query, to enable a more precise and comprehensive expression of the information need in order to produce a more accurate set of retrieved documents.
As these two techniques normally happen after a first set of results has been retrieved (and clearly found not so adequate) we discuss them in this section.

These techniques have been investigated and experimented for long time in standard IR, but not so much in Mobile IR. One of the first papers presenting an effective technique of query suggestion and reformulation on standard 9-key mobile phones is by \citet{Kamvar2008}. This paper explores the usage pattern of query entry interfaces that display suggestions. The purpose is to build a usage model of query suggestions to use for the design of a user interface for mobile text prediction.  The paper presented a test of six different interfaces, differing almost exclusively in relation to the number of suggestions displayed  as the user types the query. The study looked at the number of letters entered before the full query completion appears in the suggestion list and at the workload, enjoyment, key presses and time per query, among other things. Some interesting results were found. For example, the study showed that the time to enter a query did not reduce with the decrease in key presses and that, in fact, the presence of query suggestions slowed the number of key pressed per second, as the users spend a significant fraction of their entry time processing the suggestions. This trade-off, apparent to the research, was not discernible by the majority of users, who did not engage in a cost-benefit analysis of the time/effort spent in typing a query term or processing and accepting it as a suggestion. The study also looked at the movement of suggestions on the screen (e.g., moving up in the suggestion list as they become more likely) and showed a relation between movement and acceptance likelihood. In conclusion, the study suggested two improvements to query suggestions for mobile phone: 1) replacing suggestions viewed three times, as suggestions that were not accepted after that were unlikely to be correct; 2) replacing suggestions that create a net increase in key presses if accepted, to avoid the user choosing them and spending more time on that.

A related work aimed at query expansion, rather than interactive query reformulation, is presented by \citet{Paek2009}.  The paper proposes a method with a very small memory footprint index and retrieval algorithms for real-time query expansion on the mobile phone itself. Notice that the technique is applied on the phone itself, rather than remotely. So, the device suggests auto-completion of user input by allowing the user to select individual words and complete queries that are out-of-index, but whose words are in the index. The advantage of this technique is not only in the fact that it resided on the mobile device, but also that it reduces the number of keystrokes needed to type a query (even with disabled predictive text entry). Three versions of the proposed Phrase Building system were presented, using different user interfaces to enable the user to compositionally build a very phrase. The first one shows shows suggestion as whole phrases that can be composed by the user picking the chosen alternative from the mobile phone display. The second one shows individual words, using words columns where words are listed as the user types character by character, but can be easily chosen with one single click. The third one is a classical auto-completion interface, where as the user types characters, a drop-down box appears showing suggestions for the text-so-far. Notice that this interface does not enable the user to select a whole word with single click. The extensive evaluation showed that of the three interfaces the second was a clear winner as it facilitated higher accuracy in entering  the intended queries. Nevertheless, it also highlighted how the users' lack of familiarity with the interface hindered the acceptability of the technique and deterred from putting it into production.

A more recent  study is the one by \citet{Shokouhi2014}. By means of a  query log analysis, the paper focuses on query reformulation for mobile devices, comparing it to classical desktop query reformulation also by distinguishing between voice and text.  Reformulations are classified into various categories and it is  shown that rarely switch between different input types unless they are searching for a new intent or correcting speech recognisers errors. In fact, compared to searches on desktop, users are more likely to drop words from and spell-correct their mobile queries and less likely to use abbreviations.

Finally, it should be noted that the recent advancement in speech recognition and in interactive dialogues system (e.g., SIRI or Cortana) is likely to make work on interfaces for query expansion and reformulation useless. The ease of interaction afforded by voice, although with know limitation (see Section~\ref{sec:spoken}), is likely to direct the work in this area mostly to algorithm, rather than interfaces.

\section{Conclusions}
\label{sec:conclusions-6}

This chapter presented a long but still only partial presentation of the many issues related to the interface of a Mobile IR device.
We highlighted the key differences between a Mobile IR and a standard IR interface in relation to its main input (e.g., text, speech, and sensors) and its main output (e.g., text, summary, visual) capabilities.
We also indicated the direction of current work in this context.
However, this area is changing very fast, with mobile devices acquiring better and better capabilities every few years.
This is one reason why this area has seen a large number of papers being published and, as a consequence, this chapter is bound to become old very quickly.
Nevertheless, the lessons learned by researchers working on these topics will surely drive future research in the right direction and help avoid old mistakes.

\chapter{Context Awareness}
\label{cha:context-awareness}

The expression \emph{context-aware computing} is originally due to the seminal work of \citet{Schilit1994}, and it is quickly becoming more popular every day, together with other related and well-known terms such as \emph{ubiquitous computing} and \emph{pervasive computing}.
However, context awareness is a concept that is not easy to define, as we will see in Section~\ref{sec:context}.
The perception of the user context and its use as an aid for information access are one of the major areas of research that Mobile IR has made possible. This chapter tries to explain what context is and what being aware of it implies for Mobile IR.
It also provides a set of examples of applications using context and some of its many components that a mobile device enables to capture, such as location (Section~\ref{sec:locat-based-mobile}), time (Section~\ref{sec:ltime-aware-mobile}), contextual aspects (Section~\ref{sec:context-based-mobile}), and the society (Section~\ref{sec:social-context}).

\section{Defining Context}
\label{sec:context}

Many years ago \citet{Schmidt1999} published a paper entitled ``There is more to context than location'', realising the difficulty to give a precise and suitable definition of ``context''. Indeed, since the advent of the first mobile devices, many researchers have realised that users' needs are often strictly related not only to the query and the current location, but also to other aspects and activities. For instance, \citet{Koshman2011} starts with the study of four possible ``flavours'' of the \emph{mobile context} notion, namely, spatial/location context, temporal context, social context, and access/technical context. Indeed, even if mobile user's information needs are strongly affected by where they are and by what they are doing, the notion of context cannot be always reduced to a place and an activity. In particular, that paper emphasises the role played by the access/technical component of contexts. Indeed, the activity of finding information with mobile devices is characterised by both ``itinerancy'' and ``intermittency'' (think, for example, of a user moving from an area covered by WLAN connectivity to another area where the only possibility is to connect through a cellular network, and finally to a zone where no Internet connection is available at all). Thus, a new framework is proposed, splitting mobile information requests into two subclasses: (i), ``action-based'' requests, i.e., activities carried out by users to address a given information need upon which something is pending, (ii) ``non-action based'' requests, i.e., immediate information needs whose solution will fill in a missing piece of knowledge.

\citet{Alidin2012}, besides recognising of the importance of context in Mobile IR, suggest the idea of exploiting the availability of sensors in mobile devices, in order to automatically acquire the current context and use it to provide users with ``just-in-time'' relevant information, without the need of making explicit requests. Thus, context affects mobile users' experience and it can influence technology acceptance. This possible correlation is precisely the object of the study carried out by \citet{Arning2010}, using a focus-group-interview approach. In detail, two mobile services, concerning respectively an ICT context and a medical one, are contrasted in order to isolate motivations and barriers for/against the usage of the services, acceptance patterns and the impact of other aspects such as age, gender or technical experience of the users. The experimental results show that acceptance factors are indeed both a combination of individual features/profiles and very service-specific. Thus, the usage context of a service plays a key role in its success or failure among mobile users.

Since the interplay between  context and mobile applications often implies the presence of many automatisms in the software, it is important to provide users with suitable explanations about the complex and autonomous behaviour\footnote{A common example of an autonomous behaviour is a smartphone automatically silencing incoming calls, whenever it detects that the current location is a meeting room or a quiet place like, for example, a library.} of context-aware apps. Such explanations can help users in trusting their devices, but it is still not clear how to provide them in an effective way. In order to address such task, \citet{Lim2011} build a client/server framework allowing users to share their availability status. Low-level sensor data (e.g., position coordinates, accelerometer values etc.) are collected by mobile devices which also carry out some processing in order to send the extracted features to the server-side component. Then, the latter computes some low-level contexts (namely, Place, Motion, Sound activity, phone Ringer, Schedule, and the Contactor or inquirer) which are used to synthesise an availability context. Moreover, the server supports a query mechanism to provide clients with explanations, in both textual and graphical formats, about its behaviour and decisions. Using such platform, the authors carry out a think aloud user study to isolate patterns of intelligibility use, drawing from them some useful  guidelines to build ``intelligible'' mobile context-aware apps.

In their attempt to define in a more precise way the notion of context, \citet{Bazire2005} collect from web pages a corpus of 150 definitions used in several domains of cognitive sciences and related disciplines (e.g., computer science, philosophy, economy, etc.). Then, they perform a thorough semantical analysis to such corpus, using two approaches, namely, LSA and STONE. The results clearly indicate that some definitions of context in different disciplines can be very specific and dependent on their domain, while other definitions are highly general and could be applied to a wide variety of research fields. Applying clustering to LSA, the authors find some interesting groups of definitions: (i) ``around the idea of a psychological process of
representation building'', (ii) ``about the spatial dimension of context'', and (iii) ``taken from linguistic''. The subsequent STONE analysis contributes to highlight semantic inclusion relations between concepts (called categories in \citet{Bazire2005}) emerged from the definitions. The outcome of this phase isolates six main components: constraint, influence, behaviour, nature, structure and system. Moreover, the emerging idea of the context is related to something with a dynamic nature. For instance, more than half of the definitions used in the study are about the context of a behavior, where the latter can be either an action or a cognitive activity. Hence, they propose the following characterisation:

\begin{quote}
The context acts like a set of constraints that influence the behaviour of a system (a user or a computer) embedded in a given task.
\end{quote}

It follows that the gist of this idea of context relies on the information that must be conveyed together with an object in order to build a correct representation of it. The divergence between the specific definitions of context in many disciplines are then due to their different focuses (and objects). Hence, the authors introduce a model of context which represents the ``components of a situation'' and ``the different relations between those components''. A situation is then defined as a user, an item/object (embedded in a physical environment), and an observer. Thus, there is a clear distinction between the (physical) environment and the context which may underpin some aspects of the former, but it is not limited to those features.

Mobile guides helping users to visit cities, museums, exhibitions, etc. are probably one of the most common examples of mobile applications exploiting the notion of context to provide personalised content. Research in this field counts several works, well before the diffusion of modern smartphones with large screens and touch interfaces. For instance, \citet{Choi2008} introduce MyGuide, a ``Mobile Context-Aware Exhibit Guide System'', exploiting PDAs and ``old fashioned'' mobile phones based on Java Micro Edition and using RFID technology and tags in 2008. Their underlying notion of context is based on the following components: visitor's knowledge level, facilities of the visitor's mobile phone (whence the media type to be received: text or multimedia), preferred language, and location. The latter is determined by means of a RFID tag (associated to the mobile phone at the entrance of the exhibition) which is read by suitable readers placed at the exhibits.

One of the first attempts to embed in a systematic way the context into mobile applications is represented by the MoBe framework, introduced by \citet{Coppola2005,Coppola2005b} to allow the automatic download of applications (called MoBeLets) according to the current user's context. MoBe's overall architecture requires a small ``kernel'', named MoBeSoul, which must be installed into the mobile device to exchange data with special servers connected to a common network (e.g., the Internet, a Wi-Fi LAN or even a Bluetooth PAN). The MoBeSoul module  collects and analyse data gathered from several kinds of ``sensors'': (i) physical sensors (providing data about network connection type, noise, light level, temperature, etc.), (ii) virtual sensors (providing data coming from other processes running on the same device) and (iii) MoBeContext sensors (providing explicit context information pushed by an ad-hoc MoBe Context Server, MCS for short). Other data are gathered capturing explicit user actions (e.g., setting an alarm clock) and analysing the ``context history'', i.e., the sequence of contexts detected by MoBeSoul. Indeed, the main purpose of the latter is to exploit an inferential engine abstracting from raw data to build a higher-level representation of the context the user is currently in (e.g., ``reading the agenda at home''). Such contextual information is then sent to the MoBe Descriptors Server (MDS) which in turn sends back a list of potentially related MoBeLet Descriptors. That list is filtered locally, according to the personal settings of the user, and the resulting descriptors are sent to the MoBe MoBeLet Server (MMS), which responds with the envoy of the related executable codes. Finally, the chosen MoBeLets are executed in a local sandbox managed by the MoBeSoul module. It is worth noticing that a single MoBeLets can in turn be context-aware applications that can exploit, through a suitable API, the whole MoBe infrastructure.

A paramount example of a mobile application which can benefit from the context-aware paradigm is the web browser. Indeed, the user of a mobile device may find some difficulties in performing the typical search activity  consisting in typing a query for some search engine (e.g., Google, Yahoo, etc.), submitting it, and waiting to scroll the (long) list of results, then possibly refining the query until a suitable web page is found. Obviously, there are no issues in carrying out such activity on standard desktop/laptop computers, but mobile devices have some serious ergonomic limitations, which hinder complicated and long interactions with the user interface. Indeed, it would be better if the web browser could foresee the information needs of the user and automatically provide related web pages, without requiring an explicit user interaction. This is precisely the idea of \citet{Coppola2010}: capitalising on their previous experience with the above mentioned MoBe framework, they introduce the concept of a Context-Aware Browser (CAB), which is capable of inferring and exploiting the current context and ``push'' related web contents to the user. More precisely, the CAB architecture is a three-layers software: (i) the bottom layer handles the sensors and inferential network,  sends contextual queries to ad-hoc servers, and  filters the web content descriptors received in response from such servers, (ii) the middle layer works as a bridge, and (iii) the top layer manages user interaction, leveraging on an AJAX-based engine. The filtered web content descriptors are then used to automatically retrieve the web pages related to the current user's context. The reader should grasp the analogy between the MoBeLet descriptors in the MoBe framework and the  web content descriptors in the CAB: indeed, the latter can be seen as a specialised MoBeSoul whose aim is to provide context-aware web contents instead of context-aware apps.

Independently, \citet{Gasimov2010} introduce the Context-Aware Mobile Browser (CAMB) which, differently from the CAB, delegates the inference of the current context to an external Context Estimation Server (CES). The latter receives from the mobile device some device information in a first step and user and environment information in a second step: all those data are used to predict the ``context info''. Such information is then sent back to the mobile device which embeds it (through the \texttt{meta} property) into the CSS file used to render standard web pages retrieved through regular HTTP interactions. If no context info is available, web pages are rendered in their original style. One limitation is that the whole architecture can work only if the list of contexts is bound, since developers must know the existing possibilities to provide suitable CSS files for each context. Moreover, differently from CAB where the user can decide which data can be sent to public servers (protecting sensible information which remains secured into the mobile device), the CAMB architecture straightly exposes users' information to CES servers: this opens potential privacy issues which force the authors to obfuscate devide ids and to log users only for limited periods of time.

\citet{Mizzaro2011} embed the social dynamics driving the evolution of the Web 2.0 into the CAB, developing a variant named Social CAB (SCAB for short). Here, the users become active entities, since they can interact with resources (web pages, services, applications, etc.) and contextual information. More precisely, they can directly define their current context through collaborative annotations shared with other users. For instance, they can mark a particular resource as relevant to their current context, they can associate resources to contexts, and they can list resources marked as relevant for their current context. The aim of the SCAB is to engage users into the definition of the contextual information, in order to obtain a more dynamic and personalised context representation.

The interest for context in Mobile IR together with the development of a plethora of context-aware solutions yields the problem of properly evaluating such systems. For instance, in the case of SCAB, \citet{Menegon2009a} introduce a TREC-like benchmark and they use it to evaluate several strategies for automatically building queries based on the current context of the user. In detail, the proposed benchmark maps the usual three components of TREC, namely, topics, documents, and relevance judgments to context descriptors, web pages, and relevance judgments made by a unique judge using a four level  scale.

An interesting longitudinal study  carried out  over one month by \citet{Lee2005} highlights that the use of mobile Internet services is clustered on a few key contexts. In particular, most of the time users are using Internet services in ``public place, without any social interaction, while not moving, and when off duty''. The authors argue that this is perhaps due to the ergonomic limitations of mobile devices and to the difficulty of interacting with a mobile device while moving. Moreover, the number of  services used is very low and they can be divided into utilitarian/passive (e.g., news reports) and hedonic/active (e.g., games) types. The correlations found by the authors between services and use contexts are also interesting. For instance, they highlight that active services are typically used in low distraction contexts, where the user may concentrate on the interaction with the service. On the other hand, passive information services are typically chosen in contexts where the user is bored or trying to save time.

More recently, \citet{Goker2008} address the problem of evaluating mobile information systems for tourists. They stress the importance of collecting users' judgments at the moment they are using the system, and not in a future time for the sake of a proper evaluation. However, this forces the implementors to limit the knowledge overhead on behalf of the user, since ``time is of the essence'' for travellers/tourists and too much granularity in the required judgment could hinder the evaluation process. As a consequence, during preliminary studies asking for ``relevant'' or ``not relevant'' is sufficient. Only when users can bear with more complicated assessments, a finer precision scale for the judgments can be implemented.

In the following sections, in an attempt of disentangling the quite complex scenario depicted above, we will look at the different dimensions of context in a more systematics way, exploring them detail.

\section{Context Awareness Dimensions}
\label{sec:context-dimensions}

The importance of several contextual features (location, time etc.) was already ascertained by the studies discussed in Chapter~\ref{cha:user-needs}. In this section we will resume such analysis, providing more details on the different dimensions of context awareness.

\subsection{Location Awareness}
\label{sec:locat-based-mobile}

Physical location is definitely one of the most common elements of context (in many cases location and context are considered synonyms), at least when we speak of mobile computing. Indeed, being mobile intuitively means being able to move from one location to another. Hence, it is natural to think that ``geographic information'' can influence the information needs of mobile users. Moreover, the concept of relevance in Mobile IR must also take into account the location dimension. \citet{Raper2007} has been one of the first researchers to clearly state such idea, introducing definitions and models of geographical information needs and geographic relevance (see Section~\ref{sec:relevance}). He starts observing that in the literature ``situational relevance''\footnote{Situational relevance is a key manifestation of relevance and it is usually modelled as a function depending on task and situation.} (SR) does not take into account neither mobility nor geography. In particular, the absence of the latter contradicts several user studies where geography has been recognised as an important relevance criterion. Hence, he introduces the notion of Geographic Relevance (GR). In Mobile IR systems information is geographically relevant if it meets an information need whose aim is: (i) to assist user's mobile behaviour in real time, or (ii) to plan something before an event takes place. As the author says:

\begin{quote}
Thus geographic relevance is a kind of situational relevance that arises when a user has spatiotemporally extended information needs requiring geographic information to support mobility in space and time.
\end{quote}

Geographic information seeking (GISk) is then the process of searching for geographically relevant objects as a response to a query containing some geographic references. All of this makes sense only in the presence of a model of geographic space offering two possible interpretations: (i) a proper geometric space defined by a geo-centric coordinate referencing system, and (ii) a set of places and landmarks. In particular, the second component is crucial, since human beings continuously use mental constructs like places and landmarks to represents and communicate geographic information. Thus, also geographic information needs (GINs) are rather complex; indeed, they can be classified according to two dichotomies:
(i) intentionally-driven vs. task-driven, and (ii) geo-representation vs. geo-context. A GIN is intentionally-driven if it expresses the ``need to know'' of the user, i.e., the user must ``fill a gap'' in his own cognitive map. On the other hand a task-driven GIN is the need to acquire some geographic information to accomplish some task (e.g., directions for driving a car to a new destination). The expression geo-representation means that the GIN must deal with geographic information in the form of maps, images (e.g., oriented pictures or augmented reality) or directions, while geo-context refers to places, landmarks (in the surroundings of the user) or to the activity-defined envelope. Finally, the GR of geographic information objects (GIOs) can be assessed in terms of: (i) geometry/time or locality for ``allocentric domains'' (i.e., where the user's attention is a detached overview), and (ii) spatial perception or manifestation for ``egocentric domains'' (where the user's attention is a personal view).

\citet{Cherubini2011} investigate the reasons and the barriers behind the adoption of context-aware mobile services. In their study they isolate 24 contextual needs with 8 (one third of the total)  spatio-temporal in nature, confirming the importance of location as a notion of context. The 8 contextual needs that are spatio-temporal are defined as those: 

\begin{enumerate}
\item retrieving recommended places during idle time,
\item searching or tracking the current position,
\item searching or tracking the position of friends in the surroundings,
\item searching location of static entities,
\item searching location of dynamic entities,
\item retrieve location-based advertisements,
\item finding locations of specific goods to be bought,
\item learning historical facts of the current location.
\end{enumerate}

The influence of location in Mobile IR has been highlighted in a thorough diary study by \citet{Amin2009} (see Section~\ref{sec:diary-studies} for the details), discovering how the routine behaviour of users, who tend to stick with regularly used routes and visited places, allows one to properly characterise and classify their information needs for an improved retrieval. Another interesting study about the role of location is carried out by \citet{Lymberopoulos2011}, who use it to predict clicks in mobile local search. Indeed, as already discussed in Section~\ref{sec:third-wave}, they first analyse about 2 million local search queries by US users; then, use such data to elaborate some location-aware features to implement some models for click prediction in local search on mobile devices. The outcome of their experiments highlight an increase of precision between 5\% and 47\%.

Of course, strictly identifying the notion of location (in the sense of spatial context) with the geographic position (e.g., the latitude and longitude coordinates given by a GPS device) may be too limiting. This is exactly the point of \citet{Ahlers2009} who study spatial contextual features in information systems ranging from simple position-aware solutions to fully context-aware ones. For instance, mobile users tend to move often; whence their geographical location changes quickly and this automatically adds the temporal dimension to the context. Other spatial features, besides physical position, taken into account by the authors are: viewport of a map visualisation, average and current speed, heading, current date and time, elapsed duration of a trip, history of previous locations, location of the departure point of a trip; spatial environment features (e.g., road networks, topology etc.). Such features are then used to build several kinds of spatial queries (point queries, matching point-in-polygon queries, distance and buffer zone queries, first-N or k-nearest-neighbours, path queries, and multimedia queries) and filters (spatial and temporal proximity, prediction, visibility, field of view, and viewport) for an effective selection and ranking of the search results.

Spatial filters are the main focus of another study carried out by \citet{Mountain2007a}, who highlight their role as a solution to improve relevance and to reduce the number of results to show to the mobile user, thus removing the need of a tedious manual filtering. The filters used by the author are a subset of those described by \citet{Ahlers2009}, namely, spatial and temporal filters, prediction filter, and visibility filter. The evaluation results, obtained from a user testing carried out over 100 users by means of a location-based service (LBS) implemented in the WebPark project within the Swiss National Park, confirm that spatially filtered information is considered more relevant than unfiltered information.

Another attempt of overcoming mere location in Mobile IR is represented by the work of \citet{Cheng2010} who carried out one of the first attempts to fully use the reading of a GPS sensor. GPS sensors enable not only to know the location of the user, but also  the direction the user is facing, through the direction the mobile phone is pointed to. In that paper the authors proposed a novel ranking algorithm that  leveraged  sensors readings for traditional content-based image retrieval. The readings were used to  annotate images with geolocation and compass direction that, together with image content analysis techniques enabled to improve image retrieval.

As we mentioned above, mobile users tend to change their current location quickly since they move often. Hence, it is possible to record such location changes to build paths/trajectories. Leveraging on this, \citet{Amini2012} propose TAS: a Trajectory-Aware Search application  capable of predicting user's destination in real-time and displaying search results near the predicted location. Their experiments show that their trajectory-search approach allows users to find results faster and with higher satisfaction w.r.t. conventional search. This conclusion seems to be in contradiction with the outcomes of the study carried out by \citet{Mountain2007b}, who claim that ``users find information sorted by spatial proximity to be more relevant than that sorted by a prediction surface of likely future locations''. Perhaps such divergence can be explained by analysing the different ``domains'' of the related studies and the approach used to calculate the possible future locations of a mobile user. In the former case the user's trajectory and, consequently, the predicted future location is constrained by city routes and by a grid approach in modelling the space. In the second case the prediction filter calculates prediction surfaces on the base of the recent spatial behaviour of the mobile user, exploiting speed, directional trends, and the level of sinuosity in his path.

The diffusion and popularity of context-aware search engines introduces the problem of how to evaluate them properly, that is, how to measure user satisfaction and how to compare different approaches. Indeed, traditional evaluation methodologies are hindered by  contextual aspects (user profile, current situation) which unavoidably impact on users' judgments and ratings. To overcome such issues in the evaluation of their context-aware mobile search solution, \citet{Bouidghaghen2010} propose a mixed methodology based on context simulation and user study.
In detail, they prepare a benchmark composed by a search space corpus (3.750 web pages), a set of queries (25), and a set of hypothetic geographic context situations (6). The search space corpus is obtained by retrieving from Yahoo BOSS the first 150 documents per query. Queries are then associated to the six context situations (use cases) with clear indications about the relevance of a document w.r.t. a query and a geographical context. During the user study, users are asked to give ratings to the top 20 documents retrieved by Yahoo BOSS (without any personalisation) and by their context-aware solution (more precisely users rate $\langle$query, description, document$\rangle$ triples, where description stands for context information and relevance assessment indications). Then, user profiles are automatically generated by a simulation algorithm exploiting the manual ratings of the $\langle$query, description, document$\rangle$ triples of the benchmark. The intended purpose of such profiles is to simulate user click-through data (i.e., user interactions). Final outcomes highlight an improvement of the authors' solution w.r.t. Yahoo BOSS ranging from 50,92\% to 87,50\%.

The interest in trajectories and spatial filters to properly rank and select location-aware results is justified by the results of the survey carried out by \citet{Teevan2011}. Indeed, as we saw in Section~\ref{sec:diary-studies}, interviewing 929 Microsoft employees, the authors discover that, while performing mobile local search, only 40\% of the total number of queries is about places near the current location of users. Interestingly, 68\% of queries are generated while users are in transit and they want information about their destination (27\%). Moreover, the majority of the users (63\%) are discussing in a social platform their local searches.

From a concrete point of view there is plenty of examples of solutions exploiting location-awareness. Indeed, \citet{Frohlich2009} speak about the birth of a new ``interdisciplinary research field called mobile spatial interaction (MSI)''. The underlying idea of MSI is to provide the mobile user with the capability of interacting with a world where the direct access and manipulation of spatial information is possible. For instance, \citet{DeSilva2009} take advantage of modern smartphones and cameras facilities to embed photos with geo-referenced metadata about the location where they are taken, in order to implement an automatic system capable of creating travel photo albums and travel stories. At search time, the only action required by the user is to input spatial and temporal queries making sketches on a map displayed on the device. Then the system decodes the registered sketches, performing the related queries and retrieving photos and multimedia items both from the device and from online collections.  \citet{Karpischek2009} observe that LBS solutions are usually implemented as hardware devices or installed as apps in smartphones, leading to fragmented markets and poor standardisation. Hence, they propose to empower common web browsers with location-aware enhancements, moving LBS to the server-side.

With the continuous technological improvements and the presence of embedded cameras in mobile devices, mobile location search based on visual recognition is quickly becoming a popular service. Ideally, the user can take a photo of his surroundings and sends it to a suitable server, this tries to find the best match of the received photo with a local dataset of other views and sends back related location-information to the client. The main issue is that such solutions still have about 0.4--0.7 average precision and so, in some situations (e.g., ``urban jungles'') less than 50\% queries succeeds. As we already saw in Section~\ref{sec:zero-term-queries}, to overcome such problem, \citet{Yu2011,Yu2011a} propose a technique called Active Query Sensing (AQS) to intelligently provide the user with directions to take a more discriminative photo when the first query fails. Their experimental evaluation shows that the failure rate drops to 12\% after submitting the second visual query.

Coming to more peculiar but still relevant examples, \citet{Braunhofer2013} design and implement a music recommendation system suggesting the user an itinerary from the current location and playing music tracks related to the visited places of interest (POIs). Such relationship is built by asking volunteers to manually tag items (i.e., music tracks and POIs) using a controlled
vocabulary of adjectives from the Geneva Emotional Music Scale (GEMS) whose model features nine categories of emotions (each of which with 2€"-4 distinct emotional tags). Another interesting application of movement data to mobile security is represented by the work of~\citet{Lin2015}. The authors propose to identify illegitimate users of mobile devices exploiting anonymised location data for inferring mobility behaviours patterns for each single user.

Since nowadays the web traffic generated by smartphones and mobile devices is definitely prevailing, there are many economic interests in developing effective advertising mechanisms. Obviously, there are crucial differences w.r.t. standard advertising techniques used for desktop web surfing. For instance, ads need to be selected and shown accurately to prevent user's cognitive fatigue (for example, mobile ads are shown sequentially over time, due to screen size limitations). Moreover, user's movements can be tracked in mobile devices and such information may be exploited to implement an effective advertisement solution. This is precisely the focus of~\citet{Gatti2014}, who implement a system providing a location-aware advertisement plan. The authors start from the following argument: the same ad can affect users in different ways, according to their current position and path which may reveal user's intentions. Indeed, an ad of a shop far away has a lower probability of interest the user seeing it than a shop located along the path which he is currently following. However, given the current location of a user, there can be many paths the user could follow in the near future. An algorithm addressing the solution via two different strategies favouring either efficiency or user satisfaction is proposed.

\subsection{Time Awareness}
\label{sec:ltime-aware-mobile}

Time is undoubtedly another important element of the context. Indeed, in their four-week diary study of mobile information needs, \citet{Church2009} report that 8.4\% of mobile queries include direct temporal references. Moreover, time information usually goes together with other contextual features like, fir example, location. For instance, we recall from  Section~\ref{sec:locat-based-mobile} how \citet{DeSilva2009} allow users to input queries making sketches on a map displayed on the device. Such queries convey spatial and temporal information about the context related to the photos and multimedia items to retrieve on behalf of the user. Indeed, as we already noticed, the idea of being mobile naturally implies a change of position during an interval of time.

\citet{Church2008a} are among the first researchers recognising that search queries coming from mobile devices must be augmented with contextual information in order to allow search engines to provide a list of relevant results. They include time, besides location and community preferences, in their notion of context. Indeed, the user can see on a map of his current location, in addition to  queries and results made by other users of the community. The presence of two sliders allows the user to filter the set of queries and results visualised on the map. One of the sliders affects time ranging from ``now'' to ``earlier'' and it is set by default on ``now''. Thus, by default the user can see all recent queries and results. Moving the slide towards the ``earlier'' position, the user ``goes back in time'', as the interface  displays queries and results generated at an earlier time. Such feature turns out to be very useful to find temporal patterns of queries in a given location, and to monitor their evolution over time, highlighting some possible users' routines.

Moreover, as highlighted by \citet{Sohn2008} (see Section~\ref{sec:diary-studies} for the details of their study), time is particularly important also in deciding ``when to address an information need''. For instance, ``urgent needs'' are usually generated by contextual factors, e.g., the user's current activity, and they require immediate attention since they are time critical (e.g., getting directions while driving). In other cases, when time is not critical, information needs may be satisfied in a later moment or in other ways, e.g., by asking another physical person in the surroundings, after evaluating the costs of the expensive available network connection. 

\citet{Bouidghaghen2009} propose to extend OWL-Time ontology with the following classes: time of day, day of week and season. Indeed, they introduce a situation-aware personalised search approach, where: (i) semantic (i.e., high-level) situations are inferred from low-level location and time data, (ii) user's interests are extracted from  search history and related to  given situations, and (iii) a case-based reasoning (CBR) approach is applied to select a suitable profile according to the current situation, in order to provide personalised search results. The interesting outcome of their work is that, exactly like the model of location must go beyond the mere GPS coordinates, also the representation of time  must go beyond the pure timestamps provided by the system clock.

Some years later, a similar approach is exploited by \citet{Sassi2012} who model user's situation by means of three components: location, season and time of the day. Hence, there are two temporal components, over a total of three, to model the context of the user. Such choice is justified by the will of taking into account the user's interests in their information retrieval system. Indeed, the current season and time of the dayallow to connect the user's query  to his routine, revealing possible activities and interests usually carried out at that time. For instance, if the location is ``mountain'', the season is ``winter'' and the time of the day is ``morning'', perhaps the user current activity and interest is ``skiing''. In their experiments the authors enrich the original query with inferred user's interests, obtaining an improvement over simple queries carried out with Google search.

The impact of time in mobile web search practices is confirmed by the already cited survey of~\citet{Teevan2011} (see Section~\ref{sec:locat-based-mobile}). Indeed, while ``only'' 8\% of mobile queries directly involves time, the latter has some interesting consequences in the user's search experience. For instance, the search for a place is aimed for a visit in the same day in 89\% of the cases: moreover, in 44\% of these situations users claim to do it as soon as possible, while in another 29\% of these cases users plan to do it as part of their current path. In general mobile users would like to find search results  near in terms of time. Only 9\% of users does not care about the time required to reach the destination result. 

However, \citet{Lymberopoulos2011,Lv2012} show that in Mobile IR features based on location seems to be more important  than temporal features, at least for mobile click prediction in local search. Indeed, the experimental evaluations carried out by the above mentioned authors show that dropping the traveling distance feature causes a major loss in effectiveness, while a similar effect does not happen when removing other properties (including temporal features).

Moreover temporal scores are often too difficult to assess, and they may be removed from future editions of the TREC Contextual suggestion track (see Section~\ref{sec:locat-based-mobile}). Nevertheless, time information still plays an important role as a context component, in particular when users exhibit routine behaviours. Finally, as we will see in Section~\ref{sec:social-context}, mobile social search (i.e., mobile search made with other people) may lead to ``merging'' space and time together. For instance, in the survey carried out by \citet{Church2012} (see Section~\ref{sec:diary-studies} for the details), it turns out that at the second place among the information needs of mobile social search is ``Finding'' (28\%) which is not limited to finding the name or location of someone/something, but it is strictly related to the time it takes to travel from the current position to reach the destination.

Space and time are also jointly used by \citet{Gidofalvi2012}, who propose a complex statistical approach based on an ``inhomogeneous continuous-time Markov model'' to predict when a given user will leave his current location and where he will go. In detail, the proposed solution first analyzes GPS data to infer some grid-based staytime statistics. Then, it extracts the regions frequently visited by the user, and finally it applies the Markov model to perform the predictions. It is easy to see how such predictions could be useful for Mobile IR. Empirical evaluations, using a number of long, real world trajectories from a large data set, showed that the proposed method outperforms a state-of-the-art, rule-based trajectory predictor, both in terms of temporal and spatial prediction accuracy.

We conclude this section mentioning an original study of  temporal context to improve the user experience of mobile devices. Indeed, \citet{Pekhimenko2015} address the problem generated by the so-called trending search topics, which can hinder the end-user search experience introducing longer delays because of their ``unpredictable query load spikes''. The authors propose a new system, named PocketTrend, which is able to detect trending topics in real time, identifying the related search contents and smartly pushing such content to clients. The positive consequences are that the workloads on the datacenters are reduced and client search engines can respond to users' queries instantly.

\subsection{Beyond Location and Time Awareness}
\label{sec:context-based-mobile}

In general, the evolution of mobile devices and the widespread of different kinds of sensors on mobile platforms (see Section~\ref{sec:mobile-phone-sensors}) led researchers and programmers to exploit the latter to provide a broader notion of context, encompassing several environment features and contributing to the already cited claim ``There's more to context than location'' by~\citet{Schmidt1999}. Such features are potentially useful, from a Mobile IR point of view, to better represent user's information needs. This is not surprising: as user's location and time of the day can be exploited to better understand user's needs, it is quite likely that a more general representation of user's context could be useful to some extent.

Indeed, besides Mobile IR, a rich notion of context (beyond the ``classic'' dimensions of location and time) is generally useful in mobile usage. For instance, as reported in Section~\ref{sec:diary-studies}, \citet{Sohn2008} find that 72\% of users' needs are originated from contextual factors, including ongoing conversations and the current activity.

Of course, different situations can reveal different information needs. Since it is difficult to perform direct observations of users' behaviour, qualitative studies are useful tools to carry out a thorough analysis of such needs, and to point out the roles of different aspects of context. For instance, as we have already seen in Section~\ref{sec:diary-studies}, \citet{Chen2010a} focus their diary study (based on 200 entries made by 14 users in a period of 11 days) on information needs emerging during leisure traveling, that is, in a situation quite different from usual daily routine. However, even in this different scenario, location (53\% of entries), activity (17\% of entries), and time (10\% of entries) are found to influence quite strongly the user needs. Indeed, they represent  80\% of question types. The real difference from the daily life emerges when the intents behind the information needs are taken into consideration. Indeed, during leisure travel the geographical intent is dominant (60.5\%), followed by the informational intent (34.8\%) and by personal information management (PIM: 4.7\%). Instead, during daily life we have at the first position the informational intent (58.3\%), followed by the geographical intent (31.1\%) and by PIM (10.6\%). Hence, context is always important in a mobile setting, but the increase of mobility in leisure traveling ``activates'' the geographical intent more often than in daily life. This is also confirmed by the topics involved in the analysed diary entries, namely, places and mobility information (restaurants, sights, shops, hotels and transportation). These kinds of studies and the related results can be very useful for designers and implementors of Mobile IR applications for tourists, like mobile guides. Another interesting diary study, carried out by \citet{Hinze2010} on 12 users for a total of 220 entries over a week, confirms the influence of both location and activity in determining user's information needs. In particular, the authors find that both query words and desired answers vary according to location and activity. For instance, queries made in shopping locations tend to use the word ``how'' in order to refer to ``pricing and selection of goods'', while queries made in public locations use ``where'' since they are usually related to geographical directions.

As we anticipated at the beginning of this section,  context-aware computing is definitely influenced and made possible by hardware advancements in mobile devices. This fact is witnessed by the rise of  frameworks that process raw data coming from device sensors, in order to extract useful information and to synthesise an abstract notion of context. The main difference across the proposed architectures  are about (i) the range and type of sensors, (ii) the context inference engine, and (iii) the services supported by the middleware. For instance, \citet{Santos2009} introduce a platform integrating sensors already embedded in the mobile device with external ones connected via Bluetooth (in the paper a vest and backpack prototypes are presented). The framework allows the plug-in of different external inference engines as well. Another interesting feature is that there are both a local context inference on the mobile device (obviously related to a specific user) and a global context inference which is carried out at server-side, where data related to different users are aggregated in order to obtain more refined information.

We cite again the excellent survey of mobile phone sensing capabilities carried out by \citet{Lane2010}, observing that the notion of sensor can be abstracted to include \emph{virtual (or software) sensors} taking their input from Internet services and social networks, rather than from physical devices (e.g., accelerometers, gyroscopes, etc.). In that case the resulting framework may offer a great variety of features to synthesise a richer notion of context. For instance, in the case of the ``I'm feeling LoCo'' system by \citet{Savage2012}, discussed also in Section~\ref{sec:social-context}, the authors combine information inferred from the user's social network profile and  the user's mobile phone sensors devoted to location inference. The resulting recommendation framework greatly benefits from such synergy of information sources, allowing to determine the vehicle used for transportation.

The importance of being capable to represent a rich notion of context (not limited to the current location) is also testified by the birth of several context-aware frameworks like, e.g., Hapori, proposed by \citet{Lane2010a}. Indeed, the authors find that the effectiveness of local search engines is dependent on several contextual aspects, namely, time of the day, day of the week, user's own profile, weather, and location (distance of the user from the clicked location).

A set of context-aware prototypes supporting a rich notion of context is presented in a series of papers by \citet{Coppola2005,Coppola2005a,Coppola2005b,Coppola2010,Mizzaro2011}. A common framework leverages on a Bayesian inferential network as the context generator engine (called \emph{MoBeSoul}) which is then embodied by three different prototypes. The first prototype, MoBe is introduced by \citet{Coppola2005,Coppola2005a,Coppola2005b}, and aims at proactively retrieving and filtering, on the basis of the current context, mobile applications (named \emph{MoBeLets}) on one's own mobile device. Notice that in 2005 neither iOS nor Android were released yet, whence no app stores were available. Indeed, the idea of automatically distributing applications to mobile devices was a pioneering vision and Java Micro edition was chosen as a unifying developing platform, being at that time the most common runtime platform for smartphones and mobile devices. The set of sensors supported by MoBeSoul includes: (i) physical sensors (e.g., light, temperature, GPS, etc.), (ii) ``virtual'' sensors, namely, listeners of data from other processes running on the mobile device (e.g., an agenda, a timer, an alarm clock, etc.), (iii) MoBeContext sensors (ad-hoc servers pushing or broadcasting detailed information about the current context to the users' devices), and finally (iv) explicit user actions made through the user interface.

The second prototype named CAB (for Context-Aware Browser) \cite{Coppola2010}, leverages on technologies and habits introduced by the advent of Web 2.0 and the widespread rising of responsive web applications. Hence, instead of writing ad-hoc fully-fledged mobile applications, developers can shift to a web-oriented development style, exploiting the diffusion of mobile web browsers, and opening the way for the reuse of existing web content (whereas MoBeLets need to be developed from scratch). The only ``smart'' native application which must be installed on the device is the CAB, which embeds the MoBeSoul engine and all the data structures and mechanisms needed to communicate with sensors in order to infer the current context.

The third prototype, named SCAB (Social CAB) ~\cite{Mizzaro2011} allows users to tag both web pages or services (named ``resources'') and contexts, paving the way to a ``collaborative'' context-awareness (we will analyze such system in Section~\ref{sec:social-context}).

Another hint on the generality of the notion of context (discussed in Section~\ref{sec:context})  comes when comparing the approach of MoBe, CAB, and SCAB with the paper by \citet{Menemenis2008}, that propose AQUAM, a system to automatically derive queries to be shown to the user of a mobile device. The queries are selected by a pipeline process that starts by extracting metadata from the current web page visited by the user, and involves several steps including a semantic, an entity-based analysis, a validation activity, a querying a search engine, and a query expansion. An experimental activity is undertaken to evaluate the effectiveness of  the individual modules of the system and the overall reliability of the generated queries.

AQUAM is quite similar to MoBe, CAB and SCAB when considering its aim of automatically deriving queries from contexts. However there are some notable differences like, for example:

\begin{itemize}
\item the queries derived in AQUAM are not automatically executed but are proposed to the user for his/her selection, whereas in MoBe, CAB, and SCAB the queries are proactively executed and the results shown to the user;
\item AQUAM is justified by referring to the input and output limitations of mobile devices, whereas MoBe, CAB, and SCAB are based on the idea of exploiting the opportunities provided by context-awareness.
\end{itemize}

However, one important remark is that the notion of context is rather different in the two situations of AQUAM focuses on a very specific context feature, namely the semantic metadata on the visited web pages. On the other hand, MoBe, CAB, and SCAB envision a much wider notion of context, including data received from other running processes and sensors data.

Besides general purpose context-aware frameworks, mobile recommendation systems belong undoubtedly to a set of applications enjoying several benefits from a rich notion of context-awareness. For instance, in their recommendation system, \citet{Zhuang2011} partition the context into (i) a user context regarding ``past behaviours'' and (ii) a sensor context involving ``time and geolocation''. Hence, also in this case the authors go beyond the classical dimensions of location and  time for context. More specifically, they propose an efficient generative probabilistic framework estimating the conditional probability of suggesting either an entity type (i.e., a business category) or a specific entity (i.e., a local business site) for a given user in the current context. Such choice has the advantage of avoiding complex machine learning algorithms which could hinder the capability of the system to return an answer in real time. The interesting novelty of this framework relies on the fact that it models user intents implicitly (i.e., without resorting to explicit queries), by leveraging on a large collection of mobile click-through data from a commercial mobile search engine. Moreover, it is worth noticing that such data is related to all users, paving the way to a ``social'' notion of context along the lines discussed in~Section~\ref{sec:social-context}.

\citet{Tintarev2010} add the question of personalisation to context-awareness in their proposed POIs recommendation system. Indeed, they start from observing that common recommendation applications for tourists, even when they take contextual features into account, tend to suggest popular POIs in the surroundings of the user. This is not always the best solution, since it does not give a chance to the tourists to discover other potentially interesting POIs which, albeit not very popular, may be more suited to their interests and tastes. In order to overcome such issue the proposed recommendation system asks the users to provide as input five keywords describing their favourite kind of POIs. Hence, processing such data besides other typical contextual information (e.g., the GPS coordinates of the user), the system is capable to suggest also ``off the beaten track'' places, which can probably meet the personal tastes of tourists. Naturally, the personalisation approach used by the authors can also be seen as an additional context feature or, according to a dual perspective, the context can be partitioned into an objective component (location, time, etc.) and a subjective one (user's tastes, interests etc.).

Concluding this section, we can say that context-awareness is undoubtedly a feature tied to mobile devices and to their usage in everyday life. Moreover, such notion is enforced and enriched, beyond the mere location and time dimensions, by the user's behaviour which is usually routine. This is also true from the information discovery point-of-view. Indeed, \citet{Dumais2003} have carried out an extensive case study over more than 230 people, showing that they find more easily the needed information by re-using previously seen stuff, rather than retrieving new ones with ordinary search tools (e.g., online search engines). The likely reason of such behaviour relies on the fact that previously seen information can provide rich contextual cues, easing the users' search activity when they need to retrieve something already used in the past. Such results are taken into consideration and ``specialised'' for mobile devices by \citet{Jang2010}, when they start noticing that users tend to exploit their smartphones as ``lightweight information-capture tools'', registering information that may be re-used later. Moreover, the ergonomic limits of mobile devices play in favour of tools and interfaces requiring a minimum input effort on behalf of the user. Hence, the authors propose a proactive system, leveraging on a ``mobile hypertext'' framework to anticipate users' information needs. Indeed, their system proactively searches for content which is defined to be \emph{personal information}  including items like, e.g., e-mail or short messages, contacts etc. and items retrieved from the Internet. The above mentioned mobile hypertext can also be dynamic, including nodes that provide a ``ranked list of information item surrogates''. The framework seldom requires entering text queries. The user's interaction instead consists in a selection from a list of related information items which are proactively displayed based on  contextual data.

\subsection{Social Awareness}
\label{sec:social-context}

From the analysis of several papers taken into consideration in previous sections of this chapter and in Section~\ref{sec:diary-studies} we can infer a common underlying pattern about the behaviour of users when they try to satisfy their information needs, namely, the trend to address or involve other people either directly in a conversation or indirectly through the interface proposed by popular social networks or remote communication tools. Whence, we can deduce that users and users' actions may be part of the context as well as time, location and other features considered in  previous sections, paving the way for a \emph{social notion of context}.

Moreover, as we said in Section~\ref{sec:mobile-phone-sensors}, mobile phones are featured with many sensors which often are not all used by applications. Thus, sensors and data coming from social networks could provide much additional user contextual information that could be employed to better interpret  information needs. For instance, the readings of different sensors can be combined and used to infer user's activities and possible information needs, as proposed by \citet{Lovett2010}. In that paper the analysis is centred on the user's shared online calendar, which could prove to be a potentially valuable source of contextual information, provided, of course, the calendar represents an accurate account of real-world events. Unfortunately,  the calendar often does not represent reality well, as  events are hidden by a multitude of reminders and ``placeholders'' and  often appear in the calendar but do not actually occur.  This problem could be partially solved thorough the use of other information provided by social networks and location data, that could be ``fused'' with the  information in the calendar to significantly improve its representation of real events.

Another interesting approach to POI recommendation, exploiting ``data fusion'' with information extracted from social networks, is reported by \citet{Savage2012}. The paper tries to dispense the user from completing complicated and time consuming survey of his preferences. This information does note take into account the user's current context and is thus often inaccurate or not in sync with the current location and the timing of the recommendation. Considering this,  the interesting part of the approach reported in that paper is in the fusion of the information inferred from the user's social network profile with the user's mobile phone sensors for place discovery. In addition, the fused data enables also to discover the user's mode of transportation, thus making recommendation even more precise. Thus, this combination enables the identification of the spatiotemporal constraints that allow a more precise and timely location recommendation.

A similar approach to POI suggestion is presented in \citet{Aliannejadi2017} where the traditional notion of context as time and location is expanded to include also the social dimension. This is constituted by tags assigned to venues by the user, indicating features of the venues the user liked and disliked, which are expanded using a probabilistic generative model by tags attached to that same venues by other users, who expressed similar judgements. The approach, which makes use of data from Yelp, Foursquare and TripAdvisor, proved very successful in the 2015 and 2016  TREC Contextual Suggestion tracks \cite{Dean-Hall2015,Hashemi2017}, indicating how social information can prove to be very useful in all cases where relevance information from the user is lacking.

A more systematic approach and a case study of a fully-fledged framework, leveraging on the social dimension for the notion of context, is introduced by \citet{Mizzaro2011} under the acronym SCAB (i.e., Social CAB). In particular, such framework is an evolution of the CAB (see Section \ref{sec:context-based-mobile}) taking into account users' interaction, that is, considering other users as part of the current user's context.

The resource and context tagging activities allow to enrich the representations of the resources and of the contexts, respectively. Adding tags means to include in a context description terms that have not been inferred from the sensors, and to add to a resource description (e.g., a web page) terms not contained in it. This  allows  SCAB to performs its two-step process: first it retrieves the most relevant contexts on the basis of the current sensors data; then the context description is used to retrieve  resources that are more relevant to the current user context.

Resorting to user-generated content by means of social networks, user reviews etc., to provide personalised and more satisfactory recommendations of POIs, is the approach adopted by \citet{Biancalana2013}. In particular, they rely on two collaborative filtering algorithms: the first is based on the popular nearest neighbourhood approach, while the second is able to adapt to different situations, weighting users' ratings by means of a similarity measure between the current and past contexts. Effectively recommending POIs is also the focus of \citet{Huang2012}, namely, they show how mobile guides may benefit from context-aware collaborative filtering applied to GPS trajectories. Finally, \citet{Minsoo2014} introduce Glaucus: a social search engine for location based queries, exploiting  information coming from social networks.

We conclude this section observing that social data and mobile data in general are an interesting source of information to analyse human behavioural patterns. Moreover, from such inferred patterns which ultimately represent the normal routine of users, it is possible to detect anomalous behaviours, i.e., behaviours significantly different from the ``normal'' patterns. For instance, RMBAD (Routine Mining Based Anomaly Detection) is a generative-based system, introduced by~\citet{Qin2016}, that ``learns the pattern of normal from naturally existing human routines, free of any manually extracted features''. The  goal is to detect anomalies in mobile phone data. The same aim is the  focus of another study carried out by~\citet{Dong2015} who propose a system being able to analyse crowd behaviours in urban environments through the same kind of data exploited by~\citet{Qin2016}.

\section{Conclusions}
\label{sec:conclusions-7}

Mobile devices and applications implicitly imply a notion of dynamicity: the user is free to move in different environments and his behaviour is influenced by the surroundings in many ways. First of all, his current connection to the Internet may depend on the environment: Bluetooth tethering, Wi-Fi Hot Spot, UMTS etc. Then, the number and quality of available services may depend on the connection type (e.g., it may be unfeasible to make a good-quality Skype call with a slow GPRS connection). Finally, the environment and the user's current activity could influence the information needs (for example, the necessity to find a park in the surroundings while approaching the final destination driving a car). As a consequence, Mobile IR must take into account a notion of context, since the latter may heavily influence the relevance of the retrieved documents.

So far  there is no agreement on a precise notion of context in the research community (see Section~\ref{sec:context}). Luckily this does not mean that there are no fixed points either. Indeed, as we saw in Section~\ref{sec:context-dimensions}, there is a general consensus about including location, time, something which is amenable to the current user's activity, and social interactions into a general notion of context, at least.

In addition, context awareness has a prominent role in an emerging class of software known as Intelligent Personal Assistants (IPAs), as we anticipated in Section~\ref{sec:zero-term-queries}. Indeed, this kind of programs tend to be as much proactive as possible, in order to limit user input, automatically and partially independently searching and retrieving potentially interesting items for the user.

\chapter{Evaluation}
\label{cha:evaluation}

In this chapter we focus on Mobile IR evaluation.  After an introduction in Section~\ref{sec:importance-eval}  we report on the different approaches to evaluation.  As it is the case in the IR field, there are two classical approaches to evaluation: user studies and test collections.
At first sight, the former seems perhaps more adequate for the mobile environment, so we start with a discussion of the work based on user studies in Section~\ref{sec:user-studies}. However, also the latter has its own justifications and is perhaps gaining momentum in the last years. Section~\ref{sec:test-collection} describes the work on this direction.  We will also try to emphasise the different roles of the two approaches.

\section{The Importance of Evaluation}
\label{sec:importance-eval}

It is probably not necessary to spend much time recalling the importance of effectiveness evaluation in the IR field: most researchers, if not all of them, will agree that  evaluation is of primary importance in IR, and this is true also in a mobile setting.
Indeed, there is a quite large number of papers aiming at understanding how to evaluate the effectiveness of Mobile IR, and there is also work on evaluating its different aspects like usability, efficiency, and so on.

It must be noted that another possible approach to evaluation is on the basis of logs.
We do not address this issue in this chapter, for two reasons.
The first is that log analysis has been discussed at length in Section~\ref{sec:logs}; although there the focus is slightly different centred on understanding users and their needs rather than evaluating systems.
The second reason is space limitations. Indeed, Mobile IR evaluation is such a rich research area, that it could perhaps deserve a book on its own, and so we are forced to focus only on some of the relevant literature.

Before delving deeply into the two approaches, we should mention \citet{Tamine2010}, a methodological paper discussing evaluation of contextual IR in general and providing many useful references, as well as making specific claims, such as: (i) the need to extend the test collection approach to make it adequate for context-aware evaluation, or (ii) the need to rely on both qualitative and quantitative evaluation methods and measures.
Still from a general point of view, \citet{Sakai2012} discusses how to break down the evaluation of  the (less than) zero query task (see Section~\ref{sec:zero-term-queries}) into ``knowing what'' (i.e., evaluating system's ability to proactively issue a good query) and ``knowing when'' (i.e., evaluating system's ability to autonomously decide the right moment for issuing the query). This is a very interesting study whose implications we believe will be felt in the future.

\section{User Studies}
\label{sec:user-studies}

In this section we focus on studies centred on  user and usability. We briefly touch upon classical studies about Mobile IR effectiveness and user satisfaction, and then we  discuss some interesting issues arising when evaluating Intelligent Personal Assistants (IPAs).

\subsection{Evaluation of Mobile IR Systems}
\label{sec:evaluation-mobile-ir}

Since many user studies are presented in papers that describe both a system and its evaluation, we have already met some of those papers in previous chapters.
For example, two studies following a classical user study approach are the evaluations of Credino and SmartCREDO \cite{Carpineto2009} as well as CloudCredo \cite{Mizzaro2012} (see Section~\ref{sec:clustering}).
We can add here that those studies emphasise how the need to take into account the new kind of device (phones, tablets, etc.)
often implies adding a new independent variable in the experimental design, that therefore increases its complexity.

Also related to  the usage of clustering to present retrieval results is the work by \citeauthor{Heimonen2012} \cite{Heimonen2012,Heimonen2012a}.
He performs an in depth evaluation, by means of a longitudinal user study, of a clustering interface.
Results confirm that clustering can be useful, especially in specific situations like, for example, when the user is uncertain on how to express the information need, or requires to narrow the search, or when the information need itself is multi-faceted and it is difficult to  express with a single, unidimensional, ranking.

Another classical study is by \citet{Jones2002} (see Section~\ref{sec:text-output}); the rapid obsolescence of the results due to the fast technology development is of course another critical aspect to be taken into account also for evaluation studies.

An evaluation specific user study is by \citet{Guo2011}, who analyse how some interaction signals (e.g., number of browsed pages, zooming, scrolling/sliding, orientation change) can improve the prediction of searcher success and satisfaction.
They conduct a user study involving ten participants who used Google on a mobile phone and had to perform eight mobile search tasks.
Results show how the accuracy of success and satisfaction significantly increases when considering such signals, up to an almost 80\% accuracy (from a baseline around 50\%).

Often, with mobile search there is the problem of detecting ``good abandonments'' among ``abandoned queries'' (i.e., queries with no clicks on the results page).
\citet{Arkhipova2014} propose a new offline metric in order to properly measure quality evaluation of mobile web search, taking good abandonment rate into consideration.
\citet{Williams2016a} present a novel approach based on gesture features in order to appropriately measure users' satisfaction and detect good abandonments.
Another recent study about this issue is reported by the same authors in~\cite{Williams2016b}.
Finally, answer-like results are studied by~\citet{Lagun2014} with a particular focus on the measurement of attention and satisfaction of users.

\subsection{Evaluation of IPAs}
\label{sec:eval-ipas:-cogn}

As we have already discussed in Sections~\ref{sec:spoken} and~\ref{sec:zero-term-queries}, using traditional input methods in mobile devices is practically unfeasible or even impossible under certain circumstances.
Hence, there is a sort of renaissance of research about speech recognition algorithms for getting input from users (in both reactive and proactive systems) as well as about audio-based presentations of search results.
The wide adoption of IPAs is a practical outcome of this research areas.
However, experimental results about the cognitive load imposed on the user as well as his satisfaction about the effectiveness of this kind of solutions are mostly characterised by uncertainty, confusion and, in some cases, negative feedback.

For example, \citet{Strayer2015} conducted a thorough set of experiments on behalf of the AAA Foundation for Traffic Safety, in order to assess the impact of voice-based user-machine interactions on the cognitive workload of car drivers with the three most popular IPAs (i.e., Siri, Google Now, and Cortana).
Contrary to intuition, they registered a high workload imposed on users during such voice-based interactions.
In particular, drivers were subject to mentally demanding workloads (similar to those relative to OSPAN tasks) when interacting with the voice-enhanced native interface of the vehicle information system.
The differences between the three used IPAs were primarily due to (i) the number of system errors, (ii) the time to complete an action, and (iii) the complexity and intuitiveness of the device.
Google Now turned out to be the less demanding system w.r.t.\
the cognitive workload imposed on the user, while Siri and Cortana were equivalent under this point of view.
Similar findings about the cognitive load and the limitations of speech-only channels are illustrated by \citet{Trippas2015}  focussing  on the impact of the length of result summaries in speech-based web search.
Indeed, at least for non ambiguous queries (i.e.~single-facet queries) users prefer shortened audio summaries, probably because they do not require a significant workload for memory skills.

An important open research issue about IPAs is finding a methodology to evaluate their performances, focussing on user satisfaction.
Since users have to pay a certain price in terms of cognitive load to interact with such assistants, it is important that their efforts are considered satisfactory.
As noticed by \citet{Jiang2015}, the problem is not trivial since IPAs are complex programs with, at least, three key features (or task types): (i) the capability of understanding and executing voice commands on behalf of the user in the style of dialog-based interactions, (ii) the possibility to search for web content, with voice input queries (and also text queries for Cortana and Google Now, for instance), and (iii) the option to chat with the user (Google Now lacks this feature so far).
A thorough evaluation methodology should measure both the overall user satisfaction and the satisfaction of the single components, taken separately.
\citeauthor{Jiang2015} are the first researchers to make such an attempt. In particular, they illustrate the development and testing of their methodology on a pre-release of Cortana, using a dataset of 70.000 sessions for the experimental evaluation.
The key idea is that the independence of the evaluation protocol from the peculiar IPA and task can be achieved only by abstracting on the semantic domains involved in users' sessions.

The study starts with the isolation of the most frequent domains of user requests from the sample data (e.g., controlling the device launching apps or playing music, communicating, asking about the weather conditions etc.).
It is obvious that, depending on the features supported by the software, such domains may vary considerably.
Hence, a traditional evaluation technique may require a lot of ground truth knowledge, and the latter has to be available for each session in every scenario.
Moreover, as the authors say ``each task may have its own procedure and a unique form of correct answers''.
Thus, in order to avoid an overwhelming amount of work, it is important to train a suitable classifier over previously acquired data, in order to make it able to separate satisfactory sessions from unsatisfactory ones.
The ground truth about a session being satisfactory or not is much easier to assess.
Such classifier will then predict user satisfaction in new sessions by analysing the related action sequences (without the need to refer to features of a peculiar domain).
In conclusions, their proposed methodology consists in the following phases: (i) classify user sessions into the above mentioned three task types: dialog-based interactions (called device+dialog function in their paper), web search, and chat tasks; (ii) train separate evaluation models for each task of the three kinds of tasks; and (iii) evaluate sessions using the proper trained model, predicting user satisfaction and the quality of automated speech recognition and intent classification.

\section{Test Collections}
\label{sec:test-collection}

As already anticipated, the user study approach to evaluation seems more natural for the Mobile IR scenario, due to the intrinsic high interactivity of Mobile IR systems, to the important role assumed by the user, and to the nature of context. However, the alternative approach to evaluation in IR, namely the test collection approach, has also been widely used, has turned out to be indeed sensible in several situations, and has recently found a sort of official recognition in international evaluation initiatives like TREC and NTCIR.

\subsection{Early attempts}
\label{sec:early-attempts}

The test collection approach for Mobile IR has been proposed by \citet{Mizzaro2008}, and used in a number of papers by the same group \cite{Menegon2009,Menegon2009a,Coppola2010,Mizzaro2011} to evaluate various context-aware retrieval prototypes MoBe, CAB, and SCAB that they developed (see Section~\ref{sec:context-based-mobile} for  detailed descriptions). There are mainly three attempts to evaluation, corresponding to the three different prototypes. In their first study \citet{Mizzaro2008}  evaluated MoBe, which aimed at proactively retrieving and filtering mobile applications (MoBeLets) on one's own device on the basis of context. To this aim, they developed a small in-house test collection made up of about 500 MoBeLet descriptors (that play the role of  documents), 30 context descriptions (topics), i.e., detailed descriptions of the context the user is in, and a set of binary relevance assessments (qrels).

Both MoBeLet and context descriptors are quite rich: they contain several structured and unstructured fields. For instance, MoBeLet descriptors contain, among others, the following fields: a MoBeLet unique identifier, MoBeLet's title, a description in free text, and a list of comments provided by the users as unstructured textual information.  The context descriptors are perhaps even more complex and contain, among others: a context unique identifier; a detailed description of the context, in free text; user's activity; user's spatial position; the user posture, like ``lying down'', ``standing'', etc.; and the likelihood of the context (some contexts are uncertain).  Of course, some of the context descriptors data can be obtained automatically (e.g., GPS, time, etc.) whereas some others are more challenging, if not impossible, with  current state of the art (e.g., posture, semantic spatial position, etc.). However, for these evaluation experiments, it was assumed that the descriptors could somehow be generated. Also, it is important to have a detailed description of the context for the human judge to correctly evaluate  relevance.

Although the results were not conclusive, given the small size of the collection due to the low numbers of both context and MoBeLets descriptors, they hinted that: some fields seem more effective than others;  structured and unstructured fields have complementary effects; and some fields, and/or the related matching function, should be revised (location, activity, etc.). In conclusions, the most  interesting contribution by this paper is probably the proposal to use the benchmark approach for Mobile IR evaluation.

In the second attempt, the same research group used the same benchmark approach in \cite{Menegon2009,Menegon2009a}, but with some differences, since the system to be evaluated is CAB (Context Aware Browser, see Section~\ref{sec:context-based-mobile}), that retrieves general Web pages and not application descriptors as MoBe.  The size of the collection is still small in number of context descriptors (only 10) and Web pages (around 3500). The context descriptor is now simpler and made up of unstructured (i.e., free text) fields only, as the example  in Figure~\ref{fig:context-descriptor} shows. The evaluation process assumes that a bag-of-words query can be automatically extracted from the ``description'' field only.  An example provided in the paper is that from the context in Figure~\ref{fig:context-descriptor}, where the query ``user just landed london heathrow international airport  looking flight timetable timetable connections london lunch time'' is derived.  The other fields are used for human relevance assessment. An ``interactive search and judge'' \cite{Cormack1998} is also used so that the document collection can be further extended in the future, and new relevance judgments are performed on unjudged documents. Judge agreement is studied and the reasonable agreement level obtained indicates that the approach appears sound and reliable.

\begin{figure}[t]
\centering
\fbox{
 \begin{minipage}{.80\columnwidth}
\small
 \texttt{$<$contextDescriptor$>$\\
    $<$title$>$ Heathrow airport $<$/title$>$\\
    $<$description$>$\\
      The user has just landed at London Heathrow international airport. He is looking at a flight timetable and at a timetable for connections to London. It is lunch time.\\
    $<$/description$>$\\
    $<$narrative$>$  \ldots  $<$/narrative$>$\\
    $<$relevance$>$\\
      A Web page is relevant: it contains information about a flight, about the means of transport to reach town, about bars and fastfoods in the airport, or it allows to book a flight. A web page that contains only one of these aspects is relevant; if it contains some links to relevant pages is partially relevant. If the judge is not able, for any reason, to judge the page, its value is ``I don't know''.\\
    $<$/relevance$>$\\
    \ldots \\
 $<$/contextDescriptor$>$
 }
 \end{minipage}
}
\caption{Part of a context descriptor (from~\cite[Figure~1]{Menegon2009a}). }
\label{fig:context-descriptor}
\end{figure}

The third and so far last attempt of this series of studies is presented by \citet{Mizzaro2011}, and aims at evaluating a social version of  CAB, named SCAB for Social CAB and described more in detail in Section~\ref{sec:context-based-mobile}. This new version exploits users's behaviour and  tagging activity to improve  context-aware retrieval of web pages.  The overall evaluation approach is similar to the previous two, and a small benchmark is  built, but with some differences. The context descriptor goes back to both structured and unstructured fields: besides a free text ``description'' field, it contains some data inferred from sensors.
The benchmark contains, besides a collection of web pages (named ``resources''), also a collection of tags, to evaluate also the capability of the ecosystem (i.e.~SCAB and its users population) of producing reasonable tags for both resources and contexts. The benchmark is constructed in two stages. In the first one the participants read the context descriptors and produce some corresponding tags for both contexts and resources. In the second stage (a session held two weeks later), each participant is shown all the tags that have been produced in the first stage and  evaluates their relevance to the corresponding context.

On the basis of these data, a battery of simulations is performed, with the aim of evaluating specific aspects of  SCAB including: cold start, overall effectiveness, how many reliable tags can be expected by a user, and the effectiveness of removing and adding tags. The results, although limited in  number of context descriptors used (only five), demonstrate a good effectiveness and  overall feasibility of the social tag-based approach.

It should be noted that evaluating systems like those discussed here is truly challenging, since the classical evaluation methodologies might be, at least in theory, inadequate for the evaluation of systems designed for novel tasks, like proactive context-aware retrieval. More generally, it might even be that the high importance given to evaluation in  IR is somehow hindering the development of novel systems. This might also explain the low number of Mobile IR studies published in  core IR literature, where usually a solid evaluation is required.

\subsection{TREC and NTCIR}
\label{sec:trec-ntcir}

The test collection approach for evaluating Mobile IR has been used in TREC (Text REtrieval Conference\footnote{\url{trec.nist.gov}})
and NTICIR (NII Testbed and Community for Information access Research\footnote{\url{research.nii.ac.jp/ntcir/index-en.html}}), two international evaluation initiatives following the most classical approach to IR effectiveness evaluation, namely the test collection approach.

The TREC Contextual Suggestion Track\footnote{\url{https://sites.google.com/site/treccontext/}} that started in 2012 \cite{Dean-Hall2012}, was run also in 2013   \cite{Dean-Hall2013},  2014 \cite{Dean-Hall2014}, 2015 \cite{Dean-Hall2015}, and 2016 \cite{Hashemi2017}. The overall aim of the track is to study evaluation of Mobile IR systems aimed at satisfying information needs that are highly dependent on context and user interests. More specifically, as stated right in the introduction of \cite{Dean-Hall2012}, ``Future information retrieval systems must anticipate user needs and respond with information appropriate to the current context without the user having to enter an explicit query\ldots In a mobile context such a system might take the form of an app that recommends interesting places and activities based on the user’s location, personal preferences, past history, and environmental factors such as weather and time\ldots''. As we can see, there is a clear relationship with Mobile IR in general and context awareness in particular (see Chapter~\ref{cha:context-awareness}).

As it is sometimes the case with TREC, the terminology is somewhat odd and needs some effort to be understood. The track general organisation is as follows. The participants to the track are given a set of \emph{example} suggestions and a set of user \emph{profiles}. Each profile corresponds to a user and contains the information on example suggestions that the user liked or disliked in the past. Participants are supposed to exploit examples and profiles to build an understanding of each user's preferences. Then, participants are given  some  descriptions of contextual information needs (named \emph{contexts}), and the task is to retrieve, for each context/profile pair, a ranked list of documents (named \emph{suggestions}). After some usual pooling and judging (since  2013, also using Amazon Mechanical Turk crowdsourcing platform), the effectiveness of each run is measured by different effectiveness metrics (i.e.~Mean Reciprocal Rank, Precision at 5, and Time Based Gain)  taking into account the relevance of the first 5 retrieved suggestions to (a subset of) profile/context pairs.

Contexts are simplified descriptions of real contextual needs. In the 2012 edition, each context represented a geo-temporal location, and it includes city, day of the week, time of day, and season. Comparing Figure~\ref{fig:context-example} with Figure~\ref{fig:context-descriptor} we can see that it is definitely simpler. Moreover, from 2013, the temporal information has been discarded, because assessors found that it made relevance judgments more problematic, and context corresponds only to a location. Finally, the location is quite coarse grained since it is corresponds to a city, and the 50 contexts are therefore simply US cities randomly chosen from  Wikipedia.

\begin{figure}[t]
\centering
\fbox{
 \begin{minipage}{0.95\columnwidth}
\small
 \texttt{$<$context number = ``1''$>$\\
    ~~$<$city$>$   New York  $<$/city$>$ \\
    ~~$<$state$>$   NY $<$/state$>$\\
    ~~$<$lat$>$   40.71427 $<$/lat$>$\\
    ~~$<$long$>$   -74.00597  $<$/long$>$ \\
    ~~$<$day$>$  weekday  $<$/day$>$\\
    ~~$<$time$>$  afternoon  $<$/time$>$\\
    ~~$<$season$>$  summer  $<$/season$>$\\
 $<$/context$>$
 }
 \end{minipage}
}
\caption{Example of a context from the TREC Contextual Suggestion Track  (from~\cite[Listing~3]{Dean-Hall2012}. }
\label{fig:context-example}
\end{figure}

The organisational workflow is quite complex and aims at taking into account both user's current context and past preferences.
The overall setting of the track  remained mainly unchanged, a clue that the organisers are satisfied with the current design, although there have been some minor variations over the years.
For example, in the last years, the collection is the Fixed TREC Contextual Suggestion corpus, instead of the live Web, with the aim of making the data more reusable. Also more evaluation metrics are used (e.g., MAP, NDCG, bpref, Rprec). Finally,  the track has been organised with a two phase setup.
The five tracks are summarised in Table~\ref{tab:CST}, together with the  numbers of participants, that has been overall increasing over the years, witnessing the good success of the track.
Track participants have exploited several knowledge sources available on the Web like Foursquare, Facebook Places, Google Places API, TripAdvisor, Yelp, and so on.

\begin{table}[tb]
\centering
\small
\begin{tabular}{l c lll c ll}
  \toprule
                    &  & \textbf{2012} & \textbf{2013} & \textbf{2014} && \textbf{2015} & \textbf{2016} \\
  \midrule
 \multicolumn{2}{l}{\textbf{Track organization}} \\
  ~Setup&&&&&&\multicolumn{2}{l}{\emph{---two phases setup---}}\\

  ~Contexts          &  & (geo + time)     & (geo only)     & (geo only) & &(geo only)     & (geo only)   \\
  \midrule
  \addlinespace
  \textbf{Collection}        &  & Web           & Web,          & Web,     && \multicolumn{2}{l}{\emph{---Fixed TREC Contextual~~~}} \\
  &  &               & ClueWeb12     & ClueWeb12   & & \multicolumn{2}{l}{\emph{~~~Suggestion Web Corpus---}}\\
    \midrule
  \addlinespace
  \textbf{Metrics}   &  & MRR, P@5 & MRR, P@5, & MRR, P@5, &&  P@5, MRR, & P@5,10, MRR, NDCG,\\
  &  &          & TBG       & TBG       & &TBG &   MAP, bpref, Rprec\\
  \midrule
  \addlinespace
  \textbf{Participants} \\
  ~Groups          &  & 14            & 19            & 17  &&    6+10    &8+13  \\
  ~Runs            &  & 27            & 34            & 31    &&   9+21  &  15+30 \\
  \bottomrule
\end{tabular}
\caption{TREC Contextual Suggestion tracks data.}
\label{tab:CST}
\end{table}

The relevance of this track to Mobile IR is clear and  explicit. However, at least in principle, mobile could have been emphasised more, for example by using a smaller granularity for locations. Also, although personalisation is of course of primary importance in Mobile IR, it would make sense to try to disentangle it from purely contextual suggestions. Finally, a richer description of the context should perhaps be considered as well.

At the same time NTCIR, the TREC-like conference focusing on East Asian languages,  launched the 1CLICK ``pilot'' task at NTCIR-9 \cite{Sakai2011}. Its aim is to satisfy the user with a ``single textual output'', right after the click on the Search button, rather than providing the usual ten blue links.
The link with summarisation (see Section~\ref{sec:summarization}), related to devices having smaller screen space is obvious.
The first edition of 1CLICK was not very successful in terms of the number of participants (only 3, with 10 runs), but the task  continued as 1CLICK-2 at NTCIR-10 (10 participants, 59 runs) \cite{Kato2013} and as MobileClick-2 Task at NTCIR-12 (12 participants, 66 runs) \cite{Kato2016}, with increasing interest.
The MobileClick-2 task is a bit different, since the aim is to provide a two-layered output: a first layer/page with the most important information, plus an outline of the additional information contained in the second layer, that can be accessed by clicking on the links in the first page.

 As for the Contextual Suggestion Track, we can observe that although the task is clearly related to Mobile IR, it focuses  only on one specific aspects, providing summarised information. In addition, it is not clear whether the available test collections are indeed representative of a new real document collections for Mobile IR (see Chapter~\ref{cha:docum-coll}).

\section{Conclusions}
\label{sec:conclusions-8}

This chapter provides a brief discussion of evaluation issues in Mobile IR.
The classical user study approach (Section~\ref{sec:user-studies}) it probably the most natural to assess both effectiveness and user satisfaction of Mobile IR users (Section~\ref{sec:evaluation-mobile-ir}).
An important aspect concerns the evaluation of novel interaction modalities like those found in IPAs (Section~\ref{sec:eval-ipas:-cogn}).
Evaluation based on test collections, although perhaps less natural, seems anyway the most productive approach (Section~\ref{sec:test-collection}).
The attempts made so far  fell short both on the scale (Section~\ref{sec:early-attempts}) and on the rather limited view of Mobile IR, interpreted either as personalisation plus location or as summarisation (Section~\ref{sec:trec-ntcir}), providing a rather limited view of Mobile IR in context (see instead Chapter~\ref{cha:context-awareness}).

\chapter{Conclusions and Outlook}
\label{cha:conclusions-outlook}

This chapter concludes this book with, in Section~\ref{sec:conclusions},  a summary of the work reported and, in Section~\ref{sec:outlook}, a brief outlook of the future of Mobile IR that is likely be become obsolete very quickly, given the speed with which research in this area is evolving. 

\section{Summary}
\label{sec:conclusions}

This book reviewed the state of the art of research in Mobile IR. 
It started with highlighting the differences between IR and Mobile IR, reviewing also the foundations of Mobile IR research. 
It then continued looking at the different kinds of documents, users, and information needs that can be found in the Mobile IR world and that characterise it as different from standard IR.
It then focussed on two important issues, namely user interfaces and context-awareness.
Finally it concluded with looking at the issues related to evaluation of Mobile IR applications. 

This is short book on a topic that would have deserved much more space. 
In fact, although we reviewed more than 200 papers, there are many topics that were only briefly covered or other that we did not cover at all, like for example issues related to devices, hardware, or privacy, just to mention a few. 
However, the purpose of this book was not to provide full coverage of all the work in this area, but simply to indicate the most interesting and influential contributions that have appeared in recent years. 
Of course, the view is only partial and biased by our experience and our editorial choice. 
Obviously, we take full responsibility of anything that has not been faithfully presented or that has been left out. 
Despite all these limitations, we believe it will provide a valuable tool for helping new and old researchers approaching this exciting area of research that will certainly grow and expand exponentially in coming years.

\section{Outlook: Mobile IR Trends}
\label{sec:outlook}

As they say, ``prediction is very difficult, especially about the future''. 
This is certainly true, but in our opinion, some Mobile IR trends seem clear. 
To mention just a few: the usage of mobile devices will likely continue to increase, when compared to classical desktop computers; more devices will come into play (starting with smart glasses, smart watches, and activity trackers, that are already on the market today); more powerful sensors will allow richer and more detailed input (paving the way to fruitful synergies with techniques borrowed from Big Data  and Machine Learning), and increase the importance of context awareness; a better interoperability among devices will allow more effective results presentation and truly cross-device search. 
It is also quite natural to imagine that real world objects will become searchable, this includes people (e.g., friends on a social network that happen to be nearby) as well as things, thus foreseeing an interesting integration between the Internet of Things world and Mobile IR; also evaluation is likely to change soon, to become more comprehensive and reliable.

Besides these,  more radical and unforeseeable evolutions are lurking out there.  We do not think that we were able to capture them in this short book, but we  hope  that this work will be useful, for future researchers, as a foundation of, and an entry point to, the interesting, lively, and very dynamic Mobile IR field.

\backmatter

\bibliographystyle{plainnatnodoi}
\bibliography{MobileIR}

\end{document}